\documentstyle[a4,epsf,12pt]{article}
\newcommand{\ep}{\varepsilon}
\newcommand{\Li}[2]{{\mbox{Li}}_{#1\!}\left(#2\right)}
\newcommand{\Cl}[2]{{\mbox{Cl}}_{#1}\left(#2\right)}
\newcommand{\Gl}[2]{{\mbox{Gl}}_{#1}\left(#2\right)}
\newcommand{\Ls}[2]{{\mbox{Ls}}_{#1}\!\left(#2\right)}
\newcommand{\LS}[3]{{\mbox{Ls}}_{#1}^{(#2)}\!\left(#3\right)}
\newcommand{\Lsc}[2]{{\mbox{Lsc}}_{#1\!}\left(#2\right)}
\newcommand{\tfrac}[2]{{\textstyle{\frac{#1}{#2}}}}

\begin{document}

\renewcommand{\thefootnote}{\fnsymbol{footnote}}
\thispagestyle{empty}

\begin{flushleft}
 {DESY 01-065} \hspace{9cm} {hep-th/0012189} \\[3mm]
 {MZ-TH/00-52} \\[3mm]
 {January 2001}
\end{flushleft}

 \vspace*{2.0cm}
 \begin{center}
 {\large \bf
 New results for the $\ep$-expansion of certain \\[2mm] 
 one-, two- and three-loop Feynman diagrams}
 \end{center}
 \vspace{1cm}
 \begin{center}
 A.~I.~Davydychev$^{a,}$\footnote{On leave from
 Institute for Nuclear Physics, Moscow State University,
 119899, Moscow, Russia. Email address:
 davyd@thep.physik.uni-mainz.de}
 \ \ and \ \
 M.~Yu.~Kalmykov$^{b,}$\footnote{On leave from BLTP, JINR,
                 141980 Dubna, Russia.
Email address: kalmykov@ifh.de}
\\
 \vspace{1cm}
$^{a}${\em
 Department of Physics,
 University of Mainz, \\
 Staudingerweg 7,
 D-55099 Mainz, Germany}
\\
\vspace{.3cm}
$^{b}${\em
 DESY--Zeuthen,
Theory Group, Platanenallee 6, \\ D-15738 Zeuthen, Germany}
\\
\end{center}
 \hspace{3in}
 \begin{abstract}
For certain dimensionally-regulated
one-, two- and three-loop diagrams, problems
of constructing the $\ep$-expansion and the analytic continuation
of the results are studied.
In some examples, an arbitrary term of the $\ep$-expansion
can be calculated. For more complicated cases, only a few 
higher terms in $\ep$ are obtained.
Apart from the one-loop two- and three-point diagrams,
the examples include two-loop (mainly on-shell) propagator-type diagrams
and three-loop vacuum diagrams. 
As a by-product, some new relations involving Clausen function,
generalized log-sine integrals and certain
Euler--Zagier sums are established, and some useful results for
the hypergeometric functions of argument $\tfrac{1}{4}$
are presented. 
\end{abstract}


\newpage

\renewcommand{\thefootnote}{\arabic{footnote}}
\setcounter{footnote}{0}

\section{Introduction}
\setcounter{equation}{0}

One of the most powerful tools used in loop calculations
is dimensional regularization~\cite{dimreg}.
In some cases, one can derive results valid for an arbitrary
space-time dimension $n=4-2\varepsilon$, usually in terms of various   
hypergeometric functions.
For practical purposes the coefficients
of the expansion in $\ep$ are important.
In particular, in multiloop calculations higher terms
of the $\ep$-expansion of one- and two-loop functions are needed, since
one can get contributions where these functions are
multiplied by poles in $\ep$.
Such poles may appear not only due to factorizable loops,
but also as a result of application of the integration by parts \cite{ibp}
or other techniques \cite{Tarasov96-rec}.   

One of the important examples are massless propagator-type
diagrams occurring in the renormalization group calculations.
By now, the structure of the terms of the $\ep$-expansion 
of such diagrams is well understood.
As a rule, the occurring transcendental numbers 
can be expressed in terms of multiple Euler--Zagier 
sums \cite{Euler-Zagier}\footnote{An on-line calculator for the 
Euler sums, with an accuracy of 100 decimals, is available at
{\tt http://www.cecm.sfu.ca/projects/EZFace/} .}, 
\begin{equation}
\zeta(s_1,\ldots, s_k; \; \sigma_1,\ldots, \sigma_k) 
= \sum_{n_1>n_2> \ldots >n_k>0}\;\;\; \prod_{j=1}^{k}
\frac{(\sigma_j)^{n_j}}{n_j^{s_j}},
\label{euler}
\end{equation}
where $\sigma_j=\pm 1$ and $s_j>0$. For lower cases, these sums
correspond to ordinary $\zeta$-functions. For studying the higher-order 
sums analytical and numerical methods have been recently 
developed~\cite{mathematics}. It was demonstrated \cite{knot}
that there is a ``link'' between the quantum field theory and the knot theory
\footnote{A collection of results related to this connection can be found on 
Dirk Kreimer's home page,
{\tt http://dipmza.physik.uni-mainz.de/$\;\widetilde{}\;$kreimer/ps/knft.html} 
or on David Broadhurst's home page
{\tt http://yan.open.ac.uk/$\;\widetilde{}\;$dbroadhu/knft.html} . }:
some Feynman diagrams can be connected with knots, so that 
the values (Euler--Zagier sums) of Feynman diagrams are also
associated with knots. 

In multiloop massive calculations in QED and QCD \cite{QED} within the 
on-shell scheme, new constants appear, which are related to polylogarithms.
A detailed description 
of the basis of this type is presented in Ref.~\cite{euler-basis}. 
Two-loop vacuum diagrams with equal masses \cite{massive,DT1} yield 
the transcendental number $\Cl{2}{\frac{\pi}{3}}$, where $\Cl{j}{\theta}$
is the Clausen function (\ref{Cl_j}).
Some useful properties of this function are collected in Appendix~A
(see also in \cite{Lewin}).
The same constant $\Cl{2}{\frac{\pi}{3}}$ 
appears in the one-loop 
off-shell three-point diagram with massless internal lines, in the
symmetric case when all external momenta squared are equal, see
in~\cite{CG}.

To classify new constants appearing in single-scale massive diagrams, 
Broadhurst has introduced in Ref.~\cite{B99} 
the ``sixth root of unity'' basis connected with 
\begin{equation}
\zeta\left(\begin{array}{rcr}s_1\;&\ldots&s_k\;\\
\lambda^{p_1}&\ldots&\lambda^{p_k}\end{array}\right) =
\sum_{n_1>n_2> \ldots >n_k>0}\;\;\; 
\prod_{j=1}^k \frac{\lambda^{p_j n_j}}{n_j^{s_j}}
\label{sixth}
\end{equation}
where $\lambda=\exp\left({\rm i}\tfrac{\pi}{3}\right)$ 
and $p_j\in\{0,1,2,3,4,5\}$. 
For particular cases $p_j \in \{0,3\}$ it coincides with 
the Euler--Zagier sums (\ref{euler}).
For both definitions (\ref{euler}) and (\ref{sixth}), the
{\em weight} can be defined as $\sum_{i=1}^{k}s_i$, whereas
the value of $k$ can be associated with 
the {\em depth} (see in \cite{euler-basis,B99}).

One of the remarkable results of Ref.~\cite{B99} is that all 
finite parts of three-loop vacuum integrals without subdivergences, 
with an arbitrary distribution of massive and massless lines, 
can be expressed in terms of four weight-{\bf 4} constants:
$\zeta_4$, $\left[\Cl{2}{\frac{\pi}{3}}\right]^2$, 
\begin{equation}
\label{U31}
U_{3,1} = 
- 2 \Li{4}{\tfrac{1}{2}} + \tfrac{1}{2} \zeta_4
- \tfrac{1}{12} \ln^4 2 + \tfrac{1}{2} \zeta_2 \ln^2 2 
\end{equation}
(see Eq.~(55) of \cite{B99}) 
and, finally,
\begin{equation}
V_{3,1} = \sum_{p>k>0}
\frac{(-1)^p}{p^3 k}\; \cos\left(\tfrac{2}{3}\pi k\right) \; ,
\label{v31}   
\end{equation}
which is an essentially new constant.
The same constants appear in the $(2-2\ep)$-dimensional Feynman integrals 
\cite{trans}, as one should expect due to algebraic relations
between diagrams with shifted dimension \cite{Tarasov96-rec}.
We also note that $\left[\Cl{2}{\frac{\pi}{3}}\right]^2$ appears
in the two-loop non-planar three-point diagram~\cite{UD3}, 
when internal lines are massless, whereas all external
momenta squared are off shell and equal.
Unfortunately, a large number of elements (more than 4000) makes it   
difficult to define the complete basis of ``sixth root of unity'' at the
weight ${\bf 4}$. So far, only the cases with the depth $k\le 2$ 
have been examined~\cite{B99}.

In Refs.~\cite{BB99,BBK99}, the {\em central binomial sums} were
considered,
\begin{equation}
S(a) \equiv \sum_{n=1}^\infty \frac{(n!)^2}{(2n)!} \frac{1}{n^a} \; .
\label{cbs}
\end{equation}
It was shown that these sums are connected with
the {\em multiple Clausen values}, 
which are defined as the real or imaginary
part of the multi-dimensional
polylogarithm \cite{MDP}, 
\begin{equation}
\Li{a_1, \ldots, a_k}{z} =
\sum_{n_1>n_2> \ldots >n_k>0}\;\;\;
\prod_{j=1}^k 
\frac{z^{n_1}}{n_1^{a_1} \ldots n_k^{a_k}} \; ,
\end{equation}
taken at $z=\exp\left({\rm i}\tfrac{\pi}{3}\right)$
(i.e., the ``sixth root of unity'')\footnote{An on-line calculator 
for the multiple Clausen values, 
with an accuracy of 100 decimals, can be found at 
{\tt http://www.cecm.sfu.ca/projects/ezface+/} .}. 
In particular (see Theorem~1 in \cite{BBK99}),
\begin{equation}
S(8) = -4\pi \mbox{Im}\left[ \Li{6,1}{e^{{\rm i}\pi/3}} \right]
+ \tfrac{3462601}{233280} \zeta_8
-\tfrac{14}{15} \zeta_{5,3} - \tfrac{38}{3} \zeta_5 \zeta_3
+ \tfrac{2}{3} \zeta_2 \zeta_3^2,
\end{equation}
where we use the following short-hand notation (see Eq.~(\ref{euler}))
\begin{equation}
\zeta_{s_1, \ldots , s_k} = \zeta(s_1, \ldots , s_k; 1, \ldots, 1) \; .
\label{shortzeta}
\end{equation}
The constant $\zeta_{5,3}$ has occurred in 
the 6-loop calculation of the $\beta$-function
in $\phi^4$-theory \cite{zeta53:1} (this was recently confirmed in 
\cite{zeta53:3}). 
At the 7-loop order, a new transcendental number arises,
$\zeta_{3,5,3}$ \cite{zeta53:1}.
The $\zeta_{5,3}$ and another constant, $\zeta_{7,3}$, appear in the 
calculation of anomalous dimensions at ${\cal O}(1/N^3)$ in the 
large-$N$ limit \cite{zeta53:2}. A remarkable property of these
constants is their connection with knots\footnote{Below we shall 
discuss connection of these constants with the {\em odd} basis.}
\cite{knot,zeta53:1,zeta53:2}: 
namely, the torus knots $8_{19}$ and $10_{124}$ are associated with 
$\left[ 29 \zeta_8 - 12 \zeta_{5,3} \right]$ and
$\left[ 94 \zeta_{10} - 793 \zeta_{7,3} \right]$, 
respectively,
whereas $\zeta_{3,5,3}$ is associated with a certain
hyperbolic knot 
\cite{zeta53:1,zeta53:2}.

To predict types of functions (and the values of their arguments)
which may appear in higher
orders of the $\ep$-expansion, a geometrical approach \cite{DD}
happens to be very useful. 
Using this approach, the results for {\em all} terms of the
$\ep$-expansion have been obtained for the one-loop two-point
function with arbitrary masses \cite{Crete,D-ep}.
Moreover, {\em all} terms have been also obtained for
the $\ep$-expansion of one-loop
three-point integrals with massless internal lines and 
arbitary (off-shell) external momenta and two-loop 
vacuum diagrams with arbitrary masses
\cite{D-ep,bastei_ep}, which are related to each other, due to 
the magic connection \cite{DT2}.
All these results have been represented in terms of the
log-sine integrals (see in \cite{Lewin} and Appendix~A.1 
of this paper), whose angular arguments have a rather
transparent geometrical interpretation (angles of certain
triangles). In more complicated cases, like, e.g., the
three-point function with general values of the momenta 
and masses, an arbitary term of the $\ep$-expansion can
be represented in terms of one-fold angular integrals
whose parameters can be related to the angles accociated
with a four-dimensional simplex.

In Ref.~\cite{FKK99}, the on-shell values of
two-loop massive propagator-type integrals have been studied,
and it was observed that the finite (as $\ep\to 0$) parts of all 
such integrals without 
subdivergences can be expressed in terms of three 
weight-{\bf 3} constants, 
two for the real part, $\zeta_3$ and $\pi\Cl{2}{\frac{\pi}{3}}$, 
and one for the imaginary part, $\pi\zeta_2$.

Furthermore, in Ref.~\cite{FK99} an ansatz was elaborated 
for constructing the ``irrationalities'' 
occurring in the $\ep$-expansion of single-scale diagrams involving
cut(s) with two massive particles. This construction is 
closely related to the geometrically-inspired all-order 
$\ep$-expansion of the one-loop 
propagator-type diagrams \cite{Crete,D-ep}, which was also 
useful when fixing
the normalization factor $\tfrac{1}{\sqrt{3}}$.
The procedure of constructing the ansatz is as follows:
for each given weight ${\bf j}$ 
the set $\{b_j\}$ of the {\em basic} transcendental numbers 
contains (i) all products of the lower-weight elements 
$\{b_{j-k} b_k\}$, $k=1,2, \cdots, j-1$ and
(ii) a set of new (non-factorizable) elements
$\{\widetilde{b}_j \}$, which are associated with the quantities
arising in the real and imaginary parts of the polylogarithms 
$\Li{j}{e^{{\rm i}\theta}}$ and $\Li{j}{1-e^{{\rm i}\theta}}$, 
with $\theta=\tfrac{\pi}{3}$ or $\theta=\tfrac{2\pi}{3}$.

The real and imaginary parts of such polylogarithms 
can be expressed in terms of the Clausen function $\Cl{j}{\theta}$
(\ref{Cl_j}),
log-sine integrals $\Ls{j}{\theta}$ (\ref{log-sin}) and
generalized log-sine integrals $\LS{j}{k}{\theta}$ (\ref{log-sin-gen})
(see also in~\cite{Lewin}). 
Note that $\Ls{2}{\theta}=\Cl{2}{\theta}$.
The relevant relations are collected in Appendix~A.1, 
Eqs.~(\ref{Cl_j_}) and (\ref{Lewin+}).
Therefore, in our case 
the non-factorizable part of the basis can be expressed
in terms of the functions (\ref{Cl_j}),
(\ref{log-sin}) and (\ref{log-sin-gen})
of two possible angles, 
$\theta=\tfrac{\pi}{3}$ and $\theta=\tfrac{2\pi}{3}$.
It should be noted that this basis is not
uniquely defined, since there are several relations between 
polylogarithmic functions $\Cl{j}{\theta}$, $\Ls{j}{\theta}$ and
$\LS{j}{k}{\theta}$ of these arguments
(see, e.g., 
Eqs.~(\ref{pi/3_zoo1})--(\ref{pi/3_zoo3}) and (\ref{pi/2_zoo}) 
of this paper).
After excluding all linearly-dependent terms, the basis 
contains the following non-factorizable constants:
$\Ls{j}{\frac{2\pi}{3}}$, where $j=3,4,5$;
$\Ls{j}{\frac{\pi}{3}}$ for  $j=2,4,5$; 
$\LS{j}{1}{\frac{2\pi}{3}}$ for $j=4,5$; 
and $\LS{5}{2}{\frac{2\pi}{3}}$. This set of elements will be called 
the {\em odd} basis \cite{FK99}. The numerical values of these 
constants are given in Appendix~A of \cite{FK99}.  

One can use the {\sf PSLQ} algorithm \cite{PSLQ} 
to search for a linear relation between the given term of
the $\ep$-expansion of the diagram of interest and the set 
of numbers $\{b_j\}$. Using this procedure, it was possible to find
results for several two- and three-loop single-scale
diagrams \cite{FK99}, some of them have been also calculated 
analytically in \cite{D-ep,ChS00}.
The constructed {\em odd} basis 
has an interesting property: the number $N_j$ of the basic
irrational constants
of a weight ${\bf j}$ satisfies a simple ``empirical'' relation $N_j = 2^j$,
which has been checked up to weight ${\bf 4}$. 
We note that the constant $V_{3,1}$ (given in Eq.~(\ref{v31}))
can be expressed in terms of the weight-{\bf 4} elements
of the {\em odd} basis, 
\[
V_{3,1} = \tfrac{1}{3} \left[ \Cl{2}{\tfrac{\pi}{3}} \right]^2
- \tfrac{1}{4} \pi \Ls{3}{\tfrac{2\pi}{3}}
+ \tfrac{13}{24} \zeta_3 \ln 3
- \tfrac{259}{108} \zeta_4
+ \tfrac{3}{8} \LS{4}{1}{\tfrac{2\pi}{3}} \; .
\]
(see Eq.~(12) in Ref.~\cite{FK99}).
For the weight ${\bf 5}$, 
32 linearly-independent elements were elaborated in Ref.~\cite{FK99}. 
It should be noted, however, that  some of these irrational numbers, 
like $\LS{5}{1}{\frac{2\pi}{3}}$,
so far have not appeared in the results for Feynman 
integrals\footnote{The only known integral containing
$\LS{4}{1}{\tfrac{2\pi}{3}}$ is ${\bf D_5}(1,1,1,1,1,0)$ \cite{FK99}. 
The integral ${\bf D_5}(1,1,1,1,1,1)$ is reduced to it by
using recurrence relations \cite{leo96}.
It is natural to  expect that its $\ep$-term may
contain $\LS{5}{1}{\tfrac{2\pi}{3}}$.}.
So, the question about the completeness of the set of
weight-${\bf 5}$ basis elements is still open.
We are going to re-analyze this basis below.

By analogy with the
{\em odd} basis introduced in~\cite{FK99}, it is possible to consider
the {\em even} basis connected with the angles $\tfrac{\pi}{2}$ 
and $\pi$. 
Apart from the well-known elements $\pi$, $\ln2$, $\zeta_j$ and the
Catalan's constant\footnote{Note that 
$G=\Cl{2}{\tfrac{\pi}{2}}=\Ls{2}{\tfrac{\pi}{2}}$, see in~\cite{Lewin}.} 
$G$, this basis also contains (up to the weight~{\bf 5}) 
$\Li{j}{\tfrac{1}{2}}$ ($j=4,5$, see also in~\cite{euler-basis}), 
$\Ls{j}{\tfrac{\pi}{2}}$ ($j=3,4,5$)\footnote{An example of a physical
calculation where the constant $\Ls{3}{\tfrac{\pi}{2}}$ arises is given in
\cite{ls3_pi/2}.}, 
$\Cl{4}{\tfrac{\pi}{2}}$ 
and $\LS{5}{2}{\tfrac{\pi}{2}}$.
Instead of $\Li{4}{\tfrac{1}{2}}$ and $\Li{5}{\tfrac{1}{2}}$, 
one could take, e.g.,
$\LS{4}{1}{\tfrac{\pi}{2}}$ and $\LS{5}{1}{\tfrac{\pi}{2}}$
(or $\LS{4}{1}{\pi}$ and $\LS{5}{1}{\pi}$), 
using the relations between these elements presented
in Appendix~A, Eq.~(\ref{pi/2_zoo}).
For the {\em even} basis, the same ``empirical''
relation $N_j=2^j$ is valid up to weight ${\bf 4}$. However, at the weight 
${\bf 5}$ only 30 independent elements have been found so far. 
This issue will be also discussed below.

Later, in Ref.~\cite{KV00}, an interesting connection 
between the function $S(a;z)$ associated with the central binomial sums 
(\ref{cbs}) and the generalized log-sine integrals was established,
\begin{equation}
S(a;z) \equiv
\sum_{n=1}^\infty  \frac{(n!)^2}{(2n)!} \frac{z^n}{n^a}
= - \sum_{j=0}^{a-2} \frac{(-2)^j}{(a-2-j)!j!}
 \,( \ln z )^{a-2-j}\, \LS{j+2}{1}{2\arcsin\frac{\sqrt{z}}{2}} \; ,
\label{binomial1}
\end{equation}
where $a\geq 2$. In particular, the sums (\ref{cbs}) can be represented as
\begin{equation}
S(a) \equiv S(a;1) =
\sum_{n=1}^\infty \frac{(n!)^2}{(2n)!} \frac{1}{n^a}
= - \frac{(-2)^{a-2}}{(a-2)!} \LS{a}{1}{\tfrac{\pi}{3}}\, ,
\label{lsJ1}
\end{equation}
whereas $S(a;3)$ is related to $\LS{a}{1}{\tfrac{2\pi}{3}}$ 
(plus a combination of lower terms with $\ln3$).
Therefore, $S(a;1)$ and $S(a;3)$ are connected with the
terms of the {\em odd} basis. Furthermore, considering
$S(a;2)$ and $S(a;4)$ we obtain $\LS{a}{1}{\tfrac{\pi}{2}}$
and $\LS{a}{1}{\pi}$ (plus a combination of lower terms with $\ln2$),
respectively, which are related to the {\em even} basis.

Moreover, in Ref.~\cite{KV00}
a more general case of multiple binomial sums 
has been examined,
\begin{equation}
\Sigma_{a_1,\ldots,a_p; 
\; b_1,\ldots,b_q;c}^{\; i_1,\ldots,i_p; \;j_1,\ldots,j_q}(k)
\equiv
\sum_{n=1}^\infty \frac{(n!)^2}{(2n)!} \frac{k^n}{n^c}
[S_{a_1}(n\!-\!1)]^{i_1}\ldots [S_{a_p}(n\!-\!1)]^{i_p}\; 
[S_{b_1}(2n\!-\!1)]^{j_1}\ldots [S_{b_q}(2n\!-\!1)]^{j_q},
\label{binsum}
\end{equation}
where $S_a(n) = \sum_{j=1}^n j^{-a}$ is the harmonic sum\footnote{Another 
notation for $S_a(n)$, used by mathematicians, is $H_n^{(a)}$. In
particular, for $a=1$, $S_1(n) = H_n$.
When there are no sums of the type $S_{a}(n-1)$ or $S_{b}(2n-1)$
on the r.h.s.\ of Eq.~(\ref{binsum}),
we shall put a ``$-$'' sign instead of the indices $(a,i)$ or $(b,j)$
of $\Sigma$, respectively. If the argument $(k)$ is omitted
in (\ref{binsum}), this
means that the case $k=1$ is understood.}.
In \cite{KV00}, mainly the case $k=1$ (related to the {\em odd} basis)
has been treated. 
A {\sf PSLQ}-based analysis 
has shown that not all of sums~(\ref{binsum}) are 
{\em separately} expressible in terms of the basis elements. 

This paper is organized as follows. In Section~2 we examine
in detail the $\ep$-expansion of the two-point function with
arbitrary masses, including the problem of analytic continuation
of the corresponding functions. In Section~3 we consider 
some examples of the $\ep$-expansion of one-loop three-point functions, 
and study which functions and transcendental constants may occur
in the cases considered. In Section~4 we consider some physically
relevant two- and three-loop diagrams. In Section~5 we discuss
the results obtained in this paper. There are also two appendices
containing some further technical details and useful formulae.
In Appendix~A we collect relevant results for the polylogarithms
and associated functions. In particular, we discuss the relation
of the Euler--Zagier sums and the generalized log-sine functions.
In Appendix~B we discuss the expansion of the occurring hypergeometric
functions with respect to their parameters.

\section{One-loop two-point function}
\setcounter{equation}{0}

\subsection{General case}

Consider the one-loop two-point function 
with the external momentum $k$ and masses $m_1$ and $m_2$,
\begin{equation}
J^{(2)}(n;\nu_1,\nu_2)\equiv \int 
\frac{\mbox{d}^n q}{\left[ q^2-m_1^2 \right]^{\nu_1}
\left[ (k-q)^2-m_2^2 \right]^{\nu_2}} \; .
\end{equation}
Using, for instance, a geometrical approach~\cite{DD,Crete},
one can obtain the following result (for unit powers of the propagators,
$\nu_1=\nu_2=1$): 
\begin{eqnarray}
\label{geom1}
J^{(2)}(4\!-\!2\varepsilon;1,1) &=& \mbox{i}\; \pi^{2-\varepsilon}
\Gamma(\varepsilon) \frac{1}{2 k^2}
\Biggl\{ 
(k^2+m_1^2-m_2^2) m_1^{-2\varepsilon}\;
_2F_1\left( \left. \begin{array}{c} {1, \; \ep} \\ 
    {\textstyle{3\over2}} \end{array} 
    \right| \cos^2\tau'_{01} \right)
\nonumber \\ &&
+ (k^2-m_1^2+m_2^2) m_2^{-2\varepsilon}\;
_2F_1\left( \left. \begin{array}{c} {1, \; \ep} \\ 
    {\textstyle{3\over2}} \end{array}
    \right| \cos^2\tau'_{02} \right) 
\Biggl\} \; ,
\end{eqnarray}
where $_2F_1$ is the Gauss hypergeometric function. The angles 
$\tau'_{0i}$
are defined (see in \cite{D-ep}) via\footnote{
They are related to the angles $\tau_{0i}$ used in Ref.~\cite{DD}
as $\tau'_{0i}=\tfrac{\pi}{2}-\tau_{0i}$.}
\begin{equation}
\label{two-point}
\cos\tau'_{01} = \frac{k^2+m_1^2-m_2^2}{2m_1 \sqrt{k^2}} \; , \quad
\cos\tau'_{02} = \frac{k^2-m_1^2+m_2^2}{2m_2 \sqrt{k^2}} \; .
\label{tau_0i}
\end{equation}
In particular, for angle 
$\tau_{12}$ (defined so that  $\tau_{12}+\tau'_{01}+\tau'_{02}=\pi$) we have
\begin{equation}
\cos\tau_{12} = \frac{m_1^2+m_2^2-k^2}{2m_1 m_2} \; , \quad
\sin\tau_{12} = \frac{\sqrt{\Delta(m_1^2,m_2^2,k^2)}}{2 m_1 m_2} \; .
\end{equation}
Here the ``triangle'' function $\Delta$ is defined as
\begin{eqnarray}
\label{Delta}   
\Delta(x,y,z) & = & 2xy+2yz+2zx-x^2-y^2-z^2 
=-\lambda(x,y,z) \; ,
\end{eqnarray}
where $\lambda(x,y,z)$ is the well-known K\"allen function.
The result (\ref{geom1}) can be related to those presented in Ref.~\cite{DD}
by a simple transformation of the occurring $_2F_1$ functions.  

Using Kummer relations for the contiguous $_2F_1$ functions, we get
\begin{eqnarray}
&&
(1- 2 \ep) \;
_2F_1 \left(\begin{array}{c|} 1, \; \ep  \\
\frac{3}{2} \end{array} ~z \right) =
1 - 2 \ep (1-z) \;
_2F_1 \left(\begin{array}{c|} 1, \; 1+\ep  \\
\frac{3}{2} \end{array} ~z \right) .
\label{Kummer1}
\end{eqnarray}
The resulting $_2F_1$ function can be represented as
\begin{equation}
\label{f_ep_1}
_2F_1 \left(\begin{array}{c|} 1, 1+\ep  \\
\frac{3}{2} \end{array} \sin^2\theta \right)
= \frac{1}{\sin\theta (\cos\theta)^{1+2\ep}}\; f_{\ep}(\theta) \; ,
\end{equation}
with (see in \cite{Lewin,D-ep})
\begin{eqnarray}
\label{f_ep}
f_{\ep}(\theta) &\equiv& \int\limits_0^{\theta} 
\mbox{d}\phi (\cos\phi)^{2\ep}
= 2^{-1-2\ep} \sum_{j=0}^{\infty} \frac{(2\ep)^j}{j!}
\left[ \Ls{j+1}{\pi-2\theta} - \Ls{j+1}{\pi} \right] \; .
\end{eqnarray}
As a result, we reproduce the $\varepsilon$-expansion
of the two-point integral obtained in Ref.~\cite{D-ep},
\begin{eqnarray}
\label{2pt_res2}
J^{(2)}(4\!-\!2\varepsilon;1,1) &=& \mbox{i}\pi^{2-\varepsilon}
\frac{\Gamma(1+\varepsilon)}{2(1-2\varepsilon)}
\bigg\{ \frac{m_1^{-2\varepsilon} \!+\! m_2^{-2\varepsilon}}{\varepsilon}
+ \frac{m_1^2\!-\!m_2^2}{\varepsilon \; k^2}
\left( m_1^{-2\varepsilon} \!-\! m_2^{-2\varepsilon} \right)
\nonumber \\ 
&& 
+ \frac{\left[\Delta(m_1^2,m_2^2,k^2)\right]^{1/2-\varepsilon}}
       {(k^2)^{1-\varepsilon}}
\sum_{j=0}^{\infty} \frac{(2\varepsilon)^j}{j!}
\sum_{i=1}^{2}
\left[ {\mbox{Ls}}_{j+1}(\pi) - {\mbox{Ls}}_{j+1}(2\tau'_{0i})
\right] \bigg\} . 
\hspace*{5mm}
\end{eqnarray}
The expansion (\ref{2pt_res2}) is directly applicable
in the region where $\Delta(m_1^2,m_2^2,k^2)\geq 0$, i.e.\ when
$(m_1-m_2)^2\leq k^2\leq(m_1+m_2)^2$. To obtain results valid 
in the region $\Delta(m_1^2,m_2^2,k^2)\leq 0$ (i.e.,
$\lambda(m_1^2,m_2^2,k^2)\geq 0$)
the proper analytic continuation of the occurring $\Ls{j}{\theta}$ 
should be constructed. 

An important special case of Eqs.~(\ref{f_ep_1})--(\ref{f_ep})
is $\theta=\tfrac{\pi}{6}$ ($z=\tfrac{1}{4}$).
It corresponds to the on-shell value $k^2=m^2$ of the integral
(\ref{2pt_res2}) with $m_1=m_2\equiv m$. 
In this case, we obtain $\Ls{j+1}{\tfrac{2\pi}{3}}$
which correspond to the {\em odd} basis discussed in the
introduction.
Expansion of more general $_2F_1$ functions of argument
$z=\tfrac{1}{4}$ is discussed in Appendix~B.2.

We shall also need another representation 
for the two-point function (see, e.g., in~\cite{BDS}), 
\begin{eqnarray}
\label{2pt2}
J^{(2)}(4\!-\!2\varepsilon;1,1) &=&
\mbox{i}\pi^{2-\varepsilon}\; \Gamma(\ep) \frac{1}{k^2}
\left\{ \frac{\Gamma^2(1-\ep)}{\Gamma(2-2\ep)} \;
\lambda^{1/2-\ep} (k^2)^\ep  e^{ {\rm i} \pi \ep}
\right.
\nonumber \\ &&
+ \frac{m_1^{-2\ep}}{2(1-\ep)} \; 
\left[ k^2+m_1^2-m_2^2-\sqrt{\lambda(m_1^2,m_2^2,k^2)} \right]\;
_2F_1\left(
\left. \begin{array}{c} {1, \; \ep} \\ {2-\ep} \end{array} 
\right| z_2 \right)
\nonumber \\ &&
+ 
\left.
\frac{m_2^{-2\ep}}{2(1-\ep)} \; 
\left[ k^2-m_1^2+m_2^2-\sqrt{\lambda(m_1^2,m_2^2,k^2)} \right]\;
_2F_1\left( \left.
\begin{array}{c} {1, \; \ep} \\ {2-\ep} \end{array} \right| z_1\right)
\right\} ,
\nonumber \\ 
\end{eqnarray}
where 
\begin{equation}
z_1 = 
\frac{\left[\sqrt{\lambda(m_1^2,m_2^2,k^2)}+m_1^2-m_2^2-k^2\right]^2} 
     {4m_2^2 k^2}, \quad
z_2 = 
\frac{\left[\sqrt{\lambda(m_1^2,m_2^2,k^2)}-m_1^2+m_2^2-k^2\right]^2}
     {4m_1^2 k^2} ,
\end{equation}
with $\lambda(m_1^2,m_2^2,k^2)$ defined in Eq.~(\ref{Delta}).

Employing Kummer's relations for
contiguous functions, one can transform the
$_2F_1$ function from Eq.~(\ref{2pt2}) into
\begin{eqnarray}
\left.
_2F_1\left( \begin{array}{c} 1, \; \ep \\
                 2-\ep \end{array} \right| z  \right) =
\frac{1-\ep}{2(1\!-\!2\ep)z}
\left\{ 1+z - (1\!-\!z)^2 
\left.
_2F_1\left( \!\! \begin{array}{c} 1, \; 1\!+\!\ep \! \\
                 1-\ep \end{array} \right| z \right)
\right\} \; .
\end{eqnarray}
The resulting $_2F_1$ function can be expressed 
in terms of a simple one-fold parametric integral,
\begin{eqnarray}
\left.
_2F_1\left( \!\! \begin{array}{c} 1, \; 1\!+\!\ep \! \\
                 1-\ep \end{array} \right| z \right) 
=
(1-z)^{-1-2\ep} \left\{ 1
- \ep \int\limits_0^1 \frac{{\mbox{d}}t}{t}\; t^{-\ep}\;
\left[ (1\!-\!zt)^{2\ep} \!-\! 1 \right] \right\} \; .
\end{eqnarray}
Expanding the integrand in $\ep$, we get
\begin{equation}
\label{Aux}
\left. _2F_1\left( \begin{array}{c} 1, \; \ep \\
                 2-\ep \end{array} \right| z  \right) =
\frac{1-\ep}{2(1\!-\!2\ep)z}   
\Biggl\{ 1+z - (1\!-\!z)^{1 - 2\ep}
- 2 (1\!-\!z)^{1- 2\ep} \ep
\sum_{j=1}^\infty \ep^j \sum_{k=1}^{j} (-2)^{j-k} S_{k,j-k+1}(z)
\Biggl\} ,
\end{equation}
where $S_{a,b}(z)$ is the Nielsen polylogarithm 
(see, e.g., in Ref.~\cite{Nielsen}), whose definition (\ref{Sab})
and some properties are collected in Appendix~A.1.

\subsection{Analytic continuation}

To relate results in different regions, it is convenient to introduce 
the variables
\begin{equation}
\label{def_z} 
z_j \equiv e^{ {\rm i} \sigma \theta_j}, \hspace{5mm}
\ln(-z-{\rm i}\sigma 0) = \ln(z) - {\rm i} \sigma \pi ,
\end{equation}
where  we put $\theta_j  \equiv 2\tau'_{0j}$, 
and the choice of the sign $\sigma=\pm 1$ is related to the
causal ``+i0'' prescription for the propagators.
Whenever possible, we shall keep $\sigma$ undetermined,
since one may need different signs in different situations.

Let us note that, transforming from variable $z$ to $1/z$,
we get the same $_2F_1$ function,
\begin{eqnarray}
\label{z_to_1/z}
{}_2F_1  
\left(\begin{array}{c|} 1, \ep \\ 2 - \ep \end{array} ~z\right)
=  \frac{1}{z}~~ 
{}_2F_1 \left(\begin{array}{c|} 1, \ep \\ 2 - \ep  
\end{array} ~ \frac{1}{z} \right)
+ \frac{(1-\ep)\; \Gamma^2(1-\ep)}{\Gamma(2-2\ep)}
(-z)^{-\ep} \left(1 - \frac{1}{z} \right)^{1-2\ep} \; .
\end{eqnarray}
In particular, we can replace each $_2F_1$ function in Eq.~(\ref{2pt2})
by a linear combination of l.h.s.\ and r.h.s.\ of Eq.~(\ref{z_to_1/z}),
with the sum of the corresponding coefficients equal to one
(say, $\rho_{1,2}$ and $(1-\rho_{1,2})$),
\begin{eqnarray}
&& 
J^{(2)}(4\!-\!2\varepsilon;1,1) = \mbox{i}\pi^{2-\varepsilon}\; 
\frac{\Gamma(1+\ep)}{2(1-2\ep)} \;
\Biggr ( \frac{m_1^{-2\varepsilon} \!+\! m_2^{-2\varepsilon}}{\varepsilon}
+ \frac{m_1^2\!-\!m_2^2}{\varepsilon \; k^2}
\left( m_1^{-2\varepsilon} \!-\! m_2^{-2\varepsilon} \right)
\nonumber \\ && 
+  {\rm i} \sigma 
\frac{\left[\Delta(m_1^2,m_2^2,k^2)\right]^{1/2-\varepsilon}}
       {(k^2)^{1-\varepsilon}}
\Biggl \{ 
\frac{2 \Gamma^2(1-\ep)}{\ep \Gamma(1-2\ep)} \; (1-\rho_1-\rho_2)
+ \sum_{i=1}^2  \frac{1}{\ep} \Biggl [ \rho_i (-z_i)^{-\ep} - (1-\rho_i)(-z_i)^{\ep} 
\Biggr ]
\nonumber \\ && 
+ 2  \sum_{i=1}^2 \Biggl [
\rho_i (-z_i)^{-\ep} 
\sum_{j=1}^\infty \ep^j \sum_{k=1}^{j} (-2)^{j-k} S_{k,j-k+1}(z_i)
\nonumber \\ && 
- (1-\rho_i)(-z_i)^{\ep} 
\sum_{j=1}^\infty \ep^j \sum_{k=1}^{j} (-2)^{j-k} S_{k,j-k+1}(1/z_i)
\Biggr ] \Biggr \} \Biggl ) \; .
\label{general}
\end{eqnarray}

Comparing (\ref{2pt_res2}) and (\ref{general}) for $\rho_1=\rho_2=\frac{1}{2}$, 
we arrive at the following analytic continuation of the functions involved
in the $\ep$-expansion: 
\begin{eqnarray}
\label{Ls<->S}
&& \hspace*{-7mm}
{\rm i} \sigma \left[ \Ls{j}{\pi}  - \Ls{j}{\theta} \right]
= \frac{1}{2^j j}  \ln^j (-z) \left[ 1 - (-1)^j \right] 
\nonumber \\ && 
+ (-1)^j (j-1)! 
\sum_{p=0}^{j-2} \frac{\ln^p (-z)}{2^p p!} 
\sum_{k=1}^{j-1-p} (-2)^{-k}
\left[ S_{k,j-k-p}(z) - (-1)^p S_{k,j-k-p}(1/z) \right].
\end{eqnarray}
where $z_i$ and $\sigma$ are defined in (\ref{def_z}).
In particular, 
\begin{eqnarray}
 && \hspace*{-7mm}
{\rm i} \sigma 
\sum_{j=0}^\infty  \frac{(2\ep)^j}{j!}
\Biggl [ \Ls{j+1}{\pi} - \Ls{j+1}{\theta} \Biggr ] =
\frac{1}{2\ep} \Biggl [ (-z)^\ep - (-z)^{-\ep} \Biggr ]
\nonumber \\ && 
-(-z)^{-\ep} \sum_{j=1}^\infty \ep^j \sum_{k=1}^{j} (-2)^{j-k} S_{k,j-k+1}(z)
+(-z)^\ep \sum_{j=1}^\infty \ep^j \sum_{k=1}^{j} (-2)^{j-k} S_{k,j-k+1}(1/z) \; .
\end{eqnarray}

Since ${\mbox{Ls}}_1(\theta)=-\theta$, ${\mbox{Ls}}_2(\pi)= 0$ we get
\begin{equation}
{\rm i} \sigma \left[ \Ls{1}{\pi}  - \Ls{1}{\theta} \right]
= \ln (-z) , \quad
{\rm i}  \sigma \left[ \Ls{2}{\pi} \! -\! \Ls{2}{\theta} \right]  =
- {\textstyle{1\over2}} \left[ \Li{2}{z} \!-\! \Li{2}{1/z} \right ] .
\label{Ls2}
\end{equation}

We note that for higher values of $j$ the number of generalized 
polylogarithms involved in Eq.~(\ref{Ls<->S}) can be reduced.
For even $j=2l$ ($l\ge2$), we have 
\begin{eqnarray}
&& \hspace*{-7mm}
{\rm i}  \sigma \left[ \Ls{2l}{\pi}  - \Ls{2l}{\theta} \right]  = 
\nonumber \\ && \hspace*{-7mm}
-\frac{(2l-1)!}{2}  \sum_{p=0}^{l-2} \frac{\ln^p (-z)}{p!} 
\sum_{k=0}^{l-p-2} (-1)^k  
a_{l,p+k} \left[ S_{k+1,2l-k-p-1}(z) - (-1)^p S_{k+1,2l-k-p-1}(1/z) \right]
\nonumber \\ && \hspace*{-7mm}
+ (-1)^{l} \frac{(2l-1)!}{2^{2l-1}} \sum_{k=0}^{2l-2} 
\frac{(-1)^k}{k!} \ln^k (-z) \left[ \Li{2l-k}{z} - (-1)^k \Li{2l-k}{1/z} \right] \; ,
\label{even}
\end{eqnarray}
whereas for odd $j=2l+1$ we have 
\begin{eqnarray}
&& \hspace*{-7mm}
{\rm i}  \sigma \left[ \Ls{2l+1}{\pi}  - \Ls{2l+1}{\theta} \right]  = 
\nonumber \\ && \hspace*{-7mm}
\frac{(2l)!}{2}  \sum_{p=0}^{l-1} \frac{\ln^p (-z)}{p!} 
\sum_{k=0}^{l-p-1} (-1)^k  
a_{l+1,p+k}\left[ S_{k+1,2l-k-p}(z) - (-1)^p S_{k+1,2l-k-p}(1/z) \right]
\nonumber \\ && \hspace*{-7mm}
- \frac{1}{4^{2l}} \frac{(2l)!}{l!}  \sum_{k=0}^{j} 
\frac{(-2)^k (2l-k)!}{k!(l-k)!} \ln^k (-z) 
\left[ \Li{2l-k+1}{z} - (-1)^k \Li{2l-k+1}{1/z} \right] \; ,
\label{odd}
\end{eqnarray}
where 
\begin{equation}
a_{l,p} = \frac{p}{4^p} \;\; \sum_{q=0}^{l-p-2}
\frac{(p+2q-1)!}{4^q q! (p+q)!}, \quad \quad
a_{l,0} = 1 \; .
\label{coef}
\end{equation}

\subsection{Massless limit}

As a simple example we can consider the limit when one of the
masses vanishes, $m_1=0$ ($m_2\equiv m$).
Using hypergeometric representation (see, e.g., Eq.~(10) of \cite{BD-TMF}),
\begin{eqnarray}
\label{BD}  
&& \hspace{-7mm}
\left. J^{(2)}(4\!-\!2\varepsilon;1,1)
\right|_{m_1=0, \; m_2\equiv m} =
\mbox{i}\pi^{2-\varepsilon} m^{-2\ep}
\frac{\Gamma(1+\ep)}{\ep(1-\ep)}
\left. _2F_1\left( \begin{array}{c} 1, \; \ep \\
                  2-\ep \end{array} \right| \frac{k^2}{m^2} \right)
\end{eqnarray}
and Eq.~(\ref{Aux}), the result for an arbitrary term of the $\varepsilon$-expansion 
can be obtained \cite{bastei_ep}
\begin{eqnarray}
\label{m2=0}
&& \hspace{-8mm}
\left. J^{(2)}(4\!-\!2\varepsilon;1,1)
\right|_{m_1=0, \; m_2\equiv m} =
\mbox{i}\pi^{2-\varepsilon} m^{-2\ep}
\frac{\Gamma(1+\varepsilon)}{(1-2\varepsilon)}
\nonumber \\[-2mm] && \hspace*{-6mm}
\times
\Biggl\{ \frac{1}{\ep} - \frac{1-u}{2u\ep}
\left[ (1-u)^{-2\ep} - 1 \right]
- \frac{(1\!-\!u)^{1\!-\!2\ep}}{u}
\sum_{j=1}^\infty \ep^j \sum_{k=1}^{j} (-2)^{j-k}
S_{k,j-k+1}(u)
\Biggr\} ,
\end{eqnarray}
with $u = k^2/m^2$.

Note that for this limit the terms up to order $\varepsilon^3$ can 
be extracted
from Eq.~(A.3) of Ref.~\cite{FJTV}. Our expressions are in agreement with
their results.
We would like to mention, that each term of the expansion (\ref{m2=0}) 
can be obtained from general results, (\ref{2pt_res2}) 
or (\ref{general}).  Expanding them with respect to  
$m_1^2$ and $\ln(m_1^2)$ (see details in \cite{bastei_ep}), 
we reveal, that there is a limit 
$m_1^2 = 0$.  This procedure allows to check the 
unambiguity of $\sigma$: the limit $m_1^2 = 0$ must exist in each
order of $\ep$. So, the finite part gives relation between 
(\ref{def_z}) and  $\ln(-m^2/k^2)$, whereas  linear one 
gives rise to correlation between sign of the analytical continuation 
of function $\lambda(m_1^2,m_2^2,k^2)$ and $\ln(-m^2/k^2)$. 
Since, for Feynman propagator, $\ln(-m^2/k^2) = \ln(m^2/k^2) + {\rm i}\pi$, 
we have $\sigma=-1$.

\section{Three-point examples}
\setcounter{equation}{0}
\subsection{General remarks}

In this section we shall consider $\ep$-expansion of
one-loop three-point integrals, shown in Fig.~1,
\begin{equation}
\label{defJ3}
J^{(3)}(n; \nu_1, \nu_2, \nu_3)
\equiv \int \frac{\mbox{d}^n q}
{\big[ (p_2-q)^2-m_1^2 \big]^{\nu_1}
\big[ (p_1+q)^2-m_2^2 \big]^{\nu_2}
\big[ q^2 - m_3^2 \big]^{\nu_3}} \; .
\end{equation}

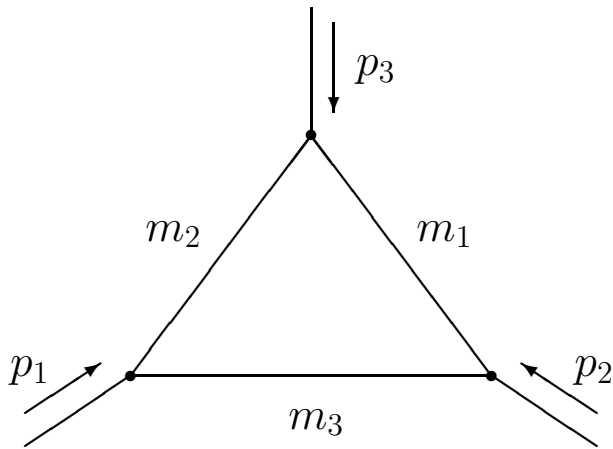
\begin{figure}[hb]
\label{oneloopvertex} 
\setlength {\unitlength}{1mm}
\begin{center}
\begin{picture}(150,70)(0,-5)
\put (50,10) {\circle*{1.5}}
\put (74,42) {\circle*{1.5}}
\put (98,10) {\circle*{1.5}}
\thicklines
\put (50,10) {\line(3,4){24}}
\put (50,10) {\line(1,0){48}}
\put (98,10) {\line(-3,4){24}}
\put (74,42) {\line(0,1){17}}
\put (50,10) {\line(-3,-2){14}}
\put (98,10) {\line(3,-2){14}}
\put (36,5) {\vector(3,2){10}}
\put (112,5) {\vector(-3,2){10}}
\put (77,57) {\vector(0,-1){12}}
\put (34,10) {\makebox(0,0)[bl]{\Large $p_1$}}
\put (109,10) {\makebox(0,0)[bl]{\Large $p_2$}}
\put (80,50) {\makebox(0,0)[bl]{\Large $p_3$}}
\put (71,3) {\makebox(0,0)[bl]{\Large $m_3$}}
\put (88,28) {\makebox(0,0)[bl]{\Large $m_1$}}
\put (52,28) {\makebox(0,0)[bl]{\Large $m_2$}}
\end{picture}
\end{center}
\caption{One-loop three-point function with masses $m_i$
and momenta $p_i$ ($p_1 + p_2 + p_3 = 0$)}
\end{figure}

We shall mainly be interested in the case of positive integer powers
of the propagators $\nu_i$.
Using the integration-by-parts approach \cite{ibp} (for details, see 
in \cite{JPA}), all such integrals 
can be algebraically reduced to $J^{(3)}(n;1,1,1)$ and 
two-point integrals.
Therefore, we shall concentrate on the case $\nu_1=\nu_2=\nu_3=1$.

To construct terms of the $\varepsilon$-expansion 
of $J^{(3)}(n;1,1,1)$
with general masses and external momenta,
the geometrical description seems to be  
rather instructive. The geometrical approach to the
three-point function is discussed in section~V of \cite{DD}
(see also in \cite{Crete}). This function can be represented
as an integral over a spherical (or hyperbolic)
triangle, as shown in Fig.~6 of \cite{DD},
with a weight factor $1/\cos^{1-2\ep}\theta$
(see eqs.~(3.38)--(3.39) of \cite{DD}).
This triangle 123 is split into three triangles 012, 023
and 031. Then, each of them is split into two
rectangular triangles, according to Fig.~9 of \cite{DD}.
We consider the contribution of one of the six resulting triangles,
namely the left rectangular triangle in Fig.~9.
Its angle at the vertex~0 is denoted as
${\textstyle{1\over2}}\varphi_{12}^{+}$, whereas the
height dropped from the vertex~0 is denoted $\eta_{12}$.

The remaining angular integration is (see eq.~(5.16) of \cite{DD})
\begin{equation}
\label{last}
\frac{1}{2\ep} \int\limits_0^{\varphi_{12}^{+}/2} \mbox{d}\varphi
\left[ 1 -
\left( 1 + \frac{\tan^2\eta_{12}}{\cos^2\varphi} \right)^{-\ep} \right]
= {\textstyle{1\over2}} \sum\limits_{j=0}^{\infty}\frac{(-\ep)^j}{(j+1)!} \!
\int\limits_0^{\varphi_{12}^{+}/2} \!\! \mbox{d}\varphi
\ln^{j+1}\left( 1\! +\! \frac{\tan^2\eta_{12}}{\cos^2\varphi} \right) .
\end{equation}
First of all, we note that the l.h.s. of Eq.~(\ref{last}) yields 
a representation valid for
an arbitrary $\varepsilon$ (i.e., in any dimension).
To get the result for the general three-point function,
we need to consider a sum of six such integrals.
The resulting representation is closely related
to the representation in terms of hypergeometric functions
of two arguments \cite{bastei_tar} (see also in \cite{SRSL} for
some special cases).

In the limit $\ep\to0$ we get a combination
of $\mbox{Cl}_2$ functions\footnote{For $\ep=\tfrac{1}{2}$
($n=3$) we reproduce the well-known result of~\cite{Nickel}
in terms of elementary functions
(for further details, see Section~VA of~\cite{DD}).},
eq.~(5.17) of \cite{DD}.
Collecting the results for all six triangles,
we get the result for the three-point function with
arbitrary masses and external momenta, corresponding
(at $\ep=0$) to the analytic continuation of the
well-known formula presented in \cite{`tHV-79}.
The higher terms of the $\ep$-expansion correspond to the
angular integrals on the r.h.s.\ of Eq.~(\ref{last}).
We note that the $\ep$-term of the three-point function with general
masses has been calculated in \cite{NMB} in terms of
$\mbox{Li}_3$.

An important special case is when all internal masses 
are equal to zero, whereas the external momenta are
off shell.
This case was considered in detail in Refs.~\cite{D-ep,bastei_ep}.
The following result was obtained:
\begin{eqnarray}
\label{conjecture}
\left.
J^{(3)}(4-2\varepsilon;1,1,1)\right|_{m_1=m_2=m_3=0} 
&=& 2 \pi^{2-\varepsilon}\;
\mbox{i}^{1+2\varepsilon}\;
\frac{\Gamma(1\!+\!\varepsilon)\Gamma^2(1\!-\!\varepsilon)}
     {\Gamma(1-2\varepsilon)}\;
\frac{\left[ \Delta(p_1^2,p_2^2,p_3^2)\right]^{-1/2+\varepsilon}}
     {(p_1^2 p_2^2 p_3^2)^{\varepsilon}}
\nonumber \\ && \hspace*{-5mm} \times
\sum_{j=0}^{\infty} \frac{(-2\varepsilon)^j}{(j\!+\!1)!}
\biggl[ {\mbox{Ls}}_{j+2}(\pi) - \sum_{i=1}^{3}
\big[ {\mbox{Ls}}_{j+2}(\pi) \!-\! {\mbox{Ls}}_{j+2}(2\phi_i) \big]
\biggr] , \hspace*{7mm}
\end{eqnarray}
where the angles
$\phi_i$ ($i=1,2,3$) are defined via
\begin{equation}
\cos\phi_1=\frac{p_2^2+p_3^2-p_1^2}{2\sqrt{p_2^2 p_3^2}}, \quad
\cos\phi_2=\frac{p_3^2+p_1^2-p_2^2}{2\sqrt{p_3^2 p_1^2}}, \quad
\cos\phi_3=\frac{p_1^2+p_2^2-p_3^2}{2\sqrt{p_1^2 p_2^2}}, 
\end{equation}
so that $\phi_1+\phi_2+\phi_3=\pi$. Therefore,
the angles $\phi_i$
can be understood as the angles of a triangle
whose sides are $\sqrt{p_1^2}$, $\sqrt{p_2^2}$ and $\sqrt{p_3^2}$,
whereas its area is $\tfrac{1}{4}\sqrt{\Delta(p_1^2,p_2^2,p_3^2)}$.
Introducing $\theta_i=2\phi_i$, we can use the same procedure
of analytic continuation as discussed in Section~2.2 
(with $\sigma=1$), in terms of the Nielsen polylogarithms 
$S_{a,b}(z)$. 

Note that the integral (\ref{conjecture}) is closely related
to the two-loop vacuum integral with different masses,
due to the magic connection \cite{DT2}. The result for
this integral, which is called $I(n;\nu_1,\nu_2,\nu_3)$, 
is also presented in Refs.~\cite{D-ep,bastei_ep},
in an arbitrary order in $\ep$. 
For the analytic continuation, we can also apply the procedure
of Section~2.2 (with $\sigma= -1$).
Such integrals will also occur in Section~4.

\subsection{On-shell triangle with two different masses}

Consider a three-point integral (\ref{defJ3}) with one massless propagator
($m_3=0$) and two adjacent legs on shell ($p_2^2=m_1^2, \; p_1^2=m_2^2$).
Such integrals are important, e.g., for studying corrections 
to the the muon decay in the Fermi model.
Using Feynman parametric representation, one can show that 
in the $\nu_1=\nu_2=\nu_3=1$ case this integral 
can be reduced to a two-point integral with the masses
of internal particles $m_1$ and $m_2$ and the external
momentum $k\equiv p_3$, 
\begin{eqnarray}
\left. J^{(3)}(4-2\ep;1,1,1)\right|_{m_3=0, \; p_2^2=m_1^2, \; 
p_1^2=m_2^2} = \frac{\pi}{2\ep} J^{(2)}(2-2\ep;1,1) \; .
\end{eqnarray}
Note that all powers of propagators are equal to one, whereas
the space-time dimension is $(2-2\ep)$ in the two-point integral.

Using Eq.~(6) of \cite{PLB'91}, we can represent it in terms of
the $(4-2\ep)$-dimensional integrals,
\begin{equation}
J^{(2)}(2-2\ep;1,1) = -\frac{1}{\pi}
\left[ J^{(2)}(4-2\ep;2,1) + J^{(2)}(4-2\ep;1,2) \right] \; .
\end{equation}
Then, using the integration-by-parts technique \cite{ibp}
(see also Eq.~(A.17) of \cite{BDS}), we can express the 
integrals with second power of one of the propagators in
terms of $J^{(2)}(4-2\ep;1,1)$ and tadpole integrals.
As a result, we get
\begin{eqnarray}
J^{(2)}(2-2\ep;1,1) &=& -\frac{1}{\pi \Delta(m_1^2,m_2^2,k^2)}
\biggl\{ 2(1-2\ep) k^2 J^{(2)}(4-2\ep;1,1) 
\nonumber \\ 
&& -\mbox{i} \; \pi^{2-\ep} \; \Gamma(\ep) \;
\left[ (k^2+m_1^2-m_2^2) m_1^{-2\ep} 
     + (k^2-m_1^2+m_2^2) m_2^{-2\ep} \right]
\biggl\} \; , \hspace*{5mm}
\end{eqnarray}
with $\Delta(m_1^2,m_2^2,k^2)$ defined in Eq.~(\ref{Delta}).

In particular, the $\ep$-expansion of this integral is
\begin{equation}
J^{(2)}(2-2\ep;1,1) = -\mbox{i}\; \pi^{1-\ep}\; 
\frac{ \Gamma(1+\ep)\; (k^2)^{\ep}}
{ \left[ \Delta(m_1^2,m_2^2,k^2) \right]^{1/2+\ep}}\;
\sum_{j=0}^{\infty} \frac{(2\ep)^j}{j!}
\sum_{i=1}^{2} 
\left[ \mbox{Ls}_{j+1}(\pi)-\mbox{Ls}_{j+1}(2\tau'_{0i}) \right] .
\end{equation}
This gives us the $\ep$-expansion of the on-shell triangle diagram
considered.
Extracting the infrared (on-shell) singularity, it can
be presented as
\begin{eqnarray}
&& 
\left. J^{(3)}(4-2\ep;1,1,1)\right|_{m_3=0, \; p_2^2=m_1^2, \;
p_1^2=m_2^2} =
\mbox{i}\; \pi^{2-\ep}\; \Gamma(1+\ep)\; 
\frac{(k^2)^{\ep}}{\left[\Delta(m_1^2,m_2^2,k^2)\right]^{1/2+\ep}}\;
\nonumber \\ &&
\hspace{35mm}
\times \biggl\{ \frac{\tau_{12}}{\ep}
- \sum_{j=0}^{\infty} \frac{(2\ep)^j}{(j+1)!}   
\sum_{i=1}^{2}
\left[ \mbox{Ls}_{j+2}(\pi)-\mbox{Ls}_{j+2}(2\tau'_{0i}) \right] 
\biggl\} \; ,
\end{eqnarray}
where $\tau_{12}=\pi-\tau'_{01}-\tau'_{02}$.

Since the functions are of the same type as in the two-point example,
we can use Eq.~(\ref{Ls<->S}) for the analytic continuation.

\subsection{A more complicated example}

Let us consider triangle integral (\ref{defJ3}) with
$m_1=m_2=m_3\equiv m$, $p_1^2=p_2^2=0$, with an arbitrary (off-shell)
value of $p_3^2$. Such diagrams occur, for example, in Higgs decay
into two photons or two gluons via a massive quark loop.
Following the notation of Ref.~\cite{BD-TMF}, we shall denote
this integral (with unit powers of propagators) as
$J_3(1,1,1; m)$. According to Eq.~(40) of Ref.~\cite{BD-TMF}, the result
in an arbitrary space-time dimension $n=4-2\ep$ is
\begin{equation}
J_3(1,1,1; m) \left. \right|_{p_1^2 = p_2^2 = 0} =
- \frac{1}{2} \mbox{i} \pi^{2-\ep} (m^2)^{-1-\ep} \Gamma(1+\ep)\;
{}_{3}F_2 \left(\begin{array}{c|} 1, 1, 1+\ep \\
\frac{3}{2}, 2 \end{array} ~\frac{p_3^2}{4m^2} \right) \; .
\label{Hgg}
\end{equation}

For the occurring $_3F_2$ function, various one-fold
integral representations can be constructed 
(see, e.g., in \cite{PBM3}). Below we list some of them: 
\begin{eqnarray}
{}_{3}F_2 \left(\begin{array}{c|} 1, 1, 1+\ep \\
\frac{3}{2}, 2 \end{array} ~z \right) &=&
-\frac{\Gamma(1-\ep)}{2 z \Gamma(1+\ep) \Gamma(1-2\ep)}
\int\limits_0^1 \frac{\mbox{d}t}{t\sqrt{1-t}} \ln (1 - tz)
\left[ \frac{t}{4(1-t)} \right]^\ep \; , \hspace*{8mm}
\\ &=&
\frac{1}{z\ep}\int\limits_0^{\infty}
\mbox{d}\tau \left[ \left( 1-\frac{z}{\cosh^2\tau}\right)^{-\ep}
-1 \right] \; ,
\\ &=&
\frac{1}{2 z\ep}\int\limits_0^1
\frac{\mbox{d} x}{x} \left\{ [1- 4zx(1-x)]^{-\ep} -1 \right\} \; .
\label{int3_hgg}
\end{eqnarray}

In fact, to get representations of the terms of the
$\ep$-expansion in terms of Clausen
and log-sine functions, the following two-fold angular
integral representation appears to be rather convenient:
\begin{equation}
\label{def_hj}
\sin^2\theta \; {}_{3}F_2 \left(\begin{array}{c|} 1, 1, 1+\ep \\
\frac{3}{2}, 2 \end{array} ~\sin^2\theta \right) =
2 \int\limits_0^\theta \mbox{d} \phi (\cos \phi)^{-2\ep}
\int\limits_0^\phi \mbox{d} \widetilde{\phi} 
(\cos \widetilde{\phi})^{2\ep}
\equiv \sum_{j=0}^{\infty} \ep^j h_j(\theta)
\end{equation}
It is easy to see that
\begin{eqnarray}
\label{h_01}
h_0(\theta) &=& \theta^2, 
\\
h_1(\theta) &=& 2 \Cl{3}{\pi-2\theta} - 2 \Cl{3}{\pi} 
- 2 \theta \Cl{2}{\pi-2\theta} \; .
\end{eqnarray}

Moreover, in terms of the function $f_{\ep}$ whose definition
and the $\ep$-expansion are given in Eq.~(\ref{f_ep}),  
we can represent the integral in Eq.~(\ref{def_hj}) as
\begin{eqnarray}
2 \int\limits_0^\theta \mbox{d} \phi (\cos \phi)^{-2\ep}
\int\limits_0^\phi \mbox{d} \widetilde{\phi} 
(\cos \widetilde{\phi})^{2\ep}
&=& 2 \int\limits_0^{\theta} f_{\ep}(\phi)\; \mbox{d} f_{-\ep}(\phi)
= 2 f_{\ep}(\theta) f_{-\ep}(\theta) 
- 2 \int\limits_0^{\theta} f_{-\ep}(\phi)\; \mbox{d} f_{\ep}(\phi) 
\nonumber \\ && \hspace*{-15mm}
= f_{\ep}(\theta) f_{-\ep}(\theta)
+ \int\limits_0^{\theta} \left[ f_{\ep}(\phi)\; \mbox{d} f_{-\ep}(\phi)
- f_{-\ep}(\phi)\; \mbox{d} f_{\ep}(\phi) \right] \; .
\hspace*{8mm}
\end{eqnarray}
The last representation is nothing but splitting the function
(\ref{def_hj}) into the even and odd parts.
The first term, $f_{\ep}(\theta) f_{-\ep}(\theta)$, is symmetric with
respect to $\ep\leftrightarrow -\ep$ and therefore contains only
even powers of $\ep$. The second (integral) term is antisymmetric
and contains only odd powers.

Thus we have obtained results for all $h_j(\theta)$ with even values 
of $j$ ($j=2l$).
One needs just to pick up the $\ep^{2l}$ term of the expansion
of $f_{\ep}(\theta) f_{-\ep}(\theta)$, with $f_{\pm \ep}$ given 
explicitly in Eq.~(\ref{f_ep}). For example, 
\begin{eqnarray}
h_2(\theta) &=& -\left[ \Ls{2}{\pi-2\theta} \right]^2
+ 2\theta \left[ \Ls{3}{\pi-2\theta}-\Ls{3}{\pi} \right] \; ,
\\ 
h_4(\theta) &=&
\left[ \Ls{3}{\pi-2\theta} - \Ls{3}{\pi}\right]^2
- {\textstyle{4\over3}} \Ls{2}{\pi-2\theta}
\left[ \Ls{4}{\pi-2\theta} - \Ls{4}{\pi}\right]
\nonumber \\ &&
+ {\textstyle{2\over3}} 
\theta \left[\Ls{5}{\pi-2\theta} - \Ls{5}{\pi} \right] \; .
\end{eqnarray}

However, the calculation of the odd terms of
the expansion ($j=2l+1$) is less trivial, starting from $j=3$
(the result for $h_1(\theta)$ is given in Eq.~(\ref{h_01})).
It appears that $h_3(\theta)$ (and the higher odd functions) cannot 
be expressed in terms of the $\mbox{Ls}_j$ functions and their simple 
generalizations. 

Of course, for a given value of $\theta$ there is no problem 
to calculate $h_j(\theta)$ with very high precision, using
integral representation (\ref{int3_hgg}).
As an illustration, let us consider a special value of $p_3^2$,
$p_3^2=m^2$ ($z={\textstyle{1\over4}}$, $\theta={\textstyle{\pi\over6}}$).
Then we have
\begin{equation}
h_j\left({\textstyle{\frac{\pi}{6}}}\right)
= \frac{ (-1)^{j+1}}{ 2 (j+1)!} 
\int\limits_0^1 \frac{\mbox{d} x}{x} \; 
\ln^{j+1}(1-x+x^2) \; .
\end{equation}
In particular, for $h_3({\textstyle{\pi\over6}})$ we obtain a 
number which, with the help of the {\sf PSLQ} program \cite{PSLQ}
can be identified as 
\begin{equation}
h_3({\textstyle{\pi\over6}}) =
{\textstyle{\frac{1}{12}}} \chi_5
+ {\textstyle{\frac{71}{324}}} \zeta_2 \zeta_3
+ {\textstyle{\frac{401}{324}}} \zeta_5
- {\textstyle{\frac{23}{243}}} \pi \Ls{4}{{\textstyle{\frac{\pi}{3}}}}
+ {\textstyle{\frac{1}{81}}} \pi \zeta_2 
  \Ls{2}{{\textstyle{\frac{\pi}{3}}}} \; ,
\end{equation}
where 
\begin{eqnarray}
\label{sigma_32}
\chi_5 \equiv \Sigma_{1;-;2}^{3;-}(1) 
& = &  \sum_{n=1}^\infty \frac{(n!)^2}{(2n)!} \; 
\frac{1}{n^2}\;
\left[ S_1(n-1) \right]^3 
= \sum_{n=1}^\infty \frac{(n!)^2}{(2n)!} \;
\frac{1}{n^2} \;
\left[ \psi(n)+\gamma \right]^3 
\nonumber \\  
& \simeq & 0.0678269619272092908692628002256569482253389\ldots \; 
\end{eqnarray}
is a special case of binomial sum \cite{BBK99,KV00}. 
Therefore, the obtained $\ep$-expansion of the $_3F_2$ function at 
$z={\textstyle{1\over4}}$ is
\begin{eqnarray}
{}_{3}F_2 \left(\begin{array}{c|} 1, 1, 1\!+\!\ep \\
\frac{3}{2}, 2 \end{array} ~\frac{1}{4} \right) \!\! &=& \!\!
{\textstyle{\frac{2}{3}}} \zeta_2  
+ \ep \biggl\{
{\textstyle{\frac{22}{9}}} \zeta_3 
- {\textstyle{\frac{8}{9}}} \pi \Ls{2}{{\textstyle{\frac{\pi}{3}}}}
\biggl\}
+ \ep^2 \biggl\{ 10 \zeta_4  
+ {\textstyle{\frac{4}{3}}} \pi \Ls{3}{{\textstyle{\frac{2\pi}{3}}}}
- {\textstyle{\frac{16}{9}}} \left[\Ls{2}{{\textstyle{\frac{\pi}{3}}}}  
\right]^2 \biggl\}
\nonumber \\ &&
+ \ep^3 \biggl\{ {\textstyle{\frac{1}{3}}} \chi_5
+ {\textstyle{\frac{71}{81}}} \zeta_2 \zeta_3
+ {\textstyle{\frac{401}{81}}} \zeta_5
- {\textstyle{\frac{92}{243}}} \pi \Ls{4}{{\textstyle{\frac{\pi}{3}}}}
+ {\textstyle{\frac{4}{81}}} \pi \zeta_2 \Ls{2}{{\textstyle{\frac{\pi}{3}}}}
\biggl\}
\nonumber \\ &&
+ \ep^4 \biggl\{
4 \left[ \Ls{3}{{\textstyle{\frac{2\pi}{3}}}} \right]^2
- {\textstyle{\frac{32}{9}}} 
  \Ls{2}{{\textstyle{\frac{\pi}{3}}}} \Ls{4}{{\textstyle{\frac{2\pi}{3}}}}
+ {\textstyle{\frac{4}{9}}} \pi \Ls{5}{{\textstyle{\frac{2\pi}{3}}}}
\nonumber \\ &&
+ 4 \pi \zeta_2 \Ls{3}{{\textstyle{\frac{2\pi}{3}}}}
+ {\textstyle{\frac{16}{3}}} \pi \zeta_3 \Ls{2}{{\textstyle{\frac{\pi}{3}}}}
+ {\textstyle{\frac{119}{2}}} \zeta_6
\biggl\}
+ {\cal O}(\varepsilon^5).
\label{Im_F10101}
\end{eqnarray}
In Appendix~B we present several terms of the $\ep$-expansion of
the $_3F_2$ function with more general parameters.

For another special case, $p_3^2=3m^2$ ($z=\tfrac{3}{4}$, 
$\theta=\tfrac{\pi}{3}$), we find
\begin{equation}
h_3\left(\tfrac{\pi}{3}\right) 
= -\tfrac{2}{9} \pi \zeta_2 \Ls{2}{\tfrac{\pi}{3}} 
  - \tfrac{4}{9} \pi \Ls{4}{\tfrac{\pi}{3}}
  + \tfrac{4}{3} \zeta_2 \zeta_3  
  + \tfrac{2}{3} \zeta_5 \; .
\end{equation}
Note that here we do not need that extra basis element $\chi_5$.

Let us discuss the appearance of the constant $\chi_5$ 
(\ref{sigma_32})
and its relation to transcendental numbers examined in
\cite{FK99}.  A detailed analysis of hypergeometric
functions of argument $z={\textstyle{1\over4}}$ (see Appendix~B) 
and Feynman diagrams 
(see Eqs.~(\ref{Hgg}), (\ref{f10101_re}) and (\ref{d4_ep}))
allows us to state that at the level ${\bf 5}$ there is a new 
(with respect to the constants defined in \cite{FK99}) independent 
irrationality related to the binomial sum (\ref{sigma_32}).
The set of sums or their linear combinations involving this
new irrationality (for a more detailed discussion, see Appendix~B.1)
consists of six elements:
$\Sigma_{1;1;2}^{2;1}$, $\Sigma_{1;1;3}^{1;1}$,
$\Sigma_{1;-;3}^{2;-}$ and three linear combinations
given in Appendix~B, Eq.~(\ref{connectedsum}).
For example,
\[
\chi_5 + 2\Sigma_{1;-;3}^{2;-}(1)
= \tfrac{4}{9} \pi \Ls{4}{\tfrac{\pi}{3}}
- \tfrac{13}{9} \zeta_2 \zeta_3 - \tfrac{47}{9} \zeta_5.
\]
Of course, instead of (\ref{sigma_32}) one could choose any other 
element of this set.
The linear independence of $\chi_5$
(\ref{sigma_32}) and other weight-${\bf 5}$
elements of the {\em odd} basis was 
established by {\sf PSLQ}-analysis of all these constants calculated 
with 800-decimal accuracy\footnote{The algorithm for high precision 
calculation of $\Ls{j}{\theta}$ and $\LS{j}{1}{\theta}$ functions  
is discussed in  Appendix~A of Ref.~\cite{KV00}. 
The corresponding programs can be found on Oleg Veretin's home page at
{\tt http://www.ifh.de/$\;\widetilde{}\;$veretin/MPFUN/mpfun.html} .}. 

Introduction of a new constant (\ref{sigma_32}) 
means that the ansatz for the {\em odd} basis of weight~${\bf 5}$
elaborated in Ref.~\cite{FK99} (see Table~1 of \cite{FK99})
should be re-analyzed.
We note that one of the elements,
$\tfrac{1}{\sqrt{3}}\Cl{5}{\frac{\pi}{3}}$,
is proportional to $\tfrac{1}{\sqrt{3}}\zeta_5$. 
Although this element is linearly independent, 
its appearance contradicts one of the mnemonic principles 
of constructing the ansatz: any non-factorizable
element may appear only once, with {\em or} without the
normalization factor $\tfrac{1}{\sqrt{3}}$. 
Since we already have $\zeta_5$ (without $\tfrac{1}{\sqrt{3}}$), 
this element can be omitted. Including, instead of it,
the new element $\chi_5$, we are able to save the rule $N_j=2^j$, 
whereas the simplicity of the ansatz gets spoiled, since,
as far as we know, the new element is not
expressible in terms of the Clausen or log-sine functions
of $\frac{\pi}{3}$ or $\frac{2 \pi}{3}$. 

Let us also consider cases when $h_j(\theta)$
can be expressed in terms of the {\em even} basis.
The threshold case, $p_3^2=4m^2$ ($z=1$,
$\theta=\tfrac{\pi}{2}$), is trivial:
\begin{equation}
{}_{3}F_2 \left(\begin{array}{c|} 1, 1, 1\!+\!\ep \\
\frac{3}{2}, 2 \end{array} ~1 \right) =
-\frac{1}{\ep} \int\limits_0^1 \mbox{d}\lambda
\frac{1\!-\!\lambda^{-2\ep}}{1-\lambda^2}
= \frac{1}{2\ep} \left[ \psi\left({\textstyle{1\over2}}\right)  
- \psi\left({\textstyle{1\over2}}\!-\!\ep\right) \right]
= \frac{1}{2} \sum_{j=0}^{\infty} \ep^j \left( 2^{j+2} \!-\! 1 \right)
\zeta_{j+2} \; .
\end{equation}
Therefore, we get
\begin{equation}
h_j\left(\tfrac{\pi}{2}\right) = \tfrac{1}{2} (2^{j+2}-1) \zeta_{j+2} .
\end{equation}
In particular, $h_3\left(\tfrac{\pi}{2}\right)=\tfrac{31}{2}\zeta_5$.

Another case of interest is $\theta=\tfrac{\pi}{4}$.
Here, the situation is rather similar to the case $\theta=\tfrac{\pi}{6}$.
Again, the calculation
of $h_3\left(\tfrac{\pi}{4}\right)$ happens to be non-trivial.
With the help of the {\sf PSLQ} program, we obtain
\begin{equation}
h_3\left(\tfrac{\pi}{4}\right) = 
\tfrac{1}{12} \tilde{\chi}_5 
- \tfrac{3}{4} \pi \Cl{4}{\tfrac{\pi}{2}}
+ \tfrac{1}{16} \pi \zeta_2 \Cl{2}{\tfrac{\pi}{2}}
+ \tfrac{1333}{512} \zeta_5 - \tfrac{91}{256} \zeta_2 \zeta_3 \; ,
\end{equation}
where
\begin{eqnarray}
\label{chi_tilde_5}
\widetilde{\chi}_5 \equiv \Sigma_{1;-;2}^{3;-}(2)
& = &  \sum_{n=1}^\infty \frac{(n!)^2}{(2n)!} \frac{2^n}{n^2} 
\left[ S_1(n-1) \right]^3
= \sum_{n=1}^\infty \frac{(n!)^2}{(2n)!} \frac{2^n}{n^2}
\left[ \psi(n)+\gamma \right]^3 \; 
\nonumber \\ & \simeq &
0.4951660770553611208168992757694958708378\ldots \; .
\end{eqnarray}
One can see that this multiple binomial sum is 
nothing but the $k=2$ version of the odd element $\chi_5$
given in Eq.~(\ref{sigma_32}).
Therefore, this element $\widetilde{\chi}_5$ should be 
added to the set of weight-{\bf 5} elements of the
{\em even} basis. A further discussion of the 
even-basis elements is given in Appendix~B.1.

\section{Two- and three-loop examples}
\setcounter{equation}{0}

\subsection{General remarks}

\begin{figure}[th]
\begin{center}
\centerline{\vbox{\epsfysize=70mm \epsfbox{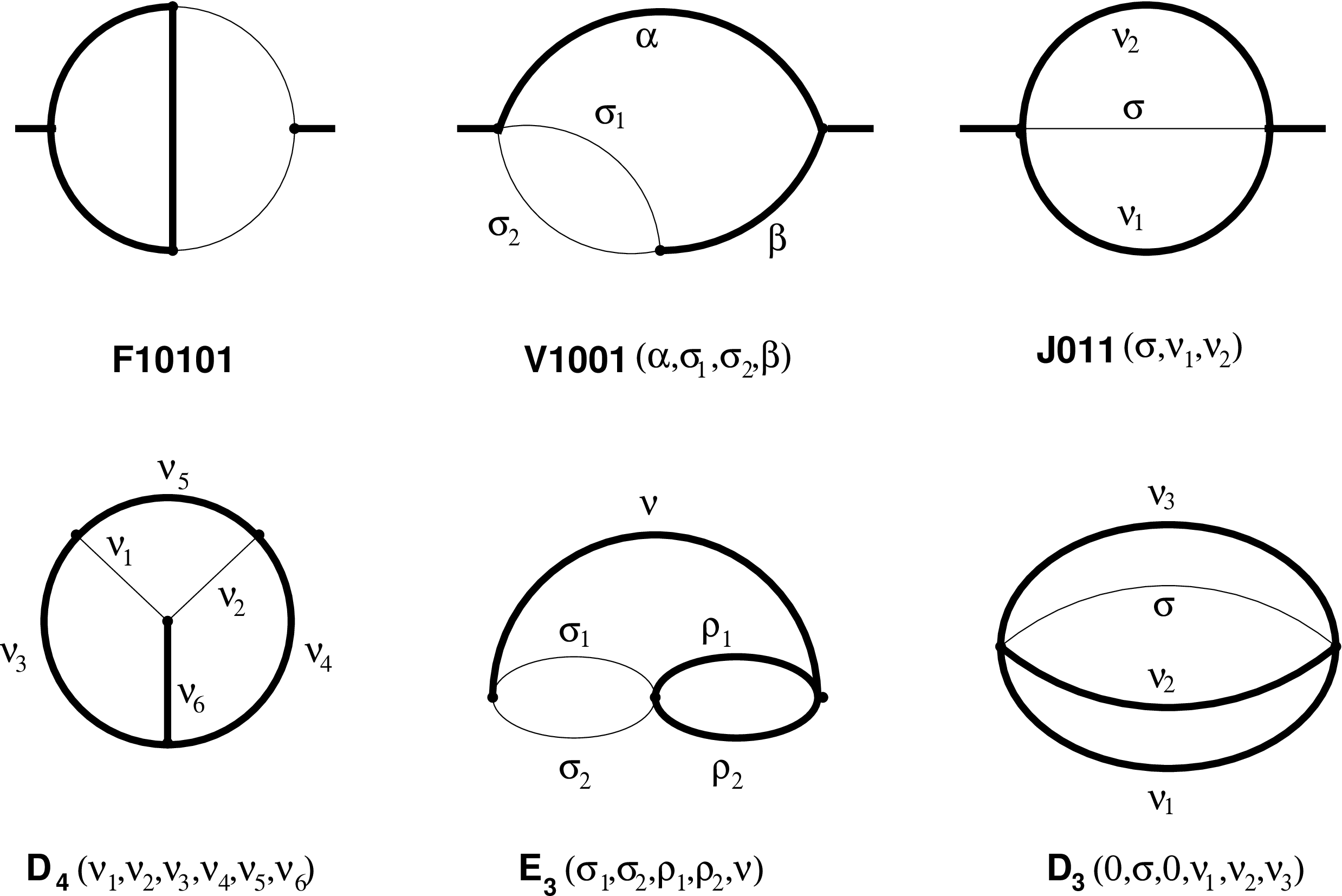}}}
\caption{\label{propagator} Two- and three-loop diagrams 
considered in the paper.
Bold and thin lines correspond to massive and
massless propagators, respectively.}
\end{center}
\end{figure}

Below we present the $\ep$-expansion of
several master integrals arising in {\sf FORM} \cite{FORM} packages 
for calculation of two-loop on-shell self-energy diagrams
\cite{onshell2} and three-loop vacuum integrals 
\cite{leo96,matad}\footnote{A review of modern computer 
packages for calculations in high-energy physics is given in~\cite{review}. 
Some useful details concerning recurrence relations for multiloop
integrals can be found in~\cite{baikov:grozin}.}. 
Our method of calculation is a combination of the Mellin--Barnes 
technique \cite{BD-TMF} and the {\sf PSLQ} analysis~\cite{PSLQ}: 
we present 
results for the {\em master} integrals in terms of 
hypergeometric functions, whose
expansions, given in Appendix~B, are obtained by means of {\sf PSLQ}.
In some cases we give the result for arbitrary mass and/or
momentum\footnote{We would like to mention that in
Ref.~\cite{three-different} divergent parts of some three-loop 
vacuum integrals with two different 
masses have been calculated,
for a special case when the ratio of the masses is $\sqrt{3}$.}. 
Sometimes, the integrals with higher powers of some propagators 
happen to be simpler from the $\ep$-expansion point of view 
(for instance, the integrals
having an ultraviolet-divergent
subgraph, like ${\bf V1001}$ or ${\bf J011}$ in 
Fig.~\ref{propagator}). In such cases we derive an expansion of these
``simple'' integrals and then reduce them to the
master integrals by means of packages \cite{onshell2} or \cite{leo96}. 

The integrals under consideration and notations (indices and mass
distributions) are shown in Fig.~\ref{propagator}. 
We are working in Minkowski space-time with dimension 
$n=4-2\varepsilon$. 
Moreover, each loop is multiplied by the normalization factor
$[{\rm i}\pi^{n/2}\Gamma(1+\varepsilon)]^{-1}$.
For example, the two-loop vacuum integrals with different masses,
which were denoted in Refs.~\cite{DT1,DT2,D-ep} as
$I(n;\nu_1,\nu_2,\nu_3;m_1,m_2,m_3)$,
are in the case of equal masses normalized as
\begin{equation}
\label{VL111}
\left. I(n;\nu_1,\nu_2,\nu_3;m,m,m)\right|_{m=1}
= - \pi^n \Gamma^2(1+\ep) \; {\bf VL111}(\nu_1,\nu_2,\nu_3)
\end{equation}
where we adopt the notation ${\bf VL111}$ from 
Ref.~\cite{FK99}\footnote{Two-loop bubble integrals are also implemented 
in the package \cite{vl111}.} (and also put $m=1$).

We can mention an example of physical calculations \cite{ChS99}, 
where the finite parts of
${\bf E_3}$, ${\bf D_3}$ (which was found in  \cite{FK99,ChS00}) 
and ${\bf D_4}$ (found in Ref.~\cite{B99}) have been used. 

\subsection{{\bf F10101}}

This integral is a good illustration of the 
application of general expressions given in Appendix~B.
The off-shell result for this integral in arbitrary 
dimension was presented in \cite{BFT93} (where it was called 
$\widetilde{I}_3$, see Eq.~(22) of \cite{BFT93}).
For unit powers of propagators, the result reads
\begin{eqnarray}
\label{tildeI3}
m^{2+4\ep}
(1-2\ep) F_{10101}(p^2,m) &=&
\frac{1}{(1-\ep^2)(1+2\ep)}
{}_{4}F_3 \left(\begin{array}{c|} 1, 1+\ep,1+\ep, 1+2\ep \\
\frac{3}{2} +\ep ,2+\ep, 2-\ep \end{array} ~\frac{p^2}{4m^2} \right)
\nonumber\\ &&
-\frac{1}{2 \ep (1+\ep)}
{}_{3}F_2 \left(\begin{array}{c|} 1, 1+\ep, 1+\ep \\
\frac{3}{2}, 2+\ep \end{array} ~\frac{p^2}{4m^2} \right)
\nonumber\\ &&
+ \frac{1}{2\ep}
\frac{\Gamma^2(1-\ep)}{\Gamma(1-2\ep)} \left(-\frac{m^2}{p^2}\right)^\ep
{}_{3}F_2 \left(\begin{array}{c|} 1, 1, 1+\ep \\
\frac{3}{2}, 2 \end{array} ~\frac{p^2}{4m^2} \right) \; .
\end{eqnarray}
The finite part of $F_{10101}$ is given in \cite{B90,FKV99}, whereas the
term linear in $\ep$ has been considered in \cite{KV00}.
An algorithm for the small momentum expansion of such diagrams 
(involving a ``zero-threshold'') has been constructed in Ref.~\cite{BDST}
where also explicit results for a few terms have been given,
for general values of the masses of massive particles, 
in the limit $\ep\to 0$. 
All these calculations agree with the result (\ref{tildeI3}). 

For $p^2>0$ this diagram 
has imaginary part, due to the
factor $(-m^2/p^2)^{\ep}$, which is related to the massless two-particle
cut. If $p^2<4m^2$, this is the only contribution to the imaginary part,
whereas for $p^2>4m^2$ we get additional contributions from the 
cuts involving two massive particles. Therefore, for $0<p^2<4m^2$ we get
\begin{equation}
m^{2+4\ep}
(1-2\ep) \mbox{Im}\; F_{10101}(p^2,m) =
\frac{\pi}{2} \frac{\Gamma(1-\ep)}{\Gamma(1+\ep) \Gamma(1-2\ep)} 
\left(\frac{m^2}{p^2}\right)^\ep
{}_{3}F_2 \left(\begin{array}{c|} 1, 1, 1+\ep \\
\frac{3}{2}, 2 \end{array} ~\frac{p^2}{4m^2} \right) \; ,
\label{ImF10101_gen}
\end{equation}
with the same $_3F_2$ function as in the three-point example in section~3.3.
This is expected, since, cutting the considered two-loop diagram
across the two massless lines, we get nothing but that triangle
diagram. Therefore, all results for the $\ep$-expansion 
($h_j(\theta)$ functions, etc.)
presented in section~3.4 are {\em directly} applicable to the
imaginary part of $F_{10101}$. 

In particular, in the on-shell
limit $p^2=m^2$ (we also put $m=1$) 
we follow the notation of Ref.~\cite{FKK99} and denote
\begin{equation}
{\bf F10101} \equiv \left. F_{10101} (p^2,m)\right|_{p^2=m^2; \;\; m=1} \; ,
\end{equation}
where it is understood that all powers of all propagators are equal to one.
This integral is used as one  of the master integrals in the
{\sf ONSHELL2} package \cite{onshell2}.

Using expressions given in Appendix~B, we can now present the 
$\ep$-expansion of the real part up to the $\ep^2$ term,
\begin{eqnarray}
(1-2\ep)\mbox{Re}\;{\bf F10101} &=&
\left[ -4 \zeta_3 + 2 \pi \Ls{2}{\tfrac{\pi}{3}}\right]
+\ep \left\{
\tfrac{16}{3} \left[ \Ls{2}{\tfrac{\pi}{3}} \right]^2
- 7 \pi \Ls{3}{\tfrac{2\pi}{3}} - \tfrac{488}{9} \zeta_4
\right\}
\nonumber \\&& \hspace*{-35mm}
+\ep^2 \Biggl\{
6 \pi \Ls{4}{\tfrac{2\pi}{3}}
- \tfrac{5}{2} \chi_5
- \tfrac{2111}{54} \zeta_2 \zeta_3 - \tfrac{521}{18} \zeta_5
+ \tfrac{56}{27} \pi \zeta_2 \Ls{2}{\tfrac{\pi}{3}}
- \tfrac{50}{81} \pi \Ls{4}{\tfrac{\pi}{3}}
\Biggr\}
+ {\cal O}(\varepsilon^3) . \hspace*{5mm}
\label{f10101_re}
\end{eqnarray}
The finite and linear in $\ep$ parts coincide with those given in 
\cite{FKK99} and \cite{KV00}, respectively\footnote{
We take into account the Erratum to \cite{FKK99} where
Eq.~(1) and Table~1 of \cite{FKK99} were corrected.
In fact, there is another misprint in Table I of 
Ref.~\cite{FKK99}: for the master integral {\bf F11111} the signs 
of the coefficients $a_1$ and $a_2$ should be changed: 
$a_1=-1$, $a_2=\tfrac{9}{2}$. The corrected result 
is presented in p.~541 of~\cite{onshell2} (see also in~\cite{BJ97}).}.
According to Eq.~(\ref{ImF10101_gen}),
the imaginary part can be extracted from Eq.~(\ref{Im_F10101}),
\begin{equation}
(1-2\ep) \mbox{Im}\;{\bf F10101} =
\frac{\pi}{2} \frac{\Gamma(1-\ep)}{\Gamma(1+\ep) \Gamma(1-2\ep)}
\left[ \mbox{r.h.s. of Eq.~(\ref{Im_F10101})} \right] \; ,
\label{f10101_im}
\end{equation}
which yields its $\ep$-expansion up to $\ep^4$.

\subsection{${\bf D_4}$}

The integral ${\bf D_4}(1,1,1,1,1,1)$ is another master integral
used in \cite{leo96,matad}.
The finite part of this integral was given in \cite{B99}.
Here we are going to obtain result for the $\ep$-term. 
It is easy to see that the integral ${\bf D_4}(1,1,1,1,1,\nu)$ can be
obtained by integrating the 
off-shell two-loop diagram $F_{10101}$ (see \cite{BFT93} and 
Eq.(\ref{tildeI3})) with a massive propagator raised to a power $\nu$. 
Using the following property (in our case $\mu^2=4m^2$)
\begin{eqnarray}
&& \hspace*{-20mm}  
\frac{1}{{\rm i} \pi^{n/2}} 
\int \frac{{\rm d}^n p}{(p^2-M^2)^\nu} 
~{}_{P}F_Q \left(\begin{array}{c|}
a_1, \cdots,  a_P \\
b_1,  \cdots, b_Q \end{array} ~ \frac{p^2}{\mu^2} \right) 
\nonumber \\ &=& 
(M^2)^{n/2-\nu}
\frac{\Gamma\left(\nu-\tfrac{n}{2}\right)}{\Gamma(\nu)}
~{}_{P+1}F_{Q+1} \left(\begin{array}{c|}
a_1, \cdots,  a_P, \tfrac{n}{2} \\
b_1,  \cdots, b_Q, \tfrac{n}{2}-\nu+1 \end{array} 
~\frac{M^2}{\mu^2}
\right) 
\nonumber \\ && 
+ (\mu^2)^{n/2-\nu}
\frac{\Gamma\left(\tfrac{n}{2}-\nu\right)}
     {\Gamma\left(\tfrac{n}{2}\right)}
\prod_{k=1}^{Q} \prod_{i=1}^{P}
\frac{\Gamma(b_k) \Gamma\left(\nu+a_i -\tfrac{n}{2}\right)}
     {\Gamma(a_i) \Gamma\left(\nu+b_k -\tfrac{n}{2}\right)}
\nonumber \\ && 
\times ~{}_{P+1}F_{Q+1} \left(\begin{array}{c|}
\nu+a_1-\tfrac{n}{2}, \ldots,  \nu+a_P-\tfrac{n}{2}, \nu \\
\nu-\tfrac{n}{2}+b_1,  \ldots, \nu-\tfrac{n}{2}+b_Q, \nu-\tfrac{n}{2}+1
 \end{array} ~\frac{M^2}{\mu^2}
\right), 
\end{eqnarray}
where $\{a_i, b_j\} \neq 0,-1,-2,\ldots$, we arrive at a result
which contains (for an arbitrary $\nu$) three $_4F_3$ functions
and three $_3F_2$ functions.  
For $\nu=1$, two of the three ${}_{4}F_3$ functions reduce to the 
${}_{3}F_2$ functions of the type considered in Appendix~B.
For some ${}_{3}F_2$ functions we use the the relation 
\begin{equation}
{}_{P}F_Q \left(\begin{array}{c|} a_1, \cdots, a_P \\ 
b_1, \cdots, b_Q \end{array} ~z \right)
= 1 + z \frac{a_1 \ldots a_P}{b_i \ldots b_Q}  
~{}_{P+1}F_{Q+1} \left(\begin{array}{c|} 1, 1+a_1,\cdots, 1+a_P \\ 
2, 1+b_1,\cdots, 1+b_Q \end{array} ~z \right) \; . 
\label{Kummer_PFQ}
\end{equation}
Then the master integral becomes  
\begin{eqnarray}
&& \hspace*{-10mm}
(1-\ep) (1-2\ep) {\bf D_4}(1,1,1,1,1,1) = 
\nonumber \\ && 
-\frac{1}{\ep(1-\ep)(1+\ep)(1+2\ep)}
~{}_{4}F_3 \left(\begin{array}{c|} 
1, 1+\ep, 1+\ep, 1+2\ep \\
\frac{3}{2}+\ep, 2+\ep, 2-\ep\end{array} ~\frac{1}{4} \right) 
\nonumber \\ && 
+ \frac{\Gamma(1-\ep) \Gamma^2(1+2\ep)\Gamma(1+3\ep)}
{\ep (1+2\ep)(1+4\ep)\Gamma^2(1+\ep) \Gamma(1+4\ep)} 
~{}_{4}F_3 \left(\begin{array}{c|} 
1, 1+2\ep, 1+2\ep, 1+3\ep \\
\frac{3}{2}+2\ep, 2, 2+2\ep\end{array} ~\frac{1}{4} \right) 
\nonumber \\ && 
+ \frac{1}{2\ep^2 (1+\ep)}
~{}_{3}F_2 \left(\begin{array}{c|}
1, 1+\ep, 1+\ep \\
\frac{3}{2}, 2+\ep\end{array} ~\frac{1}{4} \right)
-  \frac{\Gamma(1-\ep) \Gamma(1+2\ep)}{4\ep^2 \Gamma(1+\ep)}
~{}_{3}F_2 \left(\begin{array}{c|} 1, 1, 1+\ep \\
\frac{3}{2}, 2 \end{array} ~\frac{1}{4} \right)
\nonumber \\ &&
- \frac{1}{2\ep^2 (1+2\ep)^2}
~{}_{3}F_2 \left(\begin{array}{c|}
1, 1+2\ep, 1+2\ep \\
\frac{3}{2}+\ep, 2+2\ep\end{array} ~\frac{1}{4} \right)
+ \frac{\Gamma(1-\ep) \Gamma^2(1+2\ep) \Gamma(1+3\ep)}
{4\ep^4 \Gamma^2(1+\ep)\Gamma(1+4\ep)}
\nonumber \\ &&
+ 
\frac{\Gamma(1-\ep) \Gamma^2(1+2\ep)\Gamma(1+3\ep)}
{4\ep^2 (1+2\ep)(1+4\ep)\Gamma^2(1+\ep) \Gamma(1+4\ep)} 
~{}_{3}F_2 \left(\begin{array}{c|} 
1, 1+2\ep, 1+3\ep \\
\frac{3}{2}+2\ep, 2+2\ep\end{array} ~\frac{1}{4} \right) 
- \frac{1}{4\ep^4} \; .
\end{eqnarray}
Now all $_PF_Q$ functions belong to the types considered in
Appendix~B. Using those results, we obtain the following
result for the lowest terms of the $\ep$-expansion:
\begin{eqnarray}
&& \hspace*{-13mm}
(1-\ep) (1-2\ep) {\bf D_4}(1,1,1,1,1,1)
= \frac{2 \zeta_3}{\ep}
- \left\{ \tfrac{77}{12} \zeta_4 + 6\left[\Ls{2}{\tfrac{\pi}{3}} \right]^2
\right\}
\nonumber \\ && \hspace*{-7mm}
+ \ep \left\{
\tfrac{21}{8} \chi_5 - \tfrac{2615}{72} \zeta_2 \zeta_3
+ \tfrac{2047}{216} \zeta_5 + \tfrac{367}{54} \pi \zeta_2 \Ls{2}{\tfrac{\pi}{3}}
- \tfrac{161}{54} \pi \Ls{4}{\tfrac{\pi}{3}} + 7 \pi \Ls{4}{\tfrac{2\pi}{3}}
\right\}
+ {\cal O}(\ep^2) . 
\hspace*{5mm}
\label{d4_ep}
\end{eqnarray}
The finite term is in full agreement with the result of \cite{B99},
whereas the result for the $\ep$-term is new.
As we see, the new constant $\chi_5$, Eq.~(\ref{sigma_32}),
also appears in the $\ep$-part of this integral.

\subsection{${\bf V1001}$}

The off-shell integral $V_{1001}$ with two different masses and unit powers
of propagators was considered in \cite{v1001_general}. The result,
Eq.~(46) of \cite{v1001_general} is 
presented as a sum of $F_2$ and $F_4$ hypergeometric functions of two 
variables
\footnote{We note that the occurring $F_4$ function can be 
reduced to $_2F_1$ function of the type~(\ref{f_ep_1}), 
whereas the $F_2$ function can be transformed into $F_1$ function 
of two variables (see Eq.~(A.64) in \cite{Glover}).}.

Again, we define 
\begin{equation}
{\bf V1001}(\alpha,\sigma_1,\sigma_2,\beta)
\equiv \left. V_{1001}(\alpha,\sigma_1,\sigma_2,\beta;p^2,m,m) 
\right|_{p^2=m^2; \;\; m=1} \; .
\end{equation}

The finite part of on-shell master integral has been calculated 
in \cite{v1001_finite}, while the linear in $\ep$ term 
is presented in \cite{FKK99}. The on-shell diagram
${\bf V1001}(1,1,1,1)$ belongs to the set of 
master integrals used in the {\sf ONSHELL2} package \cite{onshell2}. 
Employing the Mellin--Barnes technique \cite{BD-TMF}
one can find the following result for an arbitrary set of the indices
(remember that we put $m=1$):
\begin{eqnarray}
{\bf V1001}(\alpha, \sigma_1,\sigma_2,\beta) \!\! &=& \!\!
\frac{\Gamma\left(\frac{n}{2}\!-\!\sigma_1\right) 
      \Gamma\left(\frac{n}{2}\!-\!\sigma_2\right)
      \Gamma\left(\sigma_1\!+\!\sigma_2\!-\!\frac{n}{2}\right)}
{\Gamma(\alpha) \Gamma (\sigma_1) \Gamma (\sigma_2) 
 \Gamma(n-\sigma_1-\sigma_2) \Gamma^2 \left(3-\frac{n}{2} \right)} 
\nonumber \\ && 
\times \Biggl\{ 
\frac{\Gamma(\alpha+\beta+\sigma_1+\sigma_2-n)
\Gamma(2n-2\beta-\alpha-2\sigma_1 -2 \sigma_2)}
{\Gamma\left(\frac{3}{2}n-\alpha-\beta-\sigma_1-\sigma_2\right)} 
\nonumber \\ && 
\times
~{}_{3}F_2 \left(\begin{array}{c|} 
\beta, \alpha+\beta+\sigma_1+\sigma_2-n, 
1 + \alpha+\beta+\sigma_1+\sigma_2 - \frac{3}{2} n \\
1 + \frac{\alpha}{2} + \beta+\sigma_1+\sigma_2-n,  
\tfrac{1}{2}(\alpha+1) +  \beta+\sigma_1+\sigma_2-n
\end{array} ~\frac{1}{4} \right) 
\nonumber  \\ && 
+\frac{\Gamma\left(\frac{\alpha}{2}\right) 
      \Gamma\left(\frac{\alpha}{2}+\beta+\sigma_1+\sigma_2-n\right)
      \Gamma\left(n-\sigma_1-\sigma_2-\frac{\alpha}{2}\right)}
{2 \Gamma(\beta) \Gamma\left(\frac{n-\alpha}{2}\right)} 
\nonumber \\ &&
\times
~{}_{3}F_2 \left(\begin{array}{c|}
\frac{\alpha}{2}, 1+\frac{\alpha}{2}-\frac{n}{2},
n - \sigma_1 - \sigma_2 - \frac{\alpha}{2} \\
\frac{1}{2},  1 + n - \beta - \sigma_1 - \sigma_2 - \frac{\alpha}{2}
\end{array} ~\frac{1}{4} \right)
\nonumber \\ &&
-\frac{\Gamma\left(\frac{\alpha+1}{2}\right) 
      \Gamma\left(\frac{\alpha-1}{2}+\beta+\sigma_1+\sigma_2-n\right)
      \Gamma\left(n-\sigma_1-\sigma_2-\frac{\alpha-1}{2}\right)}
{2 \Gamma(\beta) \Gamma\left(\frac{n-\alpha-1}{2}\right)} 
\nonumber \\ &&
\times
~{}_{3}F_2 \left(\begin{array}{c|}
\tfrac{1}{2}(\alpha\!+\!1), \tfrac{1}{2}(3\!+\!\alpha\!-\!n),
n - \sigma_1 - \sigma_2 + \tfrac{1}{2}(1 \!-\! \alpha) \\
\frac{3}{2},  
\frac{3}{2} - \beta - \sigma_1 - \sigma_2 - \frac{\alpha}{2} + n
\end{array} ~\frac{1}{4} \right)
\Biggr\} \; . 
\hspace*{7mm}
\label{v1001}
\end{eqnarray}

Let us consider the case $\sigma_1 = \sigma_2 =\beta = 1$.
For $\alpha=1$ (the master integral), and also for $\alpha=2$,  
the result (\ref{v1001}) can be expressed in terms of 
the $~{}_{2}F_1$ functions. 
For $\alpha=1$, they can be transformed using
the relation (\ref{Kummer1}) and another combination of Kummer relations
for contiguous functions,
\begin{equation}
_2F_1 \left(\begin{array}{c|} 1, -1+3\ep  \\
\frac{1}{2} + 2 \ep\end{array} ~z \right) =
1 - \frac{2(1-3\ep)z}{(1-2\ep)}
+ \frac{12\ep (1-3\ep) z (1-z)}{(1-2\ep) (1+4\ep)}   
~{}_{2}F_1 \left(\begin{array}{c|} 1, 1+3\ep  \\ 
\frac{3}{2} + 2 \ep\end{array} ~z \right) \; . \quad {}
\label{Kummer3}
\end{equation}
  
The integral with $\alpha=2$ can be reduced by 
the {\sf ONSHELL2} package \cite{onshell2}
to this master integral, plus some 
combination of $\Gamma$-functions: 
\begin{eqnarray}
{\bf V1001}(2,1,1,1) & = & -\tfrac{1}{3}(1-2\ep) {\bf V1001}(1,1,1,1)
+ \frac{\Gamma(1-\ep) \Gamma(1+2\ep)}{3 \ep^2 (1-2\ep)\Gamma(1 + \ep) }
\nonumber \\ && 
-\frac{(1- 4 \ep) \Gamma^2 (1 - \ep) \Gamma(1-4\ep)\Gamma(1+2\ep)}
{6 \ep^2 (1-3\ep) (1-2\ep) \Gamma(1-2\ep) \Gamma(1-3\ep) \Gamma(1 + \ep)}
\; .
\end{eqnarray}
This connection provides a non-trivial check on the results for 
the $\ep$-expansion of the occurring integrals.
The result for the master integral can be presented as 
\begin{eqnarray}
(1-2\ep)^2 {\bf V1001}(1,1,1,1)
&=&
-  \frac{3 \Gamma(1-\ep) \Gamma(1+2\ep)}{8 \ep \Gamma (1+\ep) }
~{}_{2}F_1 \left(\begin{array}{c|}
1, 1+\ep \\ \frac{3}{2}
\end{array} ~\frac{1}{4} \right)
\nonumber \\ && \hspace*{-15mm}
+ \frac{9  \Gamma^2 (1 - \ep) \Gamma(1-4\ep)\Gamma(1+2\ep)}
{ 8 \ep (1 + 4 \ep) \Gamma(1-2\ep) \Gamma(1-3\ep) \Gamma (1+\ep) }
~{}_{2}F_1 \left(\begin{array}{c|}
1, 1 + 3 \ep \\
\frac{3}{2} + 2 \ep
\end{array} ~\frac{1}{4} \right)
\nonumber \\ && \hspace*{-15mm}
+\frac{(1-\ep) \Gamma^2 (1 - \ep) \Gamma(1-4\ep)\Gamma(1+2\ep)}
{4\ep^2(1-3\ep) \Gamma(1-2\ep) \Gamma(1-3\ep) \Gamma (1+\ep)  } 
+ \frac{\Gamma(1-\ep) \Gamma(1+2\ep)}{4\ep^2 \Gamma (1+\ep)  }
\nonumber  \\ && \hspace*{-15mm}
- \frac{\pi 3^{\frac{1}{2}-\ep}\Gamma^3(1-\ep) \Gamma(1-4\ep) \Gamma(1+ 4\ep)}
{2 \ep \Gamma^3(1-2\ep) \Gamma(1+2\ep) \Gamma (1+\ep) } .
\label{V1001master}
\end{eqnarray}
Here we have two ${}_{2}F_1$ hypergeometric functions of the argument
$\tfrac{1}{4}$. 
The $\ep$-expansion of the first 
function can be extracted from \cite{D-ep} 
(see also Eqs.~(\ref{f_ep_1})--(\ref{f_ep}) of this paper),
\begin{equation}
{}_{2}F_1 \left(\begin{array}{c|} 1, 1+\ep  \\
\frac{3}{2} \end{array} ~\frac{1}{4} \right) 
= \frac{2 }{ 3^{\frac{1}{2} + \ep}}
\sum_{j=0}^\infty \frac{(2\ep)^j}{j!}   
\Biggl[ \Ls{j+1}{\tfrac{2 \pi}{3}} - \Ls{j+1}{\pi} \Biggr] \; , 
\label{2F1_1}
\end{equation}
whereas the second function can be represented as
\begin{equation}
{}_{2}F_1 \left(\begin{array}{c|} 1, 1+3\ep \\ 
\frac{3}{2} + 2 \ep \end{array} ~\frac{1}{4} \right) = 
\frac{2^{2+6\ep} (1+4\ep)}{3^{\frac{1}{2} + \ep}}
\int\limits_0^{\pi/6} {\rm d} \phi
( \sin\phi )^{4\ep} ( \cos\phi )^{2\ep} \; .
\end{equation}
When we expand in $\ep$ we get integrals of 
the products of various powers of $\ln(\sin\phi)$ and $\ln(\cos\phi)$.
An integral representation of this function for an arbitrary argument
is given in Eq.~(\ref{2f1_lsc}). Its $\ep$-expansion can be written in
terms of $\mbox{Lsc}$ function (for details, see Appendix~A.2). 
The expansion up to $\ep^4$ looks like
\begin{eqnarray} 
&& \hspace*{-7mm}
\frac{3^{\frac{1}{2} + \ep}}{4(1+4\ep)}
{}_{2}F_1 \left(\begin{array}{c|} 1, 1+3\ep \\
\frac{3}{2} + 2 \ep \end{array} ~\frac{1}{4} \right) =
\tfrac{1}{6} \pi
- (2 \ep) \tfrac{2}{3} \Ls{2}{\tfrac{\pi}{3}} 
+ (2 \ep)^2 
\left[  \tfrac{5}{72} \pi \zeta_2 
       - \tfrac{1}{2} \Ls{3}{\tfrac{2\pi}{3}} \right]
\nonumber \\ && 
+ (2 \ep)^3 \left[  \tfrac{11}{108} \pi \zeta_3 
- \tfrac{10}{27} \Ls{4}{\tfrac{\pi}{3}} 
- \tfrac{1}{6} \Ls{4}{\tfrac{2 \pi}{3}} \right]
\nonumber \\ && 
+ (2 \ep)^4 \left[  -\tfrac{1}{1152} \pi \zeta_4 
+ \tfrac{5}{4} \zeta_2 \Ls{3}{\tfrac{2 \pi}{3}}  
- \tfrac{25}{72} \Ls{5}{\tfrac{\pi}{3}} 
- \tfrac{1}{24} \Ls{5}{\tfrac{2 \pi}{3}} 
- \tfrac{5}{8} \pi \LS{4}{1}{\tfrac{2 \pi}{3}}
+ \tfrac{15}{32} \LS{5}{2}{\tfrac{2 \pi}{3}} 
\right] 
\nonumber \\ &&
+ {\cal O}(\ep^5) \; .
\hspace*{5mm}
\label{2F1_2}
\end{eqnarray}

These expressions allow us 
to fix an error in the $\ep$ part, presented in Eq.~(2)
of~\cite{FKK99}\footnote{
In Eq.~(2) and Table~2 of~\cite{FKK99}, results for two
integrals, ${\bf V1111}(1,1,1,1)$ (with all massive lines)
and ${\bf V1001}(1,1,1,1)$, are combined.
The result for ${\bf V1111}(1,1,1,1)$ is correct.
In order to correct the result for the $\ep$-term of
${\bf V1001}(1,1,1,1)$,
the following changes should be done: 
(i) the term $+9 S_2 \tfrac{\pi}{\sqrt{3}}$ 
(with $S_2=\tfrac{4}{9\sqrt{3}}\Ls{2}{\tfrac{\pi}{3}}$)
should read $+\tfrac{9}{4} b_4 S_2 \tfrac{\pi}{\sqrt{3}}$,
where $b_4=0$ for ${\bf V1001}(1,1,1,1)$, according to Table~2;
(ii) the coefficient $b_3$ of the $\zeta_3$ term (given in Table~1) 
should be $-\tfrac{3}{2}$, rather than $\tfrac{13}{6}$.
The result for ${\bf V1001}(1,1,1,1)$ on p.~546 of
\cite{onshell2} should also be corrected, according to
Eq.~(\ref{V1001corrected})}.
The correct expression is: 
\begin{eqnarray}
{\bf V1001}(1,1,1,1) &=&
\frac{1}{2\ep^2} 
+ \frac{1}{\ep} \left(\frac{5}{2}-\frac{\pi}{\sqrt{3}}\right)
+\tfrac{19}{2} + \tfrac{3}{2}\zeta_2 - 4\frac{\pi}{\sqrt{3}}
- 7 \frac{\Ls{2}{\tfrac{\pi}{3}}}{\sqrt{3}}
+ \frac{\pi}{\sqrt{3}} \ln3
\nonumber \\ &&
+ \ep \Biggl\{
\tfrac{65}{2} + 8 \zeta_2 + \tfrac{3}{2} \zeta_3
- 12 \frac{\pi}{\sqrt{3}} 
- 28 \frac{\Ls{2}{\frac{\pi}{3}}}{\sqrt{3}}
+ 7 \frac{\Ls{2}{\frac{\pi}{3}}}{\sqrt{3}} \ln3
\nonumber \\ &&
+ 4 \frac{\pi}{\sqrt{3}} \ln 3
- \frac{1}{2} \frac{\pi}{\sqrt{3}} \ln^2 3
- \frac{21}{2} \frac{\pi}{\sqrt{3}} \zeta (2)
- \frac{21}{2} \frac{\Ls{3}{\frac{2\pi}{3}}}{\sqrt{3}}
\Biggl\} + {\cal O}(\ep^2) . 
\hspace*{4mm}
\label{V1001corrected}
\end{eqnarray} 
Our expressions (\ref{V1001master})--(\ref{2F1_2}) provide two new terms,
$\ep^2$ and $\ep^3$, of the $\ep$-expansion of ${\bf V1001}(1,1,1,1)$.


\subsection{${\bf E_3}$}

The three-loop vacuum integral ${\bf E_3}$
with all indices equal to one
is one of the master integrals used in Avdeev's package  
\cite{leo96} and in {\sf MATAD} \cite{matad}.
The result for general values of the indices is
\begin{eqnarray}
&& \hspace*{-7mm}
{\bf E_3}(\sigma_1,\sigma_2,\rho_1,\rho_2,\nu) =
\frac{\Gamma (\frac{n}{2}-\sigma_1) \Gamma(\frac{n}{2}-\sigma_2)
      \Gamma(\sigma_1+\sigma_2-\frac{n}{2})}
{\Gamma (\sigma_1)\Gamma (\sigma_2) \Gamma(\frac{n}{2}) \Gamma^3(3-\frac{n}{2})} 
\nonumber \\ && \hspace*{-7mm}
\times \Biggl\{ 
\frac{\Gamma (\rho_1 + \rho_2- \frac{n}{2})\Gamma(\nu+\sigma_1+\sigma_2-n)}
{\Gamma (\nu)\Gamma (\rho_1+\rho_2) } 
~{}_{4}F_3 \left(\begin{array}{c|} \rho_1, \; \rho_2, \;
\rho_1 + \rho_2-\tfrac{n}{2}, n-\sigma_1-\sigma_2 \\
\tfrac{1}{2}(\rho_1\!+\!\rho_2),  
\tfrac{1}{2}(\rho_1\!+\!\rho_2\!+\!1), 
n\!-\!\nu\!-\!\sigma_1\!-\!\sigma_2\!+\!1 
\end{array} ~\frac{1}{4} \right)
 \nonumber \\ &&  \hspace*{-7mm}
+ \frac{\Gamma (\nu\!+\!\rho_1\!+\! \sigma_1 \!+\! \sigma_2 \!-\! n)
\Gamma(\nu\!+\!\rho_2 \!+\! \sigma_1 \!+\! \sigma_2 \!-\! n)
\Gamma\left(\nu\!+\!\rho_1\!+\!\rho_2
            \!+\!\sigma_1\!+\!\sigma_2\!-\!\tfrac{3n}{2}\right) 
\Gamma(n\!-\!\nu\!-\!\sigma_1\!-\!\sigma_2)}
{\Gamma (\rho_1) \Gamma (\rho_2) 
\Gamma (2\nu+ \rho_1+\rho_2 + 2\sigma_1 + 2\sigma_2-2n) 
\Gamma(n-\sigma_1-\sigma_2)} 
\nonumber \\ && \hspace*{-7mm}
\times
{}_{4}F_3 \left(\begin{array}{c|} 
\nu, \nu+\rho_1+ \sigma_1 + \sigma_2 - n,
\nu+\rho_2 + \sigma_1 + \sigma_2 - n,
\nu+\rho_1+\rho_2+\sigma_1+\sigma_2-\frac{3n}{2} \\
\nu \!+\! \sigma_1 \!+\! \sigma_2 
\!+\! \tfrac{1}{2}(\rho_1 \!+\! \rho_2) \!-\! n,
\nu \!+\! \sigma_1 \!+\! \sigma_2 
\!+\! \tfrac{1}{2}(\rho_1 \!+\! \rho_2 \!+\! 1) \!-\! n,
\nu \!+\! \sigma_1 \!+\! \sigma_2 \!-\!n \!+\! 1 
\end{array} ~\frac{1}{4} \right) 
\Biggr\} .
\nonumber \\
\end{eqnarray}
For the purpose of expanding in $\ep$,
the integral ${\bf E_3}(1,1,1,2,1)$ is the simplest one,
\begin{eqnarray}
{\bf E_3}(1,1,1,2,1) &=&
- \frac{1}{4\ep^2(1-\ep)(1-2\ep)}
\frac{\Gamma(1-\ep) \Gamma(1+2\ep)}{\Gamma(1+\ep)}
\Biggl\{ ~{}_{2}F_1 \left(\begin{array}{c|} 1, 1 + \ep \\
\frac{3}{2} \end{array} ~\frac{1}{4} \right)
\nonumber \\ &&
-\frac{2\Gamma(1+2\ep) \Gamma(1+3\ep)}
      {3 \ep \Gamma(1+\ep) \Gamma(1+4\ep)}
~{}_{2}F_1 \left(\begin{array}{c|} 1, 3\ep \\
\frac{1}{2}+2\ep \end{array} ~\frac{1}{4} \right) 
\Biggl\} \; .
\end{eqnarray}
The $\ep$-expansion of the first $_2F_1$
function is given in Eq.~(\ref{2F1_1}), whereas the second one
can be reduced to (\ref{2F1_2}) via Kummer relation
\begin{equation}
{}_{2}F_1 \left(\begin{array}{c|} 1, 3\ep \\
\frac{1}{2}+2\ep \end{array} ~z \right)
= 1 + \frac{6\ep z}{1+4\ep}\;
{}_{2}F_1 \left(\begin{array}{c|} 1, 1+3\ep \\
\frac{3}{2}+2\ep \end{array} ~z \right) \; ,
\label{Kummer2}
\end{equation}
which is a particular case of (\ref{Kummer_PFQ}).

Then, using recurrence relations \cite{leo96}, ${\bf E_3}(1,1,1,2,1)$ 
can be related to
to the master integral ${\bf E_3}(1,1,1,1,1)$ as
\begin{eqnarray}
(1-2\ep) {\bf E_3}(1,1,1,1,1)  &=&
- 3 {\bf E_3}(1,1,1,2,1)
- \frac{\Gamma(1-\ep) \Gamma(1+2\ep)}{2\ep^3(1-\ep)(1-2\ep) \Gamma(1+\ep)}
\nonumber \\ &&
+ \frac{(1-4\ep)\Gamma(1-\ep) \Gamma^2(1+2\ep) \Gamma(1+3\ep)}
{3\ep^3 (1-\ep)(1-2\ep)(1-3\ep) \Gamma(1+4\ep) \Gamma^2(1+\ep)} \; .
\end{eqnarray}
In this way we get
\begin{eqnarray}
(1 - 2\ep)^2 {\bf E_3}(1,1,1,1,1)    &=& 
\frac{3 \Gamma(1-\ep) \Gamma(1+2\ep)}{4 \ep^2 (1-\ep) \Gamma (1+\ep)}
~{}_{2}F_1 \left(\begin{array}{c|} 1, 1 + \ep \\
\frac{3}{2} \end{array} ~\frac{1}{4} \right)
\nonumber \\ && 
-  \frac{3 \Gamma(1-\ep)\Gamma^2 (1+2 \ep) \Gamma (1+3 \ep)}
{4 \ep^2 (1-\ep) (1+4\ep) \Gamma (1+4\ep) \Gamma^2 (1+\ep)} 
~{}_{2}F_1 \left(\begin{array}{c|} 1, 1+3\ep \\
\frac{3}{2} + 2 \ep \end{array} ~\frac{1}{4} \right) 
\nonumber \\ && 
- \frac{\Gamma(1-\ep) \Gamma (1+2 \ep)}{2 \ep^3(1-\ep) \Gamma (1+\ep)}
- \frac{\Gamma(1-\ep) \Gamma^2 (1+2 \ep)\Gamma(1+3\ep)}
{6 \ep^3 (1-3\ep) \Gamma(1+4\ep)  \Gamma^2 (1+\ep)} , \quad {}
\label{E3master}
\end{eqnarray}
with the same $_2F_1$ functions as in the case of ${\bf V1001}$.
The $\ep$-expansion of these ${}_{2}F_1$ functions is given 
in (\ref{2F1_1}) and (\ref{2F1_2}). 
Thus, for this integral we have also produced two new 
($\ep^2$ and $\ep^3$) terms of the $\ep$-expansion, 
reaching the level of 6-loop calculations (see in~\cite{ChS99}). 
We also find an error in the $\ep$-part of the result given in Eq.~(10)
of \cite{FK99}, which should read (remember that we put $m=1$ and 
$\Ls{2}{\tfrac{\pi}{3}} = \tfrac{9\sqrt{3}}{4}S_2$)
\begin{eqnarray}
&& \hspace*{-7mm}
{\bf E_3}(1,1,1,1,1)  
= -\frac{2}{3\ep^3} - \frac{11}{3\ep^2}  
+ \frac{1}{\ep} \left\{ -14 + 6 \frac{\Ls{2}{\tfrac{ \pi}{3}}}{\sqrt{3}} 
- \zeta_2 \right\}
\nonumber \\ &&
+ \left\{ -\tfrac{139}{3} + 30\frac{\Ls{2}{\tfrac{ \pi}{3}}}{\sqrt{3}}
- 6 \frac{\Ls{2}{\tfrac{ \pi}{3}}}{\sqrt{3}} \ln 3 
- 5 \zeta_2 - \tfrac{1}{3}\zeta_3 
+ \tfrac{5}{3} \frac{\pi}{\sqrt{3}} \zeta_2 
+ 9 \frac{\Ls{3}{\tfrac{2\pi}{3}}}{\sqrt{3}} \right\}
\nonumber \\ &&
+\ep \Biggl\{ - \tfrac{430}{3}
+ 102 \frac{\Ls{2}{\tfrac{\pi}{3}}}{\sqrt{3}}
- 30 \frac{\Ls{2}{\tfrac{\pi}{3}}}{\sqrt{3}} \ln3    
+ 3 \frac{\Ls{2}{\tfrac{\pi}{3}}}{\sqrt{3}} \ln^2 3
\nonumber \\ && 
+ 45 \frac{\Ls{3}{\tfrac{2\pi}{3}}}{\sqrt{3}}
- 9 \frac{\Ls{3}{\tfrac{2\pi}{3}}}{\sqrt{3}} \ln 3
+ \tfrac{80}{9} \frac{\Ls{4}{\tfrac{ \pi}{3}}}{\sqrt{3}}
+ 6 \frac{\Ls{4}{\tfrac{2\pi}{3}}}{\sqrt{3}}
- 17 \zeta_2
\nonumber \\ && 
+ \tfrac{25}{3} \frac{\pi}{\sqrt{3}} \zeta_2
- \tfrac{5}{3}  \frac{\pi}{\sqrt{3}} \zeta_2 \ln 3
- \tfrac{13}{3} \zeta_3
- \tfrac{94}{9} \frac{\pi}{\sqrt{3}} \zeta_3
+ \tfrac{3}{2} \zeta_4
+ 4 \zeta_2 \frac{\Ls{2}{\tfrac{\pi}{3}}}{\sqrt{3}} 
\Biggl\}
+ {\cal O}(\ep^2) \; . 
\end{eqnarray}


\subsection{${\bf D_3}$ and ${\bf J011}$ }

The off-shell result for the sunset-type integral $J_{011}$ with arbitrary
powers of propagators has been obtained in \cite{D91,BFT93}, by using
the Mellin--Barnes technique \cite{BD-TMF}:
\begin{eqnarray}
J_{011}(\sigma,\nu_1,\nu_2;p^2,m) \!\! &=& \!\! 
(m^2)^{n-\sigma-\nu_1-\nu_2}
\frac{
\Gamma(\nu_1\!+\!\nu_2\!+\!\sigma\!-\!n)
\Gamma\left(\frac{n}{2}\!-\!\sigma\right)
\Gamma\left(\nu_2\!+\!\sigma\!-\!\frac{n}{2}\right)
\Gamma\left(\nu_1\!+\!\sigma\!-\!\frac{n}{2}\right)}
{\Gamma(\nu_1)\Gamma(\nu_2)\Gamma\left(\frac{n}{2}\right)
\Gamma(\nu_1+\nu_2+2\sigma-n) \Gamma^2 \left(3-\frac{n}{2}\right)}
\nonumber\\ && \times
{}_{4}F_3 \left( \begin{array}{c|} 
\sigma,\nu_1+\nu_2+\sigma-n,
\nu_2+\sigma-\tfrac{n}{2}, \nu_1+\sigma-\tfrac{n}{2} \\
\tfrac{n}{2}, \sigma + \tfrac{1}{2}(\nu_1+\nu_2-n), 
\sigma+ \tfrac{1}{2}(\nu_1+\nu_2+1-n)
\end{array}~\frac{p^2}{4 m^2} \right) .
\label{j011_off-shell}
\end{eqnarray}
The three-loop vacuum integral ${\bf D_3}$ can be obtained by integrating
(\ref{j011_off-shell}) over $p$,
\begin{equation}
{\bf D_3}(0,\sigma,0,\nu_1, \nu_2, \nu_3)
= \frac{1}{{\rm i} \pi^{n/2}\Gamma\left(3-\frac{n}{2} \right)} 
\int \frac{{\rm d}^n p}{(p^2-m^2)^{\nu_3}}\left.
J_{011}(\sigma,\nu_1,\nu_2;p^2,m)\right|_{m=1} \; .
\end{equation}
In such a way we find
\begin{eqnarray}
&& \hspace*{-10mm}
{\bf D_3}(0,\sigma,0,\nu_1, \nu_2, \nu_3) = 
\frac{\Gamma\left(\tfrac{n}{2}-\sigma\right)}
{\Gamma (\nu_1)\Gamma (\nu_2) \Gamma(\nu_3) \Gamma\left(\tfrac{n}{2}\right)
\Gamma^3 \left(3-\frac{n}{2}\right)} 
\nonumber \\ && \hspace*{-7mm}
\times \Biggl\{ 
\frac{\Gamma\left(\nu_1 + \sigma - \tfrac{n}{2}\right) 
      \Gamma\left(\nu_2 + \sigma - \tfrac{n}{2}\right)
\Gamma(\nu_1 + \nu_2 + \sigma -n) \Gamma\left(\nu_3 - \tfrac{n}{2}\right)}
{\Gamma (\nu_1+\nu_2 + 2 \sigma - n) } 
\nonumber \\ && \hspace*{-7mm}
\times
~{}_{4}F_3 \left(\begin{array}{c|} 
\sigma, \nu_1 + \sigma - \tfrac{n}{2}, \nu_2 + \sigma - \tfrac{n}{2}, 
\nu_1 + \nu_2 + \sigma - n \\
\tfrac{n}{2}-\nu_3+1, \sigma + \tfrac{1}{2}(\nu_1 + \nu_2 -n ), 
\sigma + \tfrac{1}{2}(\nu_1 + \nu_2 -n + 1)
\end{array} ~\frac{1}{4} \right) 
\nonumber \\ && \hspace*{-7mm}
+\frac{\Gamma\left(\tfrac{n}{2}-\nu_3\right)
\Gamma\left(\nu_3+ \sigma-\tfrac{n}{2}\right)
\Gamma(\nu_1\!+\!\nu_3\!+\!\sigma\!-\!n)
\Gamma(\nu_2 \!+\! \nu_3 \!+\! \sigma \!-\! n)
\Gamma(\nu_1 \!+\! \nu_2 \!+\! \nu_3 \!+\! \sigma \!- \tfrac{3n}{2})}
{\Gamma(\sigma) \Gamma (\nu_1 + \nu_2 + 2 \nu_3 + 2 \sigma - 2n) } 
\nonumber \\ && \hspace*{-7mm}
\times
~{}_{4}F_3 \left(\begin{array}{c|} 
\nu_3 + \sigma - \tfrac{n}{2}, \nu_1 + \nu_3 + \sigma - n,
\nu_2 + \nu_3 + \sigma - n, 
\nu_1 + \nu_2 + \nu_3 + \sigma - \tfrac{3n}{2} \\
\nu_3 - \tfrac{n}{2} +1,
\nu_3 + \tfrac{1}{2}(\nu_1 + \nu_2) + \sigma -n,
\nu_3 + \tfrac{1}{2}(\nu_1 + \nu_2 + 1) + \sigma -n
\end{array} ~\frac{1}{4} \right) 
\Biggr\} . 
\hspace*{5mm}
\label{D3-general}
\end{eqnarray}
The integral ${\bf D_3}(0,1,0,1,1,1)$ is also one of the 
master integrals
used in the packages \cite{leo96,matad}. The result including
the $\ep$-term is available in \cite{FK99} and \cite{ChS00}. 
Here we are going to provide some further 
terms of the $\ep$-expansion.

Again, when using the general result (\ref{D3-general}) it is simpler 
to consider an integral with shifted indices, namely
${\bf D_3}(0,1,0,2,2,2)$, which is related to the master one
through a simple relation \cite{leo96}
\begin{equation}
9 {\bf D_3}(0,1,0,2,2,2) =
- \frac{(4-15\ep)}{\ep^3 (1- \ep) }
+ 2 (1-3\ep) (1- 2\ep) (2-3\ep) {\bf D_3}(0,1,0,1,1,1) \; .
\label{D3red}
\end{equation}
Considering Eq.~(\ref{D3-general}) at $\sigma=1$, $\nu_1=\nu_2=\nu_3=2$,
we get
\begin{eqnarray}
(1-\ep) {\bf D_3}(0,1,0,2,2,2) 
& = & \frac{1}{ \ep(1+2\ep)}
~~{}_{3}F_2 \left(\begin{array}{c|} 1, 1 + \ep, 1 + 2 \ep \\
 \frac{3}{2} + \ep, 1-\ep \end{array} ~\frac{1}{4} \right)
\nonumber \\ && \hspace*{-25mm}
- \frac{1}{\ep(1+4\ep)} 
\frac{\Gamma(1-\ep)\; \Gamma^2(1+2\ep)\; \Gamma(1+3\ep)}
       { \Gamma^2(1+\ep)\; \Gamma(1+4\ep)}\;
{}_{2}F_1 \left(\begin{array}{c|} 1 + 2 \ep , 1 + 3 \ep \\
 \frac{3}{2} + 2 \ep  \end{array} ~\frac{1}{4} \right) \; . 
\hspace*{7mm}
\label{D3_1222}
\end{eqnarray}
Moreover, the occurring $_2F_1$ function can be reduced to
a product of $\Gamma$-functions\footnote{Eq.~(\ref{cubic1})
can be obtained via a simple transformation of Eq.~(31)
on p.~495 of \cite{PBM3}. A standard way to obtain
formulae of such type is to use Bayley's cubic transformations
of hypergeometric functions.}
\begin{equation}
{}_{2}F_1 \left(\begin{array}{c|} 1+2\ep, 1+ 3 \ep\\
\frac{3}{2} + 2 \ep \end{array} ~\frac{1}{4} \right) =
\frac{2\pi}{3^{\frac{3}{2}+3\ep}}
\frac{\Gamma(2+4 \ep)}{\Gamma(1+2\ep)\Gamma^2(1+\ep)} \; ,
\label{cubic1}
\end{equation}
see (\ref{2F1_D3}). 
The $_3F_2$ function in Eq.~(\ref{D3_1222}) belongs to
one of the types considered in Appendix~B. Its $\ep$-expansion
is given by
\begin{eqnarray}
&& \hspace*{-7mm}
{}_{3}F_2 \left(\begin{array}{c|} 1, 1+\ep, 1 + 2 \ep \\
\frac{3}{2} + \ep, 1-\ep \end{array} ~\frac{1}{4} \right)  
= \frac{1+2\ep} {3^{\frac{1}{2}+3 \ep} }
\biggl\{ \tfrac{2}{3} \pi  + 4 \ep \Ls{2}{\tfrac{\pi}{3}}
\nonumber \\ &&
+  \ep^2 \left[\tfrac{28}{3} \pi \zeta_2 + 18 \Ls{3}{\tfrac{2\pi}{3}} \right]
-  \ep^3 \left[\tfrac{112}{3} \pi \zeta_3 + \tfrac{32}{3} \Ls{4}{\tfrac{\pi}{3}}
- 36 \Ls{4}{\tfrac{2\pi}{3}} \right]
\nonumber \\ &&
+  \ep^4 \left[216 \zeta_2 \Ls{3}{\tfrac{2\pi}{3}}
+ \tfrac{679}{6} \pi \zeta_4
- \tfrac{196}{3} \Ls{5}{\tfrac{\pi}{3}}
+ 54 \Ls{5}{\tfrac{2 \pi}{3}}
- 108 \pi \LS{4}{1}{\tfrac{2\pi}{3}}
+ 81 \LS{5}{2}{\tfrac{2\pi}{3}}
\right]
\nonumber \\ &&
+ {\cal O} (\ep^5)
\biggr\} .
\label{mixing}
\end{eqnarray}   

Finally, we obtain for the master integral:
\begin{eqnarray}
2 \ep (1-\ep) (1- 2\ep) (1-3\ep) (2-3\ep)  
{\bf D_3}(0,1,0,1,1,1) &=&
\frac{9}{(1+ 2\ep)}\;
{}_{3}F_2 \left(\begin{array}{c|} 1, 1 + \ep, 1 + 2 \ep \\
 \tfrac{3}{2} + \ep, 1-\ep \end{array} ~\frac{1}{4} \right)
\nonumber \\ && \hspace*{-40mm}
+ \frac{(4-15\ep)}{\ep^2 }
-  \frac{6\pi}{3^{\frac{1}{2}+3\ep}}
\frac{\Gamma(1+2 \ep)\Gamma(1-\ep) \Gamma(1+3\ep)}
{\Gamma^4(1+\ep)} \; . 
\hspace*{10mm}
\label{D3master}
\end{eqnarray}

Considering result (\ref{D3-general}) with a different
set of indices, $\sigma=\nu_1=1$ and
$\nu_2=\nu_3=2$, we get
\begin{eqnarray}
{\bf D_3}(0,1,0,1,2,2) &=&
- \frac{1}{(1-\ep)^2 (1+ 2\ep) \ep} \;
{}_{3}F_2 \left(\begin{array}{c|} 1, 1 + \ep, 1 + 2 \ep \\
\frac{3}{2} + \ep, 2 - \ep \end{array} ~\frac{1}{4} \right)
\nonumber \\ && 
+  \frac{1}{3 (1-\ep) \ep^3 }
\frac{\Gamma^2(1+2 \ep)\Gamma(1-\ep) \Gamma(1+3\ep)}
{\Gamma(1+4\ep)  \Gamma^2(1+\ep)}\;
{}_{2}F_1 \left(\begin{array}{c|} 2 \ep , 3 \ep \\
\frac{1}{2} + 2 \ep  \end{array} \frac{1}{4} \right) .
\hspace*{7mm}
\end{eqnarray}  
On the other hand, using the recurrence package \cite{leo96}
we obtain a much simpler result for the same integral,
\begin{equation}
{\bf D_3}(0,1,0,1,2,2) = \frac{1}{3 \ep^3 (1-\ep)} \; .
\end{equation}
Therefore, we arrive at the following
non-trivial relation between hypergeometric functions:
\begin{equation}
{}_{3}F_2 \left(\begin{array}{c|} 1, 1\!+\!\ep, 1 \!+\! 2 \ep \\
\frac{3}{2} + \ep, 2-\ep \end{array} ~\frac{1}{4} \right)
= \frac{(1\!-\!\ep)(1\!+\!2\ep)}{3 \ep^2}
\Biggl[
\frac{\Gamma^2(1\!+\!2 \ep)\Gamma(1\!-\!\ep) \Gamma(1\!+\!3\ep)}
{\Gamma^2(1+\ep)\Gamma(1+4\ep) }
{}_{2}F_1 \left( \!\begin{array}{c|} 2 \ep, 3 \ep \\
\frac{1}{2}\!+\! 2 \ep \end{array} ~\frac{1}{4} \right)
-1 \Biggr] .
\label{relation1}
\end{equation}

The same relation can be found by considering the integral (A.23)
from Ref.~\cite{Vicari}, which they denote as $G_0(1;1,1;1,2)$.
This integral is nothing but ${\bf E_4}(1,1,1,2,1)$ 
in notations of Ref.~\cite{leo96}, which does not belong 
to the set of master integrals. 
By using recurrence relations, it can be reduced to the two-loop  
vacuum integral with equal masses (\ref{VL111})
and some trivial part,
\begin{equation}
{\bf E_4}(1,1,1,2,1) = - \frac{1}{3 \ep}  
{\bf VL111}(1,1,1)
- \frac{1}{3 \ep^3 (1-\ep) (1-2\ep)} \; .
\label{E_4}
\end{equation}
The result for ${\bf VL111}(1,1,1)$
in terms of hypergeometric
function is given by Eq.~(33) from Ref.~\cite{BFT93},
\begin{equation}
{\bf VL111}(1,1,1)
= - \frac{1}{(1-\ep) (1-2\ep) \ep^2}
\Biggl[
\frac{\pi \ep}{2} 3^{\frac{1}{2} - \ep}
\frac{\Gamma(1+2\ep)}{\Gamma^2(1+\ep) }
+ \frac{3}{2} (1-2\ep)
{}_{2}F_1 \left(\begin{array}{c|} 1, \ep \\
\frac{3}{2} \end{array} ~\frac{1}{4} \right)
\Biggr] \; .
\label{VL111_111}
\end{equation}
Considering the difference of the two results, one given in~\cite{Vicari}
and another presented in Eqs.~(\ref{E_4})--(\ref{VL111_111}),
we reproduce the relation (\ref{relation1}).

It is interesting that the same function (\ref{relation1}) 
can be obtained from 
Eq.~(\ref{j011_off-shell}), at special values of the indices 
$\sigma=1$, $\nu_1=\nu_2=2$:
\begin{eqnarray}
{\bf J011}(1,2,2) &=&
\frac{1}{(1-\ep) (1+2\ep) } \;
{}_{3}F_2 \left(\begin{array}{c|} 1, 1+\ep, 1 + 2 \ep \\
\frac{3}{2} + \ep, 2-\ep \end{array} ~\frac{1}{4} \right)
\nonumber \\ 
& = & \frac{1}{3 \ep^2}
\Biggl[
\frac{\Gamma^2(1+2 \ep)\Gamma(1-\ep) \Gamma(1+3\ep)}{\Gamma^2(1+\ep)\Gamma(1+4\ep) }
{}_{2}F_1 \left(\begin{array}{c|} 2 \ep, 3 \ep \\
\frac{1}{2} + 2 \ep \end{array} ~\frac{1}{4} \right)
-1 \Biggr] \; ,
\label{J011_simple}
\end{eqnarray}
where 
\begin{equation}
{\bf J011}(\sigma, \nu_1, \nu_2) \equiv 
\left. J_{011}(\sigma, \nu_1, \nu_2; p^2,m)
\right|_{p^2=m^2; \;\; m=1}\; ,
\end{equation}
and we have used relation (\ref{relation1}).
For the same integral ${\bf J011}(1,2,2)$, 
applying Pad\'e approximants calculated from the small
momentum expansion \cite{taylor}
with the {\sf PSLQ}-based analysis 
(for details, see \cite{FKK99,FK99}),
we restore several terms of the $\ep$-expansion:
\begin{eqnarray}
{\bf J011}(1,2,2) &=&
\tfrac{2}{3} \zeta_2
- \ep \tfrac{2}{3} \zeta_3 + \ep^2 3 \zeta_4
- \ep^3 \left \{ 2 \zeta_5 + \tfrac{4}{3} \zeta_2 \zeta_3 \right\}
+ \ep^4 \left \{ \tfrac{61}{6} \zeta_6 + \tfrac{2}{3} \zeta_3^2
\right\}
\nonumber \\  
&&
- \ep^5 \left \{ 6 \zeta_7 + 4 \zeta_2 \zeta_5 + 6 \zeta_3
\zeta_4 \right\}
+ \ep^6 \left \{ \tfrac{1261}{36}\zeta_8 + 4 \zeta_5 \zeta_3   
+ \tfrac{4}{3} \zeta_2 \zeta_3^2 \right\}
\nonumber \\
&&
- \ep^7
\left \{ \tfrac{170}{9} \zeta_9 + 12 \zeta_2 \zeta_7 + \tfrac{61}{3}  
\zeta_3 \zeta_6 + 18 \zeta_4 \zeta_5 + \tfrac{4}{9} \zeta_3^3
\right\}
\nonumber \\
&&
+ \ep^8
\left \{\tfrac{4977}{40} \zeta_{10} + 12 \zeta_3 \zeta_7 + 6 \zeta_5^2
+ 6 \zeta_3^2 \zeta_4 + 8 \zeta_2 \zeta_3 \zeta_5 \right\}   
+ {\cal O} (\ep^9) .
\label{numeric}
\end{eqnarray}
Using (\ref{numeric}), the result for the $_2F_1$ function occurring
in (\ref{J011_simple}) can be deduced as:
\begin{eqnarray}
{}_{2}F_1 \left(\begin{array}{c|} 2 \ep, 3 \ep \\
\frac{1}{2} + 2 \ep \end{array} ~\frac{1}{4} \right) =
\frac{\Gamma(1+\ep)\Gamma(1+4\ep)}{\Gamma(1+2\ep)\Gamma(1+3\ep) } .
\label{deduce}
\end{eqnarray}
Therefore, we get
\begin{eqnarray}
&&
\frac{1}{(1-\ep) (1+2\ep)}
{}_{3}F_2 \left(\begin{array}{c|} 1, 1+\ep, 1+2 \ep \\
\frac{3}{2} + \ep, 2-\ep \end{array} ~\frac{1}{4} \right) =
\frac{1}{3 \ep^2}
\Biggl[ \frac{\Gamma(1+2 \ep)\Gamma(1-\ep)}{\Gamma(1+\ep) } -1 \Biggr] \; ,
\end{eqnarray}
\begin{eqnarray}
{\bf J011}(1,2,2) = \frac{1}{3 \ep^2}
\Biggl[ \frac{\Gamma(1+2 \ep)\Gamma(1-\ep)}{\Gamma(1+\ep) } -1 \Biggr].
\label{122}
\end{eqnarray}
We note that the results similar to Eq.~(\ref{deduce}) are usually obtained using
Bailey's cubic transformations for hypergeometric functions.

Let us remind that for the on-shell integral ${\bf J011}$ there are two
master integrals \cite{T97a} of this type,
${\bf J011}(1,1,1)$ and 
${\bf J011}(1,1,2)$. 
Using the fact that the integral (\ref{122}) is a linear combination 
of these integrals, we obtain
\begin{eqnarray}
{\bf J011}(1,1,2)
= - \frac{\Gamma(1+2 \ep)\Gamma(1-\ep)}{6 \ep^2 (1-2\ep)  \Gamma(1+\ep)} 
-  \frac{(2-3\ep)}{3}  {\bf J011}(1,1,1) \; .
\label{j011_master}
\end{eqnarray}
Alternatively, one can consider other independent combinations
of these master integrals: for example, ${\bf J011}(1,2,2)$ and 
$[{\bf J011}(1,2,2)+2{\bf J011}(2,1,2)$] (see in \cite{thresholds}).
While ${\bf J011}(1,2,2)$ is given in Eq.~(\ref{122}), 
the second combination is less trivial and cannot be
represented in terms of the $\Gamma$-function.
Using relation (see, e.g., in \cite{PBM3})
\begin{equation}
{}_{3}F_2 \left(\begin{array}{c|} 2, 1+\ep, 1+2 \ep \\ 
\frac{3}{2} + \ep, 2-\ep \end{array} ~z \right)
-\ep {}_{3}F_2 \left(\begin{array}{c|} 1, 1+\ep, 1+2 \ep \\ 
\frac{3}{2} + \ep, 2- \ep \end{array} ~z \right)
= 
(1-\ep)
~~{}_{3}F_2 \left(\begin{array}{c|} 1, 1+\ep, 1+2 \ep \\ 
\frac{3}{2} + \ep, 1-\ep \end{array} ~z \right)
\end{equation}
we obtain
\begin{eqnarray}
&& 
{\bf J011}(1,2,2) + 2 {\bf J011}(2,1,2) 
= - \frac{1}{\ep(1+2\ep)} 
~{}_{3}F_2 \left(\begin{array}{c|} 1, 1+\ep, 1 + 2 \ep \\
\frac{3}{2} + \ep, 1-\ep \end{array} ~\frac{1}{4} \right).
\label{linear}
\end{eqnarray}
This is the same $_3F_2$ function as in the case of ${\bf D_3}$,
whose $\ep$-expansion is given in (\ref{mixing}). 
The lowest terms of the $\ep$-expansion of Eq.~(\ref{linear}) 
coincide with Eq.~(9) of Ref.~\cite{FK99}, which has been
obtained via {\sf PSLQ} analysis, based on the small momentum 
expansion.  

\section{Conclusions}
\setcounter{equation}{0}

In this paper we have studied some important issues 
related to the $\ep$-expansion of Feynman diagrams
in the framework of dimensional regularization
\cite{dimreg}. 

For the one-loop two-point function with different
masses $m_1$ and $m_2$, the analytic continuation (\ref{general})
of known results \cite{Crete,D-ep} to other regions
of interest has been constructed. In particular,
explicit formulae,~(\ref{Ls<->S}) and (\ref{even})--(\ref{coef}), 
relating the log-sine integrals
and the generalized (Nielsen) polylogarithms have
been obtained to an arbitrary order.

Then, we have examined some physically-important 
examples of the $\ep$-expansion of the
one-loop three-point function. 
For the cases of the off-shell massless triangles,
as well as for a specific on-shell triangle diagram with
two different masses, the $\ep$-expansion and 
the procedure of analytic continuation is, basically,
similar to those for the two-point function.

However, the situation for another interesting example, 
a massive triangle loop with $p_1^2=p_2^2=0$,
appears to be much more complicated. 
We have shown that all even terms of the expansion,
i.e.\ the coefficients of $\ep^{2l}$, can be presented
in terms of the log-sine integrals. In the meantime,
for the odd terms (starting from $\ep^3$) the result
does not seem to be expressible in terms of known 
functions (related to the polylogarithms), although
its one-fold integral representation looks reasonably simple.
Nevertheless, for some special values of $p_3^2$ we have 
identified the corresponding $\ep^3$ contributions
in terms of known transcendental numbers, expanding
the hypergeometric functions and using the {\sf PSLQ}
procedure \cite{PSLQ}.

In particular, we have found that one extra term should
be added to the odd weight-{\bf 5} basis considered 
in \cite{FK99}. This new constant~(\ref{sigma_32}) can be related to
a special case of the multiple binomial sums~(\ref{binsum}).
We also constructed the {\em even} basis up to weight {\bf 5}
(see Introduction and Appendix B.1).

Then, we have obtained a number of new results for the 
higher terms of the $\ep$-expansion for certain 
two-loop (mainly, on-shell) two-point
functions (see Eqs.~(\ref{f10101_re}), (\ref{V1001master}))
and three-loop vacuum diagrams,
Eqs.~(\ref{d4_ep}), (\ref{E3master}), (\ref{D3master}).
The considered examples serve as the master integrals
in the analytic computer packages \cite{leo96,onshell2,matad}.
In particular, the obtained $\ep$-terms of three-loop
vacuum diagrams are important in the four-loop-order
calculations. 

Within these calculations, a new relation~(\ref{relation1}) between 
${}_3F_2$ and ${}_2F_1$ hypergeometric functions of argument 
$\tfrac{1}{4}$ was found.
Moreover, the analytical solution Eq.~(\ref{122}) makes it possible
to express one of the on-shell integrals, ${\bf J011}(1,2,2)$, 
in terms of $\Gamma$-functions. In this way, we have reduced 
the number of {\em non-trivial} master integrals~(\ref{j011_master})
used in the package~\cite{onshell2}.

One of the interesting issues related to the {\em odd} and {\em even}
basis construction 
is the connection between the generalized 
log-sine integrals and the multiple $\zeta$ values which have
appeared earlier in Refs.~\cite{zeta53:1,zeta53:2},
in the study of the connection between knot and quantum field theory. 
Namely, our Eq.~(\ref{pi/3_zoo2}) shows 
that $\zeta_{5,3}$, $\zeta_{7,3}$ and $\zeta_{3,7,3}$
are connected with the combinations 
$\left[ \LS{j}{1}{\tfrac{\pi}{3}}-\tfrac{\pi}{3} \Ls{j-1}{\tfrac{\pi}{3}}
 \right]$ 
(plus ordinary $\zeta$-functions), where $j$ is the weight 
of the corresponding multiple $\zeta$ value. 
Besides this, we have found that the constant $U_{5,1}$ (see
in~\cite{euler-basis})
can be expressed in terms of the {\em even} basis~(\ref{u51}).

\vspace{5mm}

{\bf Acknowledgments.}
A.~D.'s research was supported by the Deutsche Forschungsgemeinschaft. 
Partial support from 
the Australian Research Council grant No.~A00000780 and 
the RFBR grant No.~98-02-16981 is also acknowledged. 
M.~K. is grateful to the ThEP group (University of Mainz) for their
hospitality during his research stay, which was supported by BMBF
under contract 05~HT9UMB~4.
We are grateful to  D.J.~Broadhurst, F.~Jegerlehner, O.V.~Tarasov
and  O.~Veretin for useful discussions. 
Especially, we would like to thank
O.~Veretin for his help in numerical calculations.

\newpage
{\bf Note added (May 9, 2017)} \\
Over the years, a few typos were found in the published version of this paper (Nucl.\ Phys.\ B {\bf 605} (2001) 266). For completeness, we list them below.
\begin{itemize}
\item
Eq.~(2.7) and Eq.~(B.11): \\
The power of $\cos \theta$ should be $1+2\ep$ rather than $1-2\ep$.\\
The correct expression has been given in Eq.~(2.10) of  \\
A.~I.~Davydychev and M.~Yu.~Kalmykov,
Nucl.\ Phys.\ B {\bf 699} (2004) 3 [hep-th/0303162].
\item
Eq.~(4.10): \\
The coefficient at $\pi \Ls{4}{\frac{\pi}{3}}$
should be $\frac{161}{54}$ instead of  $\frac{161}{154}$. \\
This typo was mentioned in footnote~1 of \\
M.~Yu.~Kalmykov,
Nucl.\ Phys.\ B {\bf 718} (2005) 276 [hep-ph/0503070].
\item
Eq.~(4.25): \\
The finite part of master-integral should include $\frac{5}{3}\frac{\pi}{\sqrt{3}} \zeta_2$ 
rather than $\frac{5}{3}\frac{\pi}{\sqrt{3}}$.\\
The correct expression has been given in Eq.~(10) of\\
J.~Fleischer and M.~Yu.~Kalmykov,
Phys.\ Lett.\ B {\bf 470} (1999) 168 [hep-ph/9910223].
\end{itemize}
We are indebted to Andrey Grozin for discussions and independent checks.

\newpage
\appendix

\section{Polylogarithms and related functions}
\setcounter{equation}{0}
\subsection{Polylogarithms and log-sine integrals}

The polylogarithm $\Li{j}{z}$ is defined as
\begin{equation}
\Li{j}{z} = \frac{(-1)^j}{(j-1)!} \int\limits_0^1
{\rm d}\xi \; \frac{\ln^{j-1}\xi}{\xi-z^{-1}}
= \sum_{k=1}^{\infty} \frac{z^k}{k^j} \; .
\end{equation}

When the argument $z$ belongs to the unit circle in the
complex $z$-plane, $z=e^{{\rm i}\theta}$, we get
(see, e.g., in \cite{Lewin})
\begin{equation}
\Li{2l}{e^{{\rm i}\theta}} = \Gl{2l}{\theta}
+ {\rm i} \Cl{2l}{\theta}, \quad \quad 
\Li{2l+1}{e^{{\rm i}\theta}} = \Cl{2l+1}{\theta}
+ {\rm i} \Gl{2l+1}{\theta},  
\label{Cl_j_}
\end{equation}
where $\Gl{j}{\theta}$ is proportional to Bernoulli
polynomial of the order $j$, $B_j(\theta/(2\pi))$
(see Eq.~(22) in p.~300 of \cite{Lewin}), whereas
$\Cl{j}{\theta}$ is the Clausen function,
\begin{eqnarray}
\Cl{2l}{\theta} & = & 
-\frac{\sin\theta}{(2l-1)!} \int\limits_0^1
\frac{{\rm d}\xi\; \ln^{2l-1}\xi}{1-2\xi\cos\theta+\xi^2}
= \sum_{k=1}^{\infty} \frac{\sin(k\theta)}{k^{2l}},
\\
\Cl{2l+1}{\theta} & = & 
-\frac{1}{(2l)!} \int\limits_0^1
\frac{{\rm d}\xi\; (\xi-\cos\theta)\; \ln^{2l}\xi}
{1-2\xi\cos\theta+\xi^2}
= \sum_{k=1}^{\infty} \frac{\cos(k\theta)}{k^{2l+1}} .
\label{Cl_j}
\end{eqnarray} 

Below we list some useful properties of the Clausen function 
(for details see in \cite{Lewin}):
\[
\Cl{2l+1}{\tfrac{\pi}{3}} =  \tfrac{1}{2} \left(1-2^{-2l} \right)
\left(1-3^{-2l} \right)\; \zeta_{2l+1}, \quad\!\!
\Cl{2l+1}{\tfrac{2 \pi}{3}} = 
- \tfrac{1}{2} \left(1-3^{-2l} \right)\; \zeta_{2l+1}, 
\]
and 
\[
\Cl{2l}{\tfrac{\pi}{3}} =  \left(1+2^{1-2l}\right)\;
\Cl{2l}{\tfrac{2\pi}{3}} \; .
\]

When we consider the imaginary and real parts of 
$\Li{j}{1\!-\!e^{{\rm i}\theta}}$, also the log-sine function,
\begin{equation}
\Ls{j}{\theta} =
- \int\limits_0^\theta {\rm d}\phi \;
      \ln^{j-1} \left| 2\sin\frac{\phi}{2}\right| \, ,
\label{log-sin}
\end{equation}
and the generalized log-sine function,
\begin{equation}
\LS{j}{k}{\theta} =   - \int\limits_0^\theta {\rm d}\phi \;
   \phi^k \ln^{j-k-1} \left| 2\sin\frac{\phi}{2}\right| \, ,
\label{log-sin-gen}
\end{equation}   
get involved,
\begin{eqnarray}
\mbox{Re}\; \Li{3}{1\!-\!e^{{\rm i}\theta}} 
\! & = & \! \tfrac{1}{2} \left[ \zeta_3 - \Cl{3}{\theta} \right] 
 + \tfrac{1}{4} \theta^2 l_{\theta}, 
\nonumber \\
\mbox{Im}\; \Li{3}{1\!-\!e^{{\rm i} \theta}} 
\! & = & \! \tfrac{1}{24} \theta^3
- \tfrac{1}{2} \theta l_{\theta}^2  
- \Cl{2}{\theta} l_{\theta} + \tfrac{1}{2} \Ls{3}{\theta},
\nonumber \\
\mbox{Re}\; \Li{4}{1\!-\!e^{{\rm i} \theta}} 
\! & = & \! \tfrac{1}{4} \LS{4}{1}{\theta}
- \tfrac{1}{4} \theta \Ls{3}{\theta}
+ \tfrac{1}{8} \theta^2 l_{\theta}^2 
+ \tfrac{1}{2} \left[\zeta_3 -  \Cl{3}{\theta}\right] l_{\theta}
- \tfrac{1}{192} \theta^4 ,
\nonumber \\
\mbox{Im}\; \Li{4}{1\!-\!e^{{\rm i} \theta}} 
\! & = & \! -\tfrac{1}{6} \Ls{4}{\theta}
+ \tfrac{1}{2} \Ls{3}{\theta} l_{\theta}
- \tfrac{1}{6} \theta l_{\theta}^3 
- \tfrac{1}{2} \Cl{2}{\theta} l_{\theta}^2 
+ \tfrac{1}{24} \theta^3 l_{\theta}
- \tfrac{1}{4} \Cl{4}{\theta} + \tfrac{1}{4} \theta \zeta_3,
\nonumber \\
\mbox{Re}\; \Li{5}{1\!-\!e^{{\rm i} \theta}} \! & = & \!
- \tfrac{1}{12} \LS{5}{1}{\theta}
+ \tfrac{1}{8} \left[ \zeta_5 - \Cl{5}{\theta} \right]
+ \tfrac{1}{4} \LS{4}{1}{\theta}  l_{\theta}
+ \tfrac{1}{4} \left[ \zeta_3 - \Cl{3}{\theta} \right] l_{\theta}^2 
\nonumber \\  &&
- \tfrac{1}{4} \theta \Ls{3}{\theta} l_{\theta}
+  \tfrac{1}{12} \theta \Ls{4}{\theta}
- \tfrac{1}{16} \theta^2 \zeta_3
+ \tfrac{1}{24} \theta^2 l_{\theta}^3 
- \tfrac{1}{192} \theta^4 l_{\theta},
\nonumber \\
\mbox{Im}\; \Li{5}{1\!-\!e^{{\rm i} \theta}}  \! & = & \!
 \tfrac{1}{24} \Ls{5}{\theta}
-\tfrac{1}{16} \LS{5}{2}{\theta}
- \tfrac{1}{4} \Cl{4}{\theta} l_{\theta}
- \tfrac{1}{6} \Ls{4}{\theta} l_{\theta}
+ \tfrac{1}{4} \Ls{3}{\theta} l_{\theta}^2 
- \tfrac{1}{6} \Cl{2}{\theta} l_{\theta}^3  
\nonumber \\ &&
+  \tfrac{1}{8} \theta \LS{4}{1}{\theta}
-  \tfrac{1}{24} \theta l_{\theta}^4 
+  \tfrac{1}{4} \theta \zeta_3 l_{\theta}
-  \tfrac{1}{16} \theta^2 \Ls{3}{\theta}
+  \tfrac{1}{48} \theta^3 l_{\theta}^2 
-  \tfrac{1}{1920} \theta^5 ,
\nonumber \\ 
\mbox{Re}\; \Li{6}{1\!-\!e^{{\rm i} \theta}} \! & = & \!
  \tfrac{1}{48} \LS{6}{1}{\theta} \!
- \tfrac{1}{96} \LS{6}{3}{\theta} \!
+ \tfrac{1}{12} \left[ \zeta_3 \!-\! \Cl{3}{\theta} \right] l_{\theta}^3\! 
+ \tfrac{1}{8} \left[ \zeta_5 \!-\! \Cl{5}{\theta} \right] l_{\theta}\!
+ \tfrac{1}{8} \LS{4}{1}{\theta}  l_{\theta}^2 
\nonumber \\ &&
- \tfrac{1}{12}  \LS{5}{1}{\theta} l_{\theta}
- \tfrac{1}{8} \theta \Ls{3}{\theta} l_{\theta}^2 
+ \tfrac{1}{12} \theta \Ls{4}{\theta} l_{\theta}
- \tfrac{1}{48} \theta \Ls{5}{\theta} 
+ \tfrac{1}{32} \theta \LS{5}{2}{\theta}
\nonumber \\ &&
+ \tfrac{1}{96} \theta^2 l_{\theta}^4 
- \tfrac{1}{16} \theta^2 \zeta_3 l_{\theta}
- \tfrac{1}{32} \theta^2 \LS{4}{1}{\theta}
+ \tfrac{1}{96} \theta^3 \Ls{3}{\theta}
- \tfrac{1}{384} \theta^4 l_{\theta}^2
+ \tfrac{1}{23040} \theta^6 ,
\nonumber \\
\mbox{Im}\; \Li{6}{1\!-\!e^{{\rm i} \theta}} \! & = & \!
  \tfrac{1}{48}  \LS{6}{2}{\theta} 
- \tfrac{1}{120} \Ls{6}{\theta}
- \tfrac{1}{16}  \Cl{6}{\theta}
+ \tfrac{1}{24} \Ls{5}{\theta} l_{\theta}
- \tfrac{1}{16} \LS{5}{2}{\theta} l_{\theta}
\nonumber \\ && 
- \tfrac{1}{12} \Ls{4}{\theta} l_{\theta}^2 
- \tfrac{1}{8} \Cl{4}{\theta} l_{\theta}^2 
+ \tfrac{1}{12} \Ls{3}{\theta} l_{\theta}^3 
- \tfrac{1}{24} \Cl{2}{\theta} l_{\theta}^4 
- \tfrac{1}{120} \theta l_{\theta}^5 
+ \tfrac{1}{8} \theta \zeta_3 l_{\theta}^2 
\nonumber \\ && 
+ \tfrac{1}{8} \theta \LS{4}{1}{\theta} l_{\theta}
- \tfrac{1}{24} \theta \LS{5}{1}{\theta} 
+ \tfrac{1}{16} \theta \zeta_5
+ \tfrac{1}{48} \theta^2 \Ls{4}{\theta}
- \tfrac{1}{16} \theta^2 \Ls{3}{\theta} l_{\theta}
\nonumber \\ && 
+ \tfrac{1}{144} \theta^3 l_{\theta}^3 
- \tfrac{1}{96} \theta^3 \zeta_3
- \tfrac{1}{1920} \theta^5 l_{\theta},
\label{Lewin+}
\end{eqnarray}
where $l_{\theta} \equiv \ln \left| 2 \sin\frac{\theta}{2} \right|$,
$-\pi\leq\theta\leq\pi$.
Results for $\Li{j}{1\!-\!e^{{\rm i} \theta}}$ with $j\leq 4$,
as well as an outline how to get results for higher $j$'s,
can be found in \cite{Lewin}\footnote{A factor ${\textstyle{1\over2}}$
is missing in front of $\Ls{3}{\theta}$ in
eq.~(49) on p.~298 of \cite{Lewin}, whereas his Eq.~(6.56) is correct.
In eq.~(7.67) of \cite{Lewin},  
as well as in eq.~(36) on p.~301, the coefficient of
$\log^2\left(2\sin{\textstyle{1\over2}}\theta\right)\Cl{2}{\theta}$
should be $-\textstyle{1\over2}$ (rather than $+\textstyle{3\over2}$).}.

According to the definition (\ref{log-sin}), the following
integral and differential relations hold:
\begin{equation}
\LS{j+k}{i+k}{\theta} = \theta^k \LS{j}{i}{\theta}
- k \int\limits_0^\theta {\rm d}\phi\;
{\phi}^{k-1} \LS{j}{i}{\phi} \; , \quad
\label{integral_ls}
\frac{{\rm d}}{{\rm d} \theta} \LS{j+k}{i+k}{\theta} =
\theta^k \frac{{\rm d}}{{\rm d} \theta} \LS{j}{i}{\theta}
\end{equation}
In addition, using the fact that $\Ls{2}{\theta}=\Cl{2}{\theta}$ and
taking into account the integration rules for the Clausen function,
\[
\Cl{2n}{\theta} =  \int\limits_0^\theta {\rm d}\phi\; \Cl{2n-1}{\phi} ,
\quad
\Cl{2n+1}{\theta} =  \zeta_{2n+1}
- \int\limits_0^\theta {\rm d}\phi\; \Cl{2n}{\phi},
\]
one can obtain Eqs.~(7.52) and (7.53) of \cite{Lewin}. In particular,
the function $\LS{k+2}{k}{\theta}$ is always expressible
in terms of Clausen functions,
\begin{eqnarray}
\LS{3}{1}{\theta}  & = & \theta \Cl{2}{\theta} + \Cl{3}{\theta} -
\zeta_3,
\nonumber \\
\LS{4}{2}{\theta}  & = & \theta^2 \Cl{2}{\theta} + 2 \theta
\Cl{3}{\theta} - 2 \Cl{4}{\theta},
\nonumber \\
\LS{5}{3}{\theta}  & = & \theta^3 \Cl{2}{\theta} + 3 \theta^2
\Cl{3}{\theta} - 6 \theta \Cl{4}{\theta} - 6 \Cl{5}{\theta} + 6
\zeta_5,
\nonumber \\
\LS{6}{4}{\theta}  & = & 24 \Cl{6}{\theta} - 24 \theta \Cl{5}{\theta}
- 12 \theta^2 \Cl{4}{\theta}  + 4 \theta^3 \Cl{3}{\theta}
+ \theta^4 \Cl{2}{\theta} ,
\nonumber
\end{eqnarray}
etc.

It is also known that the values of $\Ls{j}{\pi}$
can be expressed in terms of $\zeta$-function, for any $j$
(see in \cite{Lewin}):
\[
\Ls{2}{\pi} = 0, \quad\!\!\!
\Ls{3}{\pi} =  -\tfrac{1}{2} \pi \zeta_2, \quad\!\!\!
\Ls{4}{\pi} =  \tfrac{3}{2} \pi \zeta_3, \quad\!\!\!
\Ls{5}{\pi} = -\tfrac{57}{8} \pi \zeta_4, \quad\!\!
\Ls{6}{\pi} = \tfrac{45}{2} \pi \zeta_5
+ \tfrac{15}{2} \pi \zeta_2 \zeta_3 ,
\]
etc. Moreover, 
\[
\Ls{3}{\tfrac{\pi}{3}} = -\tfrac{7}{18} \pi \zeta_2, \quad
\LS{4}{1}{\tfrac{\pi}{3}} = -\tfrac{17}{72} \zeta_4 \; .
\]

For $\theta=\tfrac{\pi}{3}$ there are some relations 
among the Clausen function and log-sine integrals, 
\begin{eqnarray}
\label{pi/3_zoo1}
\Cl{4}{\tfrac{\pi}{3}}  \! & = & \! \tfrac{2}{9} \Ls{4}{\tfrac{\pi}{3}}
- \tfrac{1}{9} \pi \zeta_3,
\nonumber \\
\Cl{6}{\tfrac{\pi}{3}} \! & = & \! \tfrac{2}{135} \Ls{6}{\tfrac{\pi}{3}}
- \tfrac{1}{9} \pi \zeta_5
- \tfrac{7}{81} \pi \zeta_2 \zeta_3 \; .
\end{eqnarray}
The result for $\Cl{4}{\tfrac{\pi}{3}}$ was given in \cite{FK99}.
Moreover, the quantities $\LS{j}{1}{\tfrac{\pi}{3}}$ can
be expressed in terms of $\pi\Ls{j-1}{\tfrac{\pi}{3}}$
and Euler--Zagier sums, 
\begin{eqnarray}
\label{pi/3_zoo2}
\LS{5}{1}{\tfrac{\pi}{3}} \! & = & \! \tfrac{1}{3} \pi
\Ls{4}{\tfrac{\pi}{3}}
-\tfrac{19}{4} \zeta_5 - \tfrac{1}{2} \zeta_2 \zeta_3,
\nonumber \\
\LS{6}{1}{\tfrac{\pi}{3}} \! & = & \! \tfrac{1}{3} \pi \Ls{5}{\tfrac{\pi}{3}}
+ \tfrac{2029}{96} \zeta_6 + 2 \zeta_3^2,
\nonumber \\
\LS{7}{1}{\tfrac{\pi}{3}} \! & = & \! \tfrac{1}{3} \pi \Ls{6}{\tfrac{\pi}{3}}
- \tfrac{2465}{32} \zeta_7 - \tfrac{15}{2} \zeta_5 \zeta_2
- \tfrac{205}{8} \zeta_4 \zeta_3
\nonumber \\
\LS{8}{1}{\tfrac{\pi}{3}} \! & = & \! \tfrac{1}{3} \pi \Ls{7}{\tfrac{\pi}{3}}
+ \tfrac{140879}{256} \zeta_8 + \tfrac{285}{2} \zeta_5 \zeta_3
+ \tfrac{15}{2} \zeta_3^2 \zeta_2 + \tfrac{21}{2} \zeta_{5,3},
\nonumber \\
\LS{9}{1}{\tfrac{\pi}{3}} \! & = & \! \tfrac{1}{3} \pi \Ls{8}{\tfrac{\pi}{3}}
- \tfrac{487235}{192} \zeta_9 - \tfrac{945}{4} \zeta_7 \zeta_2
- \tfrac{71015}{64} \zeta_6 \zeta_3 - \tfrac{12915}{16} \zeta_5 \zeta_4
- 35 \zeta_3^3,
\nonumber \\
\LS{10}{1}{\tfrac{\pi}{3}} \! & = & \! \tfrac{1}{3} \pi \Ls{9}{\tfrac{\pi}{3}}
+ \tfrac{14059623}{512} \zeta_{10} + \tfrac{51765}{8} \zeta_7 \zeta_3
+ \tfrac{57105}{16} \zeta_5^2 + \tfrac{4305}{4} \zeta_3^2 \zeta_4
+ 630 \zeta_2 \zeta_3 \zeta_5 + \tfrac{3075}{16} \zeta_{7,3},
\nonumber \\
\LS{11}{1}{\tfrac{\pi}{3}} \! & = & \! \tfrac{1}{3} \pi \Ls{10}{\tfrac{\pi}{3}}
- \tfrac{76115403}{512} \zeta_{11} 
- \tfrac{26775}{2} \zeta_9 \zeta_2
- \tfrac{8875377}{128} \zeta_8 \zeta_3
- \tfrac{348705}{8} \zeta_7 \zeta_4
- \tfrac{1917405}{32} \zeta_6 \zeta_5
\nonumber \\ && 
- \tfrac{17955}{2} \zeta_5 \zeta_3^2
- 315 \zeta_3^3 \zeta_2
- \tfrac{1323}{2} \zeta_{5,3} \zeta_3
- \tfrac{1323}{2} \zeta_{3,5,3},
\end{eqnarray}
where $\zeta_{a,b}$ and $\zeta_{a,b,c}$ are defined in
Eq.~(\ref{shortzeta}).
The result for $\LS{5}{1}{\tfrac{\pi}{3}}$ was given in \cite{FK99}.
We also list some other results relevant for the weights~{\bf 5} and {\bf 6}:
\begin{eqnarray}
\label{pi/3_zoo3}
\LS{5}{2}{\tfrac{\pi}{3}} \! & = & \!
\tfrac{2}{3} \Ls{5}{\tfrac{\pi}{3}}
+\tfrac{253}{54} \pi \zeta_4 \; ,
\nonumber \\
\LS{6}{2}{\tfrac{\pi}{3}} \! & = & \! -\tfrac{4}{15} \Ls{6}{\tfrac{\pi}{3}}
+ \tfrac{2}{3} \zeta_2 \Ls{4}{\tfrac{\pi}{3}}
+ \tfrac{5}{6} \pi \zeta_5
+ \tfrac{11}{3} \pi \zeta_2 \zeta_3,
\nonumber \\
\LS{6}{3}{\tfrac{\pi}{3}} \! & = & \! \tfrac{2}{3} \pi \Ls{5}{\tfrac{\pi}{3}}
+ \tfrac{18887}{432} \zeta_6 + 4 \zeta_3^2 \; .
\end{eqnarray}
The results for $\LS{5}{1}{\tfrac{\pi}{3}}$ 
and $\LS{5}{2}{\tfrac{\pi}{3}}$ were given in \cite{FK99}.
All these results (\ref{pi/3_zoo1})--(\ref{pi/3_zoo3})
have been obtained by the {\sf PSLQ} procedure.

Using one-fold series representation~(\ref{lsJ1}) for 
$\LS{j}{1}{\tfrac{\pi}{3}}$ and the corresponding expression for 
$\Ls{j}{\tfrac{\pi}{3}}$ (see Appendix~A in \cite{KV00}\footnote{A general 
factor, $(-1)^n$, in definition of $\Ls{n}{\theta}$ is missing 
in Appendix~A in \cite{KV00}. }),
\begin{equation}
\Ls{j}{\tfrac{\pi}{3}} = (-1)^j (j-1)! \sum_{k=0}^\infty \frac{(2k)!}{(k!)^2}
\frac{1}{(2k+1)^j} \left( \frac{1}{16} \right)^k , 
\end{equation}
we obtain rapidly-convergent series, which can be used for the 
high-precision calculations of $\zeta_{5,3}$, $\zeta_{7,3}$ and 
$\zeta_{3,5,3}$. For example, 
\begin{eqnarray}
\label{zt53}
\zeta_{5,3} 
&=& 
-\frac{2^5 \cdot3 \cdot 5}{7} 
\left[ \frac{1}{2^6}  \sum_{k=1}^\infty \frac{(k!)^2}{(2k)!} \frac{1}{k^8}
- \frac{\pi}{3}  \sum_{k=0}^\infty \frac{1}{(2k+1)^7} \frac{(2k)!}{(k!)^2} 
\left( \frac{1}{16} \right)^k
\right]
\nonumber \\ && 
- \frac{17 \cdot 8287 \cdot \pi^8}{2^8 \cdot 3^4 \cdot 5^2 \cdot 7^2 } 
- \frac{5  \cdot 19 }{7} \zeta_5 \zeta_3 
- \frac{5 \cdot \pi^2}{2 \cdot 3 \cdot 7} \zeta_3^2 .  
\end{eqnarray}
For the calculation of $\zeta$-function with an arbitrary accuracy, 
an algorithm elaborated in Ref.~\cite{zeta} can be used.  

Considering $\LS{j}{l}{\tfrac{\pi}{2}}$ and $\LS{j}{l}{\pi}$
($j=4,5$; $l=1,2$) we find that most of them 
(except for $\LS{5}{2}{\tfrac{\pi}{2}}$
which is one of the {\em even}-basis elements)
are connected with $\Li{j}{\tfrac{1}{2}}$,   
\begin{eqnarray}
\label{pi/2_zoo}
\LS{4}{1}{\tfrac{\pi}{2}} & = & -\tfrac{5}{96} \ln^4 2
+ \tfrac{5}{16} \zeta_2 \ln^2 2 - \tfrac{35}{32} \zeta_3 \ln 2
+ \tfrac{125}{32} \zeta_4 + \tfrac{1}{2} \pi \Ls{3}{\tfrac{\pi}{2}}
- \tfrac{5}{4} \Li{4}{\tfrac{1}{2}},  
\nonumber \\
\LS{5}{1}{\tfrac{\pi}{2}} & = & -\tfrac{1}{16} \ln^5 2
+ \tfrac{5}{16} \zeta_2 \ln^3 2
- \tfrac{105}{128} \zeta_3 \ln^2 2
- \tfrac{15}{8} \Li{4}{\tfrac{1}{2}} \ln 2 - \tfrac{9}{8} \zeta_2 \zeta_3
\nonumber \\ &&
+ \tfrac{1}{2} \pi \Ls{4}{\tfrac{\pi}{2}}
- \tfrac{1209}{256} \zeta_5
- \tfrac{15}{8} \Li{5}{\tfrac{1}{2}},
\nonumber \\
\LS{4}{1}{ \pi }  & = & - \tfrac{1}{6}\ln^4 2 + \zeta_2 \ln^2 2
- \tfrac{7}{2} \zeta_3 \ln 2 + \tfrac{19}{8} \zeta_4
- 4 \Li{4}{\tfrac{1}{2}},
\nonumber \\
\LS{5}{1}{\pi} & = & -\tfrac{2}{5} \ln^5 2
+ 2 \zeta_2 \ln^3 2 - \tfrac{21}{4} \zeta_3 \ln^2 2
- 12 \Li{4}{\tfrac{1}{2}} \ln 2 + \tfrac{9}{2} \zeta_2 \zeta_3
+ \tfrac{93}{32} \zeta_5 - 12 \Li{5}{\tfrac{1}{2}},
\nonumber \\
\LS{5}{2}{\pi} & = & -\tfrac{1}{3} \pi \ln^4 2
+ 2 \pi \zeta_2 \ln^2 2 - 7 \pi \zeta_3 \ln 2
+ \tfrac{27}{4} \pi \zeta_4 - 8 \pi \Li{4}{\tfrac{1}{2}}.
\end{eqnarray}
All relations (\ref{pi/3_zoo1})--(\ref{pi/2_zoo})
have been obtained using the {\sf PSLQ} 
procedure~\cite{PSLQ}\footnote{The result for 
$\LS{4}{1}{\pi}$ coincides with 
Eq.~(7.71) of \cite{Lewin}, whereas in Eq.~(7.145) for
$\LS{5}{2}{\pi}$ the term  
$+2\pi\zeta_2 \ln^2 2$ is missing, and the term
$+37\pi^5/360$ should read $ 3 \pi^5/40$.
There is also an error in Eq.~(7.144) of \cite{Lewin}:
$\LS{5}{2}{2\pi}$ should be equal to $-13\pi^5/45$,
rather then to $7\pi^5/30$.}.
We give some relevant numerical values,
\begin{eqnarray*}
\Ls{3}{\tfrac{\pi}{2}}  & \simeq &
-2.03357650607205460091206896970\ldots \; ,
\\
\Ls{4}{\tfrac{\pi}{2}}  & \simeq &
6.003109556529006567309305614033\ldots \; ,
\\
\Cl{4}{\tfrac{\pi}{2}}  & \simeq &   
0.9889445517411053361084226332284\ldots \; ,
\\
\Ls{5}{\tfrac{\pi}{2}}  & \simeq &
-24.0143377201598359235946799181446\ldots \; ,
\\
\LS{5}{2}{\tfrac{\pi}{2}} & \simeq & 
-0.1268132428355886971002322996611\ldots \; .
\end{eqnarray*}

As an example of application of the {\em even} basis, we consider
\begin{equation}
U_{a,b} \equiv \zeta(a,b;-1,-1) \; . 
\end{equation} 
The lowest basis element 
of alternating Euler sums~\cite{euler-basis} which cannot be expressed 
in terms of $\zeta_j$, $\ln 2$ and $\Li{j}{\tfrac{1}{2}}$
or their products\footnote{Two other constants are $\zeta(5,1,-1)$ and 
$\zeta(3,3,-1)$, see Eq.~(17) in~\cite{euler-basis}.}
is $U_{5,1}$. 
This constant appears in the $\ep$-expansion of three-loop integral
${\bf B_4}$ (see details in \cite{hypergeometric})
which is connected with the charge renormalization in QED
(Eq.~(51) in~\cite{euler-basis}), as well as
in the ${\cal O}(1/N^3)$ contribution to the critical exponent 
in the large-$N$ limit in $(3-2\ep)$ dimensions (see Eq.~(23)
in~\cite{sum:even}). The {\sf PSLQ}-analysis, with an accuracy 
of 300 decimals, yields the following relation:
\begin{eqnarray}
\label{u51}
U_{5,1} & = &
\tfrac{8}{39} \LS{6}{1}{\tfrac{\pi}{2}}
- \tfrac{4}{39} \pi \Ls{5}{\tfrac{\pi}{2}}
+ \tfrac{10}{13} \Li{6}{\tfrac{1}{2}}
+ \tfrac{10}{13} \Li{5}{\tfrac{1}{2}} \ln 2
+ \tfrac{5}{13} \Li{4}{\tfrac{1}{2}} \ln^2 2
\nonumber \\ &&
+ \tfrac{5}{468} \ln^6 2
- \tfrac{1913}{208} \zeta_6 - \tfrac{31}{104} \zeta_3^2
+ \tfrac{31}{16} \zeta_5 \ln 2
+ \tfrac{35}{312} \zeta_3 \ln^3 2
- \tfrac{5}{104} \zeta_2 \ln^4 2 \; .
\end{eqnarray}

For the procedure of the analytic continuation (in Section~2.2)
we need the generalized (Nielsen) polylogarithms which are defined as 
(see, e.g., in Ref.~\cite{Nielsen})
\begin{equation}
\label{Sab}
S_{a,b}(z) = \frac{(-1)^{a+b-1}}{(a-1)! \; b!} \!\int\limits_0^1
\mbox{d} \xi\; \frac{\ln^{a-1}\!\xi \ln^b (1\!-\!z\xi)}{\xi} \; .
\end{equation}
In particular,
\begin{equation}
S_{a,1}(z) = \Li{a+1}{z}, \quad
z \frac{{\rm d}}{{\rm d} z} S_{a,b} (z) =  S_{a-1,b} (z), \quad
S_{0,b}(z) = \frac{(-1)^b}{b!} \ln^b(1-z).
\end{equation}

The following integration formula is useful:
\begin{eqnarray}
\label{intS}
&&  \hspace*{-7mm}
\int \frac{{\rm d} z}{z} \ln^p(-z) \
\Biggl [ S_{a,b}(z) - (-1)^p  S_{a,b}(1/z) \Biggr ]
\nonumber \\ && \hspace*{-7mm}
= \sum_{k=0}^p (-1)^k \ln^{p-k} (-z) \frac{p!}{(p-k)!}
\Biggl [ S_{a+1+k,b}(z) + (-1)^{p+k}  S_{a+1+k,b}(1/z) \Biggr ] \; .
\end{eqnarray}
Starting from Eq.~(\ref{Ls<->S}) and
applying relations (\ref{integral_ls}) and (\ref{intS}),
we arrive at 
\begin{eqnarray}
&&  
\LS{j+1}{1}{\theta} - \LS{j+1}{1}{\pi}
= \theta \Biggl [ \Ls{j}{\theta} - \Ls{j}{\pi}
\Biggr ]
- \frac{1}{2^j j (j+1)}  \ln^{j+1} (-z) \left[ 1 - (-1)^j \right]
\nonumber \\ && 
- (-1)^j (j-1)! \sum_{p=0}^{j-2} \frac{\ln^p (-z) }{2^p p!}
\sum_{k=1}^{j-1-p} \frac{(-1)^k}{2^k} k 
\left[ S_{k+1,j-k-p}(z) + (-1)^p S_{k+1,j-k-p}(1/z) \right ] 
\nonumber \\ &&
+ 2 (-1)^j (j-1)! \sum_{k=1}^{j-1} \frac{(-1)^k}{2^k} k  S_{k+1,j-k}(-1)
\end{eqnarray}
Substituting $\theta\to\pi-{\rm i}\sigma \ln(-z)$
(see Eq.~(\ref{def_z})) and using Eq.~(\ref{Ls<->S})
we can rewrite this result in a different form,
\begin{eqnarray}
&&
\LS{j+1}{1}{\theta} - \LS{j+1}{1}{\pi}
= \pi \Biggl [ \Ls{j}{\theta} - \Ls{j}{\pi} \Biggr ]
+ \frac{1}{2^j (j+1)}  \ln^{j+1} (-z) \left[ 1 - (-1)^j \right]
\nonumber \\ &&
+ 2 (-1)^j (j-1)! \sum_{k=1}^{j-1} \frac{(-1)^k}{2^k} k S_{k+1,j-k}(-1)
\nonumber \\ &&
+ (-1)^j (j-1)! \sum_{p=1}^{j-2} \frac{\ln^p (-z) }{2^p p!}
\sum_{k=1}^{j-1-p} \frac{(-1)^k}{2^k} (p-k)
\left[ S_{k+1,j-k-p}(z) + (-1)^p S_{k+1,j-k-p}(1/z) \right ]
\nonumber \\ &&
+ (-1)^j (j-1)! \sum_{r=1}^{j-1} \frac{\ln^r (-z) }{2^r (r-1)!}
\left[ S_{1,j-r}(z) + (-1)^r S_{1,j-r}(1/z) \right ].
\end{eqnarray}

\subsection{Auxiliary function}

It is useful to define another generalization of the log-sine
integrals as 
\begin{equation}
\Lsc{i,j}{\theta} = -\int\limits_0^\theta
{\mbox d} \phi
\ln^{i-1} \left| 2\sin\frac{\phi}{2} \right|
\ln^{j-1} \left| 2\cos\frac{\phi}{2} \right|  \; .
\end{equation}  
When $i=1$ or $j=1$, it reduces to ordinary $\mbox{Ls}$ functions, 
\begin{equation}
\Lsc{i,1}{\theta} = \Ls{i}{\theta} \; ,
\quad\;
\Lsc{1,j}{\theta} = -\Ls{j}{\pi-\theta} + \Ls{j}{\pi} \; .
\end{equation}
There exists an obvious symmetry property,
\begin{equation}
\Lsc{i,j}{\theta} = - \Lsc{j,i}{\pi-\theta} + \Lsc{j,i}{\pi} \; ,
\end{equation}
which implies that the functions $\Lsc{i,j}{\theta}$ with $i>j$
are not independent.
The values of $\Lsc{i,j}{\pi}$ can be extracted from
Lewin's book \cite{Lewin},  Eqs.~(7.114)--(7.118).
We present here some of them
\begin{equation}
\Lsc{2,2}{\pi} =   \tfrac{1}{4} \pi \zeta_2, \quad
\Lsc{2,3}{\pi} = - \tfrac{1}{4} \pi \zeta_3, \quad
\Lsc{3,3}{\pi} = -\tfrac{9}{16} \pi \zeta_4, \quad
\Lsc{2,4}{\pi} = \tfrac{21}{16} \pi \zeta_4 \; .
\end{equation}  

Using $\sin\phi=2\sin\frac{\phi}{2}\cos\frac{\phi}{2}$,
the following duplication formula can be derived:
\begin{equation}
\Ls{k}{2\theta} =
2 (k-1)! \sum_{i=0}^{k-1} \frac{1}{i!(k-1-i)!} \Lsc{i+1,k-i}{\theta} \; .
\end{equation}
For special cases $k=2,3,4,5$ it gives, respectively,
\begin{eqnarray}
&& \hspace*{-10mm}
\Ls{2}{\theta} - \Ls{2}{\pi-\theta} = \tfrac{1}{2} \Ls{2}{2\theta},
\nonumber \\ && \hspace*{-10mm}
2 \Lsc{2,2}{\theta} = \tfrac{1}{2} \Ls{3}{2\theta}
- \Ls{3}{\theta}
+ \Ls{3}{\pi-\theta} - \Ls{3}{\pi} \; ,
\nonumber \\ && \hspace*{-10mm}
3\left[ \Lsc{2,3}{\theta} - \Lsc{2,3}{\pi-\theta} \right]
= \tfrac{1}{2} \Ls{4}{2\theta} - \Ls{4}{\theta}
+ \Ls{4}{\pi-\theta}-\tfrac{1}{2}\Ls{4}{\pi} \; ,
\nonumber \\ && \hspace*{-10mm}
6 \Lsc{3,3}{\theta}
+ 4 \left[ \Lsc{2,4}{\theta} - \Lsc{2,4}{\pi-\theta} \right]
= \tfrac{1}{2}\Ls{5}{2\theta}
- \Ls{5}{\theta} + \Ls{5}{\pi-\theta}
+ \tfrac{15}{8}\pi\zeta_4 \; .
\hspace*{7mm}
\label{binom_Lsc}
\end{eqnarray}
We can see that for odd values of $k$ one would always
obtain representations of the functions
$\Lsc{(k+1)/2,\; (k+1)/2}{\theta}$
in terms of the $\mbox{Lsc}$ functions with $i\neq j$.
This means that the functions $\Lsc{i,j}{\theta}$
with $i=j$ are not independent. Therefore,
it is enough to consider only the functions with $i<j$.
In particular, up to the level $k=5$ only two new functions
are needed, in addition to ordinary $\Ls{j}{\theta}$:
$\Lsc{2,3}{\theta}$ and $\Lsc{2,4}{\theta}$.

For a particular point $\theta=\frac{\pi}{2}$,
Eqs.~(\ref{binom_Lsc}) yield
\begin{eqnarray}
\Lsc{2,2}{\tfrac{\pi}{2}} &=& -\tfrac{1}{4} \Ls{3}{\pi}
   = \tfrac{1}{8}\pi\zeta_2 \; ,
\nonumber \\
\Lsc{2,3}{\pi} &=& -\tfrac{1}{6} \Ls{4}{\pi}
   = - \tfrac{1}{4}\pi\zeta_3 \; ,
\nonumber \\
\Lsc{3,3}{\tfrac{\pi}{2}} &=& -\tfrac{2}{3} \Lsc{2,4}{\pi}
-\tfrac{1}{12} \Ls{5}{\pi} = - \tfrac{9}{32}\pi\zeta_4 \; ,
\label{Lsc_pi2}
\end{eqnarray}
whereas for $\theta=\tfrac{\pi}{3}$ we get
\begin{eqnarray}
&& \hspace*{-10mm}
\Lsc{2,2}{\tfrac{\pi}{3}} = \tfrac{3}{4} \Ls{3}{\tfrac{2\pi}{3}}
- \tfrac{1}{2} \Ls{3}{\tfrac{\pi}{3}}
+ \tfrac{1}{4}\pi\zeta_2 \; ,
\nonumber \\ && \hspace*{-10mm}
3 \left[ \Lsc{2,3}{\tfrac{\pi}{3}} - \Lsc{2,3}{\tfrac{2\pi}{3}} \right]
= \tfrac{3}{2} \Ls{4}{\tfrac{2\pi}{3}} - \Ls{4}{\tfrac{\pi}{3}}
-\tfrac{3}{4} \pi \zeta_3 \; ,
\nonumber \\ && \hspace*{-10mm}
6 \Lsc{3,3}{\tfrac{\pi}{3}}
+4\left[ \Lsc{2,4}{\tfrac{\pi}{3}} - \Lsc{2,4}{\tfrac{2\pi}{3}} \right]
= \tfrac{3}{2} \Ls{5}{\tfrac{2\pi}{3}} - \Ls{5}{\tfrac{\pi}{3}}
+\tfrac{15}{8} \pi \zeta_4 \; .
\label{Lsc_pi3}
\end{eqnarray}

For these particular values of $\theta$ the {\sf PSLQ} procedure yields
\begin{eqnarray}  
\Lsc{2,3}{\tfrac{\pi}{3} } &=& -\tfrac{59}{108} \pi \zeta_3
- \tfrac{2}{27} \Ls{4}{\tfrac{\pi}{3}} + \tfrac{1}{2} \Ls{4}{\tfrac{2\pi}{3}},
\nonumber \\ 
\Lsc{2,3}{\tfrac{2\pi}{3} } &=& -\tfrac{8}{27} \pi \zeta_3
+ \tfrac{7}{27} \Ls{4}{\tfrac{\pi}{3}},
\nonumber \\ 
\Lsc{2,4}{ \tfrac{\pi}{3} } &=&
\tfrac{595}{192} \pi \zeta_4
+ \tfrac{3}{4} \zeta_2 \Ls{3}{\tfrac{2\pi}{3}}
- \tfrac{3}{8} \pi \LS{4}{1}{\tfrac{2\pi}{3}}
- \tfrac{1}{24} \Ls{5}{\tfrac{\pi}{3}}
+ \tfrac{3}{8} \Ls{5}{\tfrac{2\pi}{3}}
+ \tfrac{9}{32} \LS{5}{2}{\tfrac{2\pi}{3}},
\nonumber \\ 
\Lsc{2,4}{\tfrac{2 \pi}{3} } &=&
\tfrac{421}{384} \pi \zeta_4
+ \tfrac{27}{8} \zeta_2 \Ls{3}{\tfrac{2\pi}{3}}
- \tfrac{27}{16} \pi \LS{4}{1}{\tfrac{2\pi}{3}}
- \tfrac{13}{48} \Ls{5}{\tfrac{\pi}{3}}
+ \tfrac{81}{64} \LS{5}{2}{\tfrac{2\pi}{3}},
\nonumber \\ 
\Lsc{3,3}{ \tfrac{\pi}{3} } &=&
-\tfrac{589}{576} \pi \zeta_4
+ \tfrac{7}{4} \zeta_2 \Ls{3}{\tfrac{2\pi}{3}}
- \tfrac{7}{8} \pi \LS{4}{1}{\tfrac{2\pi}{3}} 
- \tfrac{23}{72} \Ls{5}{\tfrac{\pi}{3}}
+ \tfrac{21}{32} \LS{5}{2}{\tfrac{2\pi}{3}},
\nonumber \\ 
\Lsc{2,3}{ \tfrac{\pi}{2} } &=& \tfrac{1}{8} \pi \zeta_3
- \tfrac{1}{3} \Ls{4}{\tfrac{\pi}{2}} + 2 \Cl{4}{\tfrac{\pi}{2}}.
\label{Lsc_zoo}
\end{eqnarray}
These results obey the conditions (\ref{Lsc_pi2})--(\ref{Lsc_pi3}).
However, no relation has been found for
\begin{equation}
\label{Lsc24}
\Lsc{2,4}{ \tfrac{\pi}{2} } \simeq 
0.30945326106363459854315773429895417071\ldots \; ,
\end{equation}
which represents a new transcendental constant appearing
at the weight-{\bf 5} level of the even basis.

If we introduce a variable $z=e^{{\rm i}\phi}$, we see that
\[
\ln \left( 2\sin \tfrac{\phi}{2} \right)
\leftrightarrow \ln(1-z) - \tfrac{1}{2} \ln (-z), \quad
\ln \left( 2\cos \tfrac{\phi}{2} \right)
\leftrightarrow \ln(1+z) - \tfrac{1}{2} \ln z,
\]
Therefore, the analytic continuation of 
$\Lsc{i,j}{\theta}$ is related to the integrals of the type
\[
\int \frac{{\rm d} z}{z} 
\ln^\alpha (1+z) \ln^\beta (1-z) \ln^\gamma z \; .
\]
This is nothing but a particular case of 
the {\em harmonic polylogarithms} \cite{RV00}.
It is known that some of such functions of weight~{\bf 4} 
cannot be expressed in
terms of polylogarithms (or Nielsen polylogarithms) 
of specific types of arguments, like $\pm z$ and $\tfrac{1}{2}(1\pm z)$
(see in Ref.~\cite{RV00}).
For example,
in our case the analytic continuation of $\Lsc{2,3}{\theta}$
involves an integral $\int\frac{{\rm d}z}{z} \ln^2(1+z) \ln(1-z)$.

\section{Hypergeometric functions}
\setcounter{equation}{0}
\subsection{Procedure of the $\ep$-expansion}

Let us briefly describe some technical details of 
obtaining terms of the $\ep$-expansion of hypergeometric functions 
$_2F_1$, $_3F_2$ and $_4F_3$ given below. 
All of them belong to the type
\begin{equation}
_PF_Q\left(\begin{array}{c|}
{\cal A}_1+a_1\ep, \ldots, {\cal A}_P+a_P\ep \\
{\cal B}+\tfrac{1}{2} + b \ep, 
{\cal C}_1+c_1\ep, \ldots, {\cal C}_{Q-1}+c_{Q-1}\ep 
\end{array} ~\frac{1}{4} \right),
\label{PFQ_gen}
\end{equation}
where ${\cal A}_j={\cal B}=1$, whereas ${\cal C}_j$ take the values 1 or 2.
More generally,
we shall assume that ${\cal A}_j$, ${\cal B}$ and ${\cal C}_j$ 
are positive and integer.
This function (\ref{PFQ_gen}) can be presented as
\begin{equation}
\sum_{j=0}^{\infty} \frac{1}{j!}\;
\frac{({\cal A}_1+a_1\ep)_j \ldots ({\cal A}_P+a_P\ep)_j}
     {({\cal C}_1+c_1\ep)_j \ldots ({\cal C}_{Q-1}+c_{Q-1}\ep)_j} \;
\frac{(2{\cal B}+1+2b\ep)_{2j}}{({\cal B}+1+b\ep)_j} \; ,
\label{sum}
\end{equation}
where $(\alpha)_j\equiv\Gamma(\alpha+j)/\Gamma(\alpha)$ is 
the Pochhammer symbol,
and we have used the duplication formula
$(2\beta)_{2j}=4^j(\beta)_j(\beta+\tfrac{1}{2})_j$.

To perform the $\ep$-expansion we use the well-known representation
\begin{equation}
(1+a\ep)_j = j! \; \exp\left[ -\sum_{k=1}^{\infty} \frac{(-a\ep)^k}{k}
S_k(j) \right] \; ,
\end{equation}
which, for ${\cal A}>1$, yields
\begin{equation}
({\cal A}+a\ep)_j = ({\cal A})_j \;
\exp\left\{ -\sum_{k=1}^{\infty} \frac{(-a\ep)^k}{k}
\left[ S_k({\cal A}+j-1) - S_k({\cal A}-1) \right] \right\} \; ,
\end{equation}
where 
$S_k(j) = \sum_{l=1}^j l^{-k}$ is the harmonic sum
satisfying the relation
$S_k(j) = S_k(j-1) + j^{-k}$.

In Ref.~\cite{KV00} it was demonstrated
that the multiple binomial sums (\ref{binsum}),
or their certain linear combinations, can be expressed in terms of
transcendental
constants elaborated in \cite{FK99}. 
The only  values of ${\cal A}$, ${\cal B}$ and ${\cal C}$ 
for which {\em each} term 
of the $\ep$-expansion of (\ref{sum}) 
can be expressed in terms of the multiple binomial sums 
(\ref{binsum}) are ${\cal B}=1$, ${\cal A}=1$~or~2,  ${\cal C}=1$~or~2.
For other values of the indices binomial sums of a
different type appear, which are connected with series like
\[
\sum_{n=1}^\infty  \frac{(n!)^2}{(2n)!} \frac{1}{(2n+p)^q}
\prod_{a,b,i,j} \left[ S_a(n-1) \right]^i \left[ S_b(2 n-1) \right]^j,
\]
where $p$ and $q$ are positive integers.
In such series, new transcendental constants appear. 
For example, for $p=1$, $q=2$, $i=j=0$ we get 
(see in \cite{RN1})
\[
\sum_{n=0}^\infty  \frac{(n!)^2}{(2n)!} 
\frac{1}{(2n+1)^2} = \tfrac{8}{3} G - \tfrac{1}{3}\pi\ln(2+\sqrt{3}),   
\]
where $G$ is the Catalan's constant.

The values of multiple binomial sums (\ref{sum}) or their
combinations up to weight ${\bf 4}$ are presented in Appendix B of 
Ref.~\cite{KV00}. Most of the sums can be expressed in terms of
the basis elements. There are, however, a few weight-{\bf 3}
and weight-{\bf 4} sums which
are not separately expressible, but only certain linear combinations
of them, namely:
\begin{equation}
\label{weight34-sum}
\Sigma_{-;1;1}^{-;2}+\Sigma_{-;2;1}^{-;1}, \quad  
\Sigma_{-;1;2}^{-;2}+\Sigma_{-;2;2}^{-;1}, \quad
\Sigma_{1;1;1}^{1;2}+\Sigma_{1;2;1}^{1;1}, \quad
\Sigma_{-;1;1}^{-;3}+3\Sigma_{-;1,2;1}^{-;1,1}
+2\Sigma_{-;3;1}^{-;1} \; .
\end{equation}
The situation at the weight-${\bf 5}$ level is similar.
Again, most of the sums can separately be expressed
in terms of the basis elements, except for a few ones,
which are expressible only in certain linear combinations. 
There are 4 linear combinations   
independent of $\chi_5$ (\ref{sigma_32}),
\begin{eqnarray}
\label{unconnectedsum}
&&
\Sigma_{1;1;1}^{2;2}+\Sigma_{1;2;1}^{2;1}, \quad
\Sigma_{2;1;1}^{1;2}+\Sigma_{2;2;1}^{1;1}, \quad
\Sigma_{1;1;1}^{1;3}+3\Sigma_{1;1,2;1}^{1;1,1}+2\Sigma_{1;3;1}^{1;1},\quad  
\nonumber \\ &&
\Sigma_{-;1;1}^{-;4}+3\Sigma_{-;2;1}^{-;2}+6\Sigma_{-;4;1}^{-;1}
+ 8\Sigma_{-;1,3;1}^{-;1,1}+6\Sigma_{-;1,2;1}^{-;2,1},
\end{eqnarray}
and 3 combinations which involve $\chi_5$,
\begin{equation}
\label{connectedsum}
\Sigma_{-;1;3}^{-;2}+\Sigma_{-;2;3}^{-;1}, \quad
\Sigma_{-;1;2}^{-;3}+2\Sigma_{-;3;2}^{-;1}+3\Sigma_{-;1,2;2}^{-;1,1}, \quad
\Sigma_{1;1;2}^{1;2}+\Sigma_{1;2;2}^{1;1} .
\end{equation}

Using relation~(\ref{binomial1}) and the results of Ref.~\cite{KV00},
is it natural to expect that the sums~(\ref{binsum}), not only with $k=1$ 
but also with $k=3$, are connected with the {\em odd} basis, whereas the
sums~(\ref{binsum}) with $k=2$ and $k=4$ 
should be associated with the {\em even} basis. 

To check this conjecture, we have performed {\sf PSLQ}-analysis of the 
sums~(\ref{binsum}) with $k=2$ up to weight ${\bf 5}$\footnote{The values 
of some sums~(\ref{binsum}) with $k=4$ up to weight ${\bf 3}$ can 
be extracted from the results of~\cite{FKV98}.}.
At the weight-{\bf 5} level, one needs to add
two new elements to the {\em even} basis of \cite{FK99}.
One of them can be associated with $\Lsc{2,4}{\tfrac{\pi}{2}}$
(see Eq.~(\ref{Lsc24}) in Appendix~A.2), whereas the second one,
$\widetilde{\chi}_5$, is given in Eq.~(\ref{chi_tilde_5}).
In this way we restore, accidentally, the ``empirical'' relation
$N_j=2^j$ for the $j=5$ level of the {\em even} basis.
Then, we find that the expressions for the sums~(\ref{binsum}) with $k=2$,
contain all elements of the {\em even} basis\footnote{
This is connected with 
the fact that both $\LS{j}{1}{\tfrac{\pi}{2}}$ and
$\LS{j}{1}{\pi}$, for $j=4,5$, are expressible in terms of
the same constants (see Appendix~A.1).}.
However, the question about possibility to express {\em all} 
elements of the {\em even} basis in terms of only binomial 
sums~(\ref{binsum}) is still open. 
As in the $k=1$ case, most of the $k=2$ sums are
{\em separately} expressible in terms of the {\em even} 
basis elements, except for the same set of sums,
which can be expressed only in certain linear combinations\footnote{
Such interrelation is not trivial. For example, for $k=3$
the number of such combinations of sums is reduced, so that the
first irreducible combination arises only at the weight {\bf 4}.
Besides this, the element $\chi_5$ is not reproduced by series like
(\ref{sigma_32}) with $k=3$.}:
namely, for the weights {\bf 3} and {\bf 4} the
combinations given in Eq.~(\ref{weight34-sum})
and at the weight {\bf 5} the familiar combinations 
(\ref{unconnectedsum}) and (\ref{connectedsum}), 
everything with $k=2$.
Worth noting is that 
the combinations~(\ref{unconnectedsum}) do not contain
$\widetilde{\chi}_5$,
whereas the sums~(\ref{connectedsum}) do not involve 
$\Lsc{2,4}{\tfrac{\pi}{2}}$.
In total, 
we have found 8 elements which involve $\tilde{\chi}_5$:
besides the combinations given in~(\ref{connectedsum}) (with $k=2$)
and the sums
$\Sigma_{1;1;2}^{2;1}(2)$, $\Sigma_{1;1;3}^{1;1}(2)$,
$\Sigma_{1;-;3}^{2;-}(2)$ (which have already occurred in the
$k=1$ case),
there are two other sums, $\Sigma_{1;-;4}^{1;-}(2)$
and $\Sigma_{-;1;4}^{-;1}(2)$
(for $k=1$ these two sums are expressible in terms of  
$\zeta_5$,  $\zeta_2 \zeta_3$ and $\pi \Ls{4}{\tfrac{\pi}{3}}$).

We would like to mention some issues related to the high-precision
calculation\footnote{All such computations have been performed
with the help of David Bailey's {\sf MPFUN} routines. 
The latest version of these codes, including the documentation and 
original papers, is available from 
{\tt http://www.nersc.gov/$\;\widetilde{}\;$dhb/mpdist/mpdist.html} .} 
of the elements of our basis, needed for the {\sf PSLQ}-analysis. 
Most of the constants of the basis up to weight {\bf 5}
can be produced, with desirable accuracy of 1000 or more digits,
using the algorithm described in Appendix~A of Ref.~\cite{KV00}. 
The $\chi_5$ and $\widetilde{\chi}_5$ constants are originally defined
in terms of rapidly-convergent series, see Eqs.~(\ref{sigma_32})
and (\ref{chi_tilde_5}). Therefore, only three constants, 
$\LS{5}{2}{\tfrac{2\pi}{3}}$, $\Lsc{2,4}{\tfrac{\pi}{2}}$ and 
$\LS{5}{2}{\tfrac{\pi}{2}}$, may require additional consideration.  
In principle, the corresponding binomial sums may be used for 
precise numerical calculation. However, specifically for
$\LS{5}{2}{\tfrac{2\pi}{3}}$, the following representation 
in terms of inverse tangent integral (see in \cite{Lewin})
happens to be very useful:
\begin{eqnarray}
{\mbox{Ti}}_5 \left( \tfrac{1}{\sqrt{3}}\right) 
&\equiv& \tfrac{1}{\sqrt{3}} \sum_{j=0}^\infty  
\frac{(-1)^j}{3^j(2j+1)^5} 
= \tfrac{1}{2304} \pi \ln^4 3
+ \tfrac{1}{18} \pi \zeta_2 \ln^2 3
+ \tfrac{59}{432} \pi \zeta_3 \ln 3
+ \tfrac{1273}{2304} \pi \zeta_4
\nonumber \\ && 
- \tfrac{5}{288} \Ls{2}{\tfrac{\pi}{3}} \ln^3 3
+ \tfrac{5}{64}  \Ls{3}{\tfrac{2\pi}{3}} \ln^2 3
- \tfrac{7}{108} \Ls{4}{\tfrac{\pi}{3}} \ln 3
- \tfrac{5}{48}  \Ls{4}{\tfrac{2\pi}{3}} \ln 3
\nonumber \\ && 
- \tfrac{7}{16} \zeta_2 \Ls{3}{\tfrac{2\pi}{3}} 
+ \tfrac{11}{288} \Ls{5}{\tfrac{\pi}{3}}
+  \tfrac{5}{96}  \Ls{5}{\tfrac{2 \pi}{3}}
+ \tfrac{7}{32} \pi  \LS{4}{1}{\tfrac{2\pi}{3}}
- \tfrac{21}{128} \LS{5}{2}{\tfrac{2\pi}{3}}.
\hspace*{6mm}
\end{eqnarray}

For numerical calculation of $\LS{j}{k}{\theta}$ the following 
algorithm is useful. 
Substituting $y=\sin\tfrac{\phi}{2}$, we get from the original integral 
representation~(\ref{log-sin-gen}) 
\[
\LS{j}{k}{\theta} = - 2^{k+1} \int\limits_0^{\sin(\theta/2)} 
\frac{{\rm d}y}{\sqrt{1-y^2}}
\left(\arcsin y \right)^k \ln^{j-k-1}(2y). 
\]
Expanding 
\[
\frac{1}{\sqrt{1-y^2}} = \sum_{r=0}^\infty \frac{(2r)!}{(r!)^2}
\frac{y^{2r}}{4^r} \; 
\]
and $\left(\arcsin y \right)^k$ (see below) in $y$ and using 
\[
\int x^a \ln^b x ~dx = (-1)^b\; b!\; x^{a+1} 
\sum_{p=0}^b \frac{\left(- \ln x \right)^{b-p}}{(b-p)!(a+1)^{p+1}} \; ,
\]
we obtain the multiple series representation suitable for 
numerical eveluation.

The expansion of $\left(\arcsin y \right)^k$
is discussed in Ref.~\cite{RN1}.
The generating expression is 
$\exp(a \arcsin y) = \sum_{p=0}^\infty (b_p y^p/p!)$,
where $b_0=1$, $b_1=a$ and 
\begin{eqnarray*}
b_{2k+1} & = & a(a^2+1)(a^2+3^2) \cdots (a^2+(2k-1)^2), \quad k\geq 1 \; ;
\nonumber \\
b_{2k} & = & a^2(a^2+2^2)(a^2+4^2) \cdots (a^2+(2k-2)^2), \quad k\geq 1 \; .
\end{eqnarray*}
Expanding $\exp\left(a\arcsin y\right)$ in a power series in $a$ 
and equating coefficients of $a^k$ on both sides, the Taylor series for 
$\left(\arcsin y\right)^k$ can be deduced. For example,
\begin{eqnarray*}
\frac{1}{2!}\left(\arcsin y \right)^2 & = & \sum_{r=0}^\infty
\frac{4^r (r!)^2 y^{2r+2}}{(2r+2)!} \; ,
\nonumber \\
\frac{1}{3!}\left(\arcsin y \right)^3 & = & \sum_{r=1}^\infty
\left( 1+ \frac{1}{3^2}+ \cdots + \frac{1}{(2r-1)^2} 
\right)
\frac{(2r)!}{(r!)^2} \frac{y^{2r+1}}{4^r (2r+1)} \; , 
\nonumber \\
\frac{1}{4!}\left(\arcsin y \right)^4 & = & \sum_{r=1}^\infty
\left( \frac{1}{2^2}+ \frac{1}{4^2}+ \cdots + \frac{1}{(2r)^2} 
\right) \frac{4^r (r!)^2 y^{2r+2}}{(2r+2)!} \; .
\end{eqnarray*}

\subsection{The $_2F_1$ function}

Here we present some relevant results for the hypergeometric 
function of the type
\begin{equation}
 {}_{2}F_1 \left(\begin{array}{c|} 1 + a_1 \ep, 1 + a_2\ep \\
\frac{3}{2} + b \ep \end{array} ~z \right) \; .
\label{2F1_app}
\end{equation}
Especially we are interested in its expansion in $\ep$,
which basically corresponds to the calculation of the derivatives
of $_2F_1$ function with respect to the parameters.

There are some special cases when the result is known exactly.
For instance, when $a_1=-a_2\equiv a$ and $b=0$ we have
(see Eq.~(94) in p.~469 of \cite{PBM3}),
\begin{equation}
{}_{2}F_1 \left(\begin{array}{c|} 1+a\ep, 1- a \ep \\
\frac{3}{2} \end{array} ~\sin^2\theta \right) =
\frac{\sin \left( 2 a \ep \theta \right)}{a\ep \sin(2\theta)} \; .
\end{equation}
When $a_1=b=0$ ($a_2\equiv a$) we get (see Eqs.~(\ref{f_ep_1})--(\ref{f_ep}))
\begin{equation}
_2F_1 \left(\begin{array}{c|} 1, 1+a\ep  \\
\frac{3}{2} \end{array} ~\sin^2\theta \right)
= \frac{1}{2^{1+2a\ep} \sin\theta (\cos\theta)^{1+2a\ep}}\;
\sum_{j=0}^{\infty} \frac{(2a\ep)^j}{j!}
\left[ \Ls{j+1}{\pi-2\theta} - \Ls{j+1}{\pi} \right] \; ,
\end{equation}
i.e., we know all orders of the $\ep$-expansion, in terms of the
log-sine
integrals (\ref{log-sin}). For $z=\tfrac{1}{4}$ ($\theta=\tfrac{\pi}{6}$)
we get Eq.~(\ref{2F1_1}).
Another interesting case is 
$a_1=0$, $a_2=2b\equiv a$, for which we find
\begin{equation}
{}_{2}F_1 \left(\begin{array}{c|} 1, 1+ a \ep \\
\frac{3}{2} + \frac{1}{2}a\ep \end{array} ~\sin^2 \theta \right) =
-  \frac{1+a\ep}{\left[ 2 \sin (2\theta) \right]^{1+a\ep}}
\sum_{j=0}^\infty \frac{(a\ep)^j}{j!} \Ls{j+1}{4\theta} \;
\end{equation}

Moreover, in a more general case, when $a_1=0$ while $a_2\equiv~a$
and $b$ are arbitrary, the $_2F_1$ function (\ref{2F1_app}) can be 
reduced to an incomplete Beta-function (see in \cite{PBM3}). 
As a result, we get the integral representation
\begin{equation}
{}_{2}F_1 \left(\begin{array}{c|} 1, 1+ a \ep \\
\frac{3}{2}+b \ep \end{array} ~\sin^2 \theta \right) =
\frac{1+ 2 b \ep}{ 
\left( \sin\theta \right)^{1+2b\ep} 
\left( \cos\theta \right)^{1+2a\ep-2b\ep}}
\int\limits_0^\theta {\rm d}\phi 
(\sin\phi)^{2b\ep} (\cos\phi)^{2a\ep-2b\ep} \; .
\label{2f1_lsc}
\end{equation}
The $\ep$-expansion can be written in terms of $\mbox{Lsc}$ function
(see Appendix~A.2). 

There are some further results available for the case when
$z=\tfrac{1}{4}$ ($\theta=\tfrac{\pi}{6}$). 
When $a_1=b\equiv~a$ and $a_2=\tfrac{3}{2}a$ we get
\begin{equation}
{}_{2}F_1 \left(\begin{array}{c|} 1+a\ep, 1+ \frac{3}{2} a \ep\\
\frac{3}{2} + a \ep \end{array} ~\frac{1}{4} \right) =
\frac{2\pi}{3^{\frac{3}{2}(1+a\ep)}}
\frac{\Gamma(2+2 a \ep)}{\Gamma(1+a \ep)\Gamma^2(1+\tfrac{1}{2}a\ep)}
\label{2F1_D3}
\end{equation}

Using the procedure described in Appendix~B.1, we 
obtain a few terms of the $\ep$-expansion for general 
values of $a_1$, $a_2$ and $b$:
\begin{eqnarray}
&& \hspace*{-7mm} 
 {}_{2}F_1 \left(\begin{array}{c|} 1 + a_1 \ep, 1 + a_2\ep \\ 
\frac{3}{2} + b \ep \end{array} ~\frac{1}{4} \right)
 = \frac{2 (1+2 b \ep)}{3^{\frac{1}{2}+A_1\ep-b\ep}} 
\nonumber \\ && \hspace*{-7mm}
\times \Biggl\{
\tfrac{1}{3} \pi
+ \ep  (2 A_1 - 5b) \sum_{j=0}^{\infty} \frac{(-2\ep)^j}{(j+1)!} (b - A_1)^j 
\left[ \Ls{j+2}{\tfrac{2\pi}{3}} -  \Ls{j+2}{\pi} \right]
\nonumber \\ && \hspace*{-7mm} 
+  \ep^2 
\left[  \tfrac{19}{18} b^2 + \tfrac{1}{54} ( A_1^2 - A_2 ) 
- \tfrac{5}{18} b A_1 \right] \pi \zeta_2 
+  \ep^3 \biggl[
\tfrac{1}{9} ( A_1^2 - A_2 )  ( 2 A_1 - 5 b ) \zeta_2 \Ls{2}{\tfrac{\pi}{3}}
\nonumber \\ && \hspace*{-7mm} 
+ \left(  \tfrac{193}{162} A_1 (A_1^2 - A_2)
- \tfrac{17}{27} b^3  - \tfrac{541}{162} b A_1^2
+ \tfrac{409}{162} b A_2 + \tfrac{13}{9} b^2 A_1   \right) \pi \zeta_3 
\nonumber \\ && \hspace*{-7mm} 
+ \left( - \tfrac{32}{81} ( 2 A_1 - 5 b ) ( A_1^2 - A_2 )
+ \tfrac{4}{81} b (6 A_2 + 9  b A_1 - 42 b^2 )
\right) \Ls{4}{\tfrac{\pi}{3}} 
\biggr]
\nonumber \\ && \hspace*{-7mm} 
+  \ep^4 \biggl[
\tfrac{1}{27} ( A_1^2 - A_2 ) (2 A_1 - 5 b )(11 A_1 - 35 b ) 
   \zeta_3 \Ls{2}{\tfrac{ \pi}{3}}
-  \tfrac{1}{27} ( A_1^2 - A_2 ) (2 A_1 - 5 b )^2 
   \pi \left[ \Ls{2}{\tfrac{ \pi}{3}} \right]^2
\nonumber \\ && \hspace*{-7mm} 
+ \tfrac{1}{3} (b - A_1)
\left(  - \tfrac{17}{2} ( 2 A_1 - 5 b ) ( A_1^2 - A_2 )
+ 6 b A_2 + 9  b^2 A_1 - 42 b^3 \right) \zeta_2 \Ls{3}{\tfrac{2\pi}{3}}
\nonumber \\ && \hspace*{-7mm} 
- \tfrac{1}{6} (b - A_1)
\left( - 8 (2 A_1 - 5 b ) ( A_1^2 - A_2 ) + 6 b A_2 + 9  b^2 A_1 - 42 b^3
\right)
\left[ \pi \LS{4}{1}{\tfrac{2\pi}{3}} - \tfrac{3}{4} \LS{5}{2}{\tfrac{2\pi}{3}}
\right]
\nonumber \\ && \hspace*{-7mm} 
+ \biggl(  \tfrac{209}{36} b^4
- \tfrac{1753}{144} b^3 A_1
+ \tfrac{10399}{648} b^2 A_2
- \tfrac{11069}{1296} b^2 A_1^2
+ \tfrac{10061}{648} b A_1^3  
- \tfrac{10799}{648} b A_1 A_2 
\nonumber \\ && \hspace*{-7mm} 
+ \tfrac{305}{72} A_1^2 A_2 
- \tfrac{5491}{1296} A_1^4 
+ \tfrac{1}{1296} A_2^2
\biggr) \pi \zeta_4 
+ \biggl( \tfrac{11}{9} b^4
- \tfrac{53}{18} b^3 A_1 
+ \tfrac{293}{81} b^2 A_2 
- \tfrac{433}{162} b^2 A_1^2
\nonumber \\ && \hspace*{-7mm} 
+ \tfrac{370}{81} b A_1^3
- \tfrac{361}{81} b A_1 A_2
- \tfrac{104}{81} ( A_1^2 - A_2 ) A_1^2
\biggr) \Ls{5}{\tfrac{\pi}{3}}
\biggr]
+ {\cal O}(\ep^5) \Biggr \} \; ,
\label{2F1_PSLQ}
\end{eqnarray}
where 
$A_j \equiv \sum_{i=1}^2 a_i^j$,
and we take into account that there are some relations among $A_j$,
\[ 
2 A_3 + A_1^3 - 3 A_2 A_1 = 0 \; , \quad \;\;
2 A_4 + A_1^4 - 2 A_1^2 A_2 - A_2^2 = 0 \; .
\]

At the weight-{\bf 5} level of the {\em odd} basis,
we have observed that the constants $\pi\LS{4}{1}{\tfrac{2\pi}{3}}$
and $\LS{5}{2}{\tfrac{2\pi}{3}}$ appear only in the combination
$\left[ 
\pi \LS{4}{1}{\tfrac{2\pi}{3}} - \tfrac{3}{4} \LS{5}{2}{\tfrac{2\pi}{3}}
\right]$. 
Moreover, this combination arises only in the sums
like ${\Sigma_{a;b;1}^{i;j}}$. It is natural to expect that similar 
``junctions'' also happen for higher generalized log-sine integrals. 

We also mention some useful results for the contiguous $_2F_1$
functions. One of them, Eq.~(\ref{deduce}), is rather essential for 
the results of this paper. Another interesting case,
\begin{equation}
{}_{2}F_1 \left(\begin{array}{c|} \ep, 1-3\ep  \\
\frac{3}{2}-2\ep \end{array} ~\frac{1}{4} \right)
= \frac{2^{1-4\ep}\;\Gamma(\frac{1}{3}) \Gamma(\frac{3}{2}-2\ep)}
     {3^{1-3\ep}\;\Gamma(\frac{1}{2})\Gamma(\frac{4}{3}-2\ep)} \; ,
\end{equation}
can be obtained by a simple transformation of Eq.~(114)
on p.~461 of \cite{PBM3}.

\subsection{The $_3F_2$ function}

Here we present some results for the hypergeometric
function of the type
\begin{equation}
 {}_{3}F_2 \left(\begin{array}{c|} 
1 + a_1 \ep, 1 + a_2\ep, 1 + a_3\ep \\
\frac{3}{2} + b \ep, {\cal C}+c\ep \end{array} ~z \right) \; ,
\label{3F2_app}
\end{equation}
with ${\cal C}=1$ or ${\cal C}=2$.

There are a few special cases, 
when the considered function (\ref{3F2_app}) 
can be expressed in terms of $_2F_1$ functions.
First of all, when ${\cal C}=1$ and $c$ is equal to one of the $a_i$
(for example $c=a_3$), we obtain the function (\ref{2F1_app}).
Another reduction case is ${\cal C}=2$ and $c=a_3=0$
(see Eq.~(7) in p.~497 of \cite{PBM3} and Eq.~(\ref{Kummer_PFQ}) 
of this paper).

There is also a less trivial relation,
valid for the case when ${\cal C}=2$ and $c=2b=2a_3=a_1+a_2$ 
(see, e.g., Eq.~(22) on p.~498 of \cite{PBM3}),
\begin{equation}
{}_{3}F_2 \left(\begin{array}{c|}
1+a_1 \ep, 1+ a_2 \ep, 1+ \tfrac{1}{2}(a_1+a_2)\ep  \\
\tfrac{3}{2}+ \tfrac{1}{2}(a_1+a_2)\ep, 2+(a_1+a_2)\ep \end{array} ~z
\right) =
(1-z) \left[
{}_{2}F_1 \left(\begin{array}{c|} 1+\tfrac{1}{2} a_1 \ep,
1+\tfrac{1}{2} a_2 \ep \\
\tfrac{3}{2}+\tfrac{1}{2}(a_1+a_2)\ep \end{array} ~z \right)
\right]^2 ,
\label{less_trivial}
\end{equation}
which reduces the $_3F_2$ function to a square of the $_2F_1$
function of the type (\ref{2F1_app}).

Another interesting relation (which follows from
Eq.~(20) on p.~498 of \cite{PBM3}) reads
\begin{eqnarray}
&& \hspace*{-10mm}
{}_{3}F_2 \left(\begin{array}{c|}   
1+a_1 \ep, 1+ a_2 \ep, 1+ \tfrac{1}{2}(a_1+a_2)\ep  \\
\tfrac{3}{2}+ \tfrac{1}{2}(a_1+a_2)\ep, 1+(a_1+a_2)\ep \end{array} ~z
\right) =  
{}_{2}F_1 \left(\begin{array}{c|} 1+\tfrac{1}{2}a_1\ep,
1+\tfrac{1}{2} a_2 \ep \\
\tfrac{3}{2}+\tfrac{1}{2}(a_1+a_2)\ep \end{array} ~z \right)
\nonumber \\ &&
\times \left\{ 1 + \frac{a_1 a_2 \ep^2 z}
{2\left[1+(a_1+a_2)\ep\right]}\;
{}_{3}F_2 \left(\begin{array}{c|}
1+\tfrac{1}{2} a_1 \ep, 1+\tfrac{1}{2} a_2 \ep, 1 \\
\tfrac{3}{2}+ \tfrac{1}{2}(a_1+a_2)\ep, \; 2 \end{array} ~z \right)
\right\} \; .
\label{even_less_trivial}
\end{eqnarray}
This relation gives a possibility to cross-check results
for the $\ep$-expansion of both $_3F_2$ functions presented
below, together with the result for the $_2F_1$ function.

We note that the results for some other values of the parameters
can be obtained using relations between contiguous $_3F_2$ functions,
like (see Eq.~(26) on p.~440 of~\cite{PBM3})
\begin{eqnarray}
(1+a_3\ep)\;
{}_3F_2\left(\begin{array}{c|}
1\!+\!a_1\ep, 1\!+\!a_2\ep, 2\!+\!a_3\ep \\
\tfrac{3}{2}+b\ep, 2+c\ep \end{array} ~z \right)
&=& (1+c\ep)\;
{}_3F_2\left(\begin{array}{c|}
1\!+\!a_1\ep, 1\!+\!a_2\ep, 1\!+\!a_3\ep \\
\tfrac{3}{2}+b\ep, 1+c\ep \end{array} ~z \right)
\nonumber \\ && \hspace*{-13mm}
+(a_3-c)\ep \;
{}_3F_2\left(\begin{array}{c|}
1\!+\!a_1\ep, 1\!+\!a_2\ep, 1\!+\!a_3\ep \\
\tfrac{3}{2}+b\ep, 2+c\ep \end{array} ~z \right) .
\hspace*{10mm}
\end{eqnarray}
This connection makes it possible to use, as a check, another
special case (see case (iii) in Table~I of \cite{karlsson}),
\begin{eqnarray}
&& \hspace*{-7mm}
{}_{3}F_2 \left(\begin{array}{c|}   
1 +  a_1 \ep, 1+ a_2 \ep, 2 - a_2 \ep \\
\tfrac{3}{2}+ \tfrac{1}{2}(a_1 \!+\! a_2)\ep, 
2+\tfrac{1}{2}(a_1\!-\!a_2)\ep \end{array} ~\frac{1}{4}
\right) =  
\frac{2^{2+\tfrac{2}{3} a_1\ep}
\Gamma\left(\tfrac{3}{2}+\tfrac{1}{2}(a_1\!+\!a_2)\ep \right)
 \Gamma\left(2+\tfrac{1}{2}(a_1\!-\!a_2)\ep \right)}
{3 a_2 \ep (1-a_2\ep) \Gamma(1+a_1\ep)}
\nonumber \\ && \hspace*{-7mm}
\times \left\{
\frac{ 2^\tfrac{2}{3}\Gamma\left(\tfrac{1}{3}+ \tfrac{1}{3}a_1\ep \right)}
      {\Gamma\left(\tfrac{2}{3}+\tfrac{1}{6}(a_1\!-\!3a_2)\ep\right)
       \Gamma\left(\tfrac{1}{6}+\tfrac{1}{6}(a_1\!+\!3a_2)\ep \right)}
-\frac{\Gamma(1+\tfrac{1}{3}a_1\ep)}
      {\Gamma\left(1+\tfrac{1}{6}(a_1\!-\!3a_2)\ep\right)
       \Gamma\left(\tfrac{1}{2}+\tfrac{1}{6}(a_1\!+\!3a_2)\ep \right)}
\right\}.
\nonumber \\ && \hspace*{-7mm}
\end{eqnarray}

Again,
using the procedure described in Appendix~A.1, we 
obtain a few terms of the $\ep$-expansion for general
values of $a_i$, $b$ and $c$. For the case ${\cal C}=2$ we find
\begin{eqnarray}
&& \hspace*{-7mm}
 {}_{3}F_2 \left(\begin{array}{c|} 1 + a_1 \ep, 1 + a_2\ep, 1 + a_3 \ep \\ 
 \frac{3}{2} + b \ep, 2 + c \ep
\end{array} ~\frac{1}{4} \right)
 = 2 (1+2 b \ep) (1 + c \ep)
\nonumber \\ && \hspace*{-7mm} 
\times \biggl\{
\tfrac{1}{3} \zeta_2 
+ \ep \biggl[
\tfrac{1}{9} \zeta_3 \left(  11 A_1 - 35 b + c \right)
- \tfrac{2}{9} \pi \Ls{2}{\tfrac{\pi}{3}}
(2 A_1  - 5 b + c )
\biggr]
\nonumber \\ && \hspace*{-7mm} 
+ \ep^2 \biggl[
\tfrac{1}{3} \pi \Ls{3}{\tfrac{2\pi}{3}}
(  2 A_1  - 5 b + c )
( A_1 - b  - c )
- \tfrac{2}{9} \left[ \Ls{2}{\tfrac{\pi}{3}} \right]^2
(  2 A_1  - 5 b + c )
(  2 A_1  - 5 b - 2c )
\nonumber \\ && \hspace*{-7mm} 
+ \zeta_4 \biggl(  
- \tfrac{170}{9} b A_1
- \tfrac{139}{54} c A_1
+ \tfrac{377}{36} b c 
+ \tfrac{160}{9} b^2  
- \tfrac{211}{108} c^2 
+ \tfrac{1085}{216} A_1^2 
- \tfrac{5}{216} A_2
\biggr)
\biggr]
\nonumber \\ && \hspace*{-7mm} 
+ \ep^3 \biggl[
\tfrac{1}{24} \chi_5  
(  2 A_1 \!- 5 b + c ) (  2 A_1 \!- 5 b - 2c ) (  A_1\!- b - c )
- \tfrac{1}{3} \pi \Ls{4}{\tfrac{2\pi}{3}}
b (  2 A_1\! - 5 b + c ) (  A_1\! - b - c )
\nonumber \\ && \hspace*{-7mm} 
- \tfrac{1}{162} \pi \zeta_2  \Ls{2}{\tfrac{\pi}{3}} 
(  2 A_1  - 5 b + c ) 
(109b^2 - 23 b A_1 + 23 b c + 2 c^2 - 2 A_2 )
\nonumber \\ && \hspace*{-7mm} 
+ \zeta_2 \zeta_3 
\biggl(  
\tfrac{437}{162} b  A_1^2
- \tfrac{2059}{648} b c A_1
- \tfrac{69}{8} b^2 A_1
+ \tfrac{13}{36} c A_1^2
+ \tfrac{23}{162} A_1 A_2
- \tfrac{167}{162} b A_2
+ \tfrac{31}{162} c A_2 
\nonumber \\ && \hspace*{-7mm} 
+ \tfrac{8}{27} A_3
- \tfrac{13}{81} c^2 A_1
- \tfrac{773}{648} b c^2 
+ \tfrac{833}{108} b^2 c 
+ \tfrac{3361}{648} b^3
- \tfrac{53}{324} c^3
\biggr)
\nonumber \\ && \hspace*{-7mm} 
+ \zeta_5
\biggl(
  \tfrac{1903}{72} b^2 A_1
- \tfrac{5695}{648} b c A_1
- \tfrac{529}{162} b A_1^2
+ \tfrac{47}{36} c A_1^2 
+ \tfrac{17}{81} c A_2
+ \tfrac{53}{162} A_1 A_2
+ \tfrac{58}{27} A_3
\nonumber \\ && \hspace*{-7mm} 
+ \tfrac{164}{81} c^2 A_1
- \tfrac{1271}{162} b A_2
- \tfrac{4217}{648} b c^2 
+ \tfrac{1445}{108} b^2 c 
- \tfrac{22907}{648} b^3
+ \tfrac{103}{324} c^3
\biggr)
\nonumber \\ && \hspace*{-7mm} 
+ \pi \Ls{4}{\tfrac{\pi}{3}}
\biggl(
 \tfrac{259}{486} b c A_1
- \tfrac{41}{18} b^2 A_1
+  \tfrac{116}{243} b A_1^2
-   \tfrac{20}{243} c^2 A_1
+ \tfrac{136}{243} b A_2
- \tfrac{8}{243} c A_2
\nonumber \\ && \hspace*{-7mm} 
- \tfrac{4}{27} A_3
- \tfrac{10}{243} A_1 A_2
- \tfrac{1}{9} c A_1^2
+ \tfrac{53}{486} b c^2 
- \tfrac{29}{81} b^2 c 
+ \tfrac{911}{486} b^3
- \tfrac{7}{243} c^3
\biggr)
\biggr]
+ {\cal O}(\ep^4)
\biggr \} \; ,
\label{3F2_PSLQ_1}
\end{eqnarray}
where $A_j\equiv\sum_{i=1}^3 a_i^j$,
while $\chi_5$ is defined in Eq.~(\ref{sigma_32}).
In particular, we have checked that 
this result obeys the condition
(\ref{less_trivial}), where the $\ep$-expansion of the
resulting $_2F_1$ function is given in Eq.~(\ref{2F1_PSLQ})).

For another case of interest, ${\cal C}=1$, we find
\begin{eqnarray}
&& \hspace*{-7mm}
 {}_{3}F_2 \left(\begin{array}{c|} 1 + a_1 \ep, 1 + a_2\ep, 1 + a_3 \ep \\ 
 \frac{3}{2} + b \ep, 1 + c \ep  
\end{array} ~\frac{1}{4} \right)
 = \frac{2 (1+2 b \ep)}{3^{\tfrac{1}{2}+A_1\ep-b\ep-c\ep}} 
\nonumber \\ && \hspace*{-7mm}
\times \biggl\{
\tfrac{1}{3} \pi
+ \ep  (2 A_1 - 5b -2 c) \sum_{j=0}^{\infty} 
\frac{(2\ep)^j}{(j+1)!} (A_1-b-c)^j 
\left[ \Ls{j+2}{\tfrac{ 2\pi}{3}} -  \Ls{j+2}{\pi} \right]
\nonumber \\ && \hspace*{-7mm} 
+  \ep^2 
\pi \zeta_2 \tfrac{1}{54} 
\left[  A_1^2-A_2 + (c-A_1)(15b+2c) + 57 b^2 \right]
\nonumber \\ && \hspace*{-7mm} 
+  \ep^3 \biggl[
\pi \zeta_3 \left(  
  \tfrac{541}{81} A_1 b c 
+ \tfrac{13}{9} A_1 b^2 
+ \tfrac{242}{81} A_1 c^2 
- \tfrac{49}{162} A_1 A_2 
- \tfrac{541}{162} A_1^2 b 
- \tfrac{145}{54} A_1^2 c 
+ \tfrac{145}{162} A_1^3 
\right.
\nonumber \\ && \hspace*{-7mm} 
\left.
- \tfrac{475}{81} b c^2 
+ \tfrac{409}{162} b A_2 
- \tfrac{13}{9} b^2 c 
- \tfrac{17}{27} b^3 
+ \tfrac{49}{162} c A_2
- \tfrac{49}{81} c^3 
- \tfrac{16}{27} A_3
\right) 
\nonumber \\ && \hspace*{-7mm} 
+ \zeta_2 \Ls{2}{\tfrac{\pi}{3}} \left(  
 \tfrac{10}{9} A_1 b c 
\!+\! \tfrac{2}{9} A_1 c^2 
\!+\! \tfrac{1}{9} A_1 A_2
\!-\! \tfrac{5}{9} A_1^2 b 
\!-\! \tfrac{1}{3} A_1^2 c 
\!+\! \tfrac{1}{9} A_1^3 
\!-\! \tfrac{10}{9} b c^2 
\!+\! \tfrac{5}{9} b A_2 
\!-\! \tfrac{1}{9} c A_2 
\!+\! \tfrac{2}{9} c^3 
\!-\! \tfrac{2}{9} A_3
\right) 
\nonumber \\ && \hspace*{-7mm} 
+ \Ls{4}{\tfrac{\pi}{3}} \left(  
\tfrac{4}{9} A_1 b^2 
- \tfrac{320}{81} A_1 b c 
- \tfrac{148}{81} A_1 c^2 
+ \tfrac{10}{81} A_1 A_2 
+ \tfrac{160}{81} A_1^2 b 
+ \tfrac{46}{27} A_1^2 c 
- \tfrac{46}{81} A_1^3 
\right.
\nonumber \\ && \hspace*{-7mm} 
\left.
+ \tfrac{296}{81} b c^2 
- \tfrac{136}{81} b A_2 
- \tfrac{4}{9} b^2 c 
- \tfrac{56}{27} b^3 
- \tfrac{10}{81} c A_2 
+ \tfrac{20}{81} c^3 
+ \tfrac{4}{9} A_3
\right) 
\biggr]
\nonumber \\ && \hspace*{-7mm} 
+  \ep^4 \biggl[
\zeta_2  \Ls{3}{\tfrac{2 \pi}{3}} (A_1-b-c) 
\left(  4 A_1^3 
- \tfrac{85}{6} A_1^2 b 
- 12 A_1^2 c 
+ \tfrac{85}{3} A_1 b c 
- 3 A_1 b^2 
\right.
\nonumber \\ && \hspace*{-7mm} 
\left.
- \tfrac{2}{3} A_1 A_2 
+ \tfrac{38}{3} A_1 c^2 
- \tfrac{79}{3} b c^2 
+ 3 b^2 c 
+ 14 b^3 
+ \tfrac{2}{3} c A_2 
+ \tfrac{73}{6} b A_2 
- \tfrac{10}{3} A_3 
- \tfrac{4}{3} c^3
\right)  
\nonumber \\ && \hspace*{-7mm} 
+ \zeta_3 \Ls{2}{\tfrac{\pi}{3}}
\left(
 \tfrac{8}{9} A_1 c^3 - \tfrac{38}{27} A_1 A_3 
- \tfrac{170}{27} A_1 b c^2 - \tfrac{40}{27} A_1 A_2 b 
- \tfrac{350}{27} A_1 b^2 c - \tfrac{34}{27} A_1 A_2 c 
\right.
\nonumber \\ && \hspace*{-7mm} 
\left.
+ \tfrac{70}{9} A_1^2 b c 
+ \tfrac{175}{27} A_1^2 b^2 
+ \tfrac{19}{27} A_1^2 c^2 
+ \tfrac{35}{27} A_1^2 A_2 
- \tfrac{70}{27} A_1^3 b 
- \tfrac{8}{9} A_1^3 c 
+ \tfrac{1}{9} A_1^4 
+ \tfrac{40}{27} b c A_2 
- \tfrac{80}{27} b c^3 
\right.
\nonumber \\ && \hspace*{-7mm} 
\left.
+ \tfrac{110}{27} b A_3 
+ \tfrac{350}{27} b^2 c^2 
- \tfrac{175}{27} b^2 A_2 
+ \tfrac{14}{27} c A_3 
- \tfrac{1}{27} c^2 A_2 
+ \tfrac{2}{27} c^4 
\right)  
\nonumber \\ && \hspace*{-7mm} 
+ \pi \zeta_4 
\left(
\tfrac{78145}{1296} A_1 b c^2 
- \tfrac{39377}{2592} A_1 A_2 b 
+ \tfrac{11069}{648} A_1 b^2 c 
- \tfrac{1753}{144} A_1 b^3 
- \tfrac{1843}{432} A_1 A_2 c 
+ \tfrac{11021}{648} A_1 c^3 
\right.
\nonumber \\ && \hspace*{-7mm} 
\left.
+ \tfrac{5279}{432} A_1 A_3 
- \tfrac{38971}{864} A_1^2 b c 
- \tfrac{11069}{1296} A_1^2 b^2
- \tfrac{6931}{288} A_1^2 c^2 
- \tfrac{1619}{288} A_1^2 A_2 
+ \tfrac{38971}{2592} A_1^3 b
+ \tfrac{9475}{648} A_1^3 c 
\right.
\nonumber \\ && \hspace*{-7mm} 
\left.
- \tfrac{6125}{2592} A_1^4 
+ \tfrac{39377}{2592} b c A_2 
- \tfrac{37901}{1296} b c^3 
- \tfrac{1273}{1296} b A_3 
- \tfrac{31867}{1296} b^2 c^2 
+ \tfrac{10399}{648} b^2 A_2 
+ \tfrac{1753}{144} b^3 c 
\right.
\nonumber \\ && \hspace*{-7mm} 
\left.
+ \tfrac{209}{36} b^4 
- \tfrac{2437}{1296} c A_3 
+ \tfrac{3665}{2592} c^2 A_2
- \tfrac{1223}{432} c^4 
+ \tfrac{5491}{1296} A_2^2 
- \tfrac{305}{36} A_4
\right)  
\nonumber \\ && \hspace*{-7mm} 
+ \pi \left[ \Ls{2}{\tfrac{\pi}{3}} \right]^2 \left(
 \tfrac{20}{27} A_1 b c^2 
+ \tfrac{10}{27} A_1 A_2 b 
+ \tfrac{50}{27} A_1 b^2 c 
+ \tfrac{7}{27} A_1 A_2 c 
- \tfrac{2}{9} A_1 c^3 
+ \tfrac{56}{81} A_1 A_3 
- \tfrac{10}{9} A_1^2 b c   
\right.
\nonumber \\ && \hspace*{-7mm} 
\left.
- \tfrac{25}{27} A_1^2 b^2 
- \tfrac{1}{27} A_1^2 c^2 
- \tfrac{16}{27} A_1^2 A_2 
+ \tfrac{10}{27} A_1^3 b 
+ \tfrac{1}{9} A_1^3 c 
+ \tfrac{4}{81} A_1^4   
- \tfrac{10}{27} b c A_2
+ \tfrac{20}{27} b c^3   
\right.
\nonumber \\ && \hspace*{-7mm} 
\left.
- \tfrac{20}{27} b A_3 
- \tfrac{50}{27} b^2 c^2 
+ \tfrac{25}{27} b^2 A_2 
- \tfrac{2}{27} c A_3 
+  \tfrac{1}{27} c^2 A_2 
-  \tfrac{2}{27} c^4 
+  \tfrac{4}{27} A_2^2 
-  \tfrac{8}{27} A_4
\right)  
\nonumber \\ && \hspace*{-7mm} 
+ \Ls{5}{\tfrac{\pi}{3}}
\left(
  \tfrac{2729}{162} A_1 b c^2 
- \tfrac{1231}{324} A_1 A_2 b 
+ \tfrac{433}{81} A_1 b^2 c 
- \tfrac{53}{18} A_1 b^3 
- \tfrac{131}{162} A_1 A_2 c 
+ \tfrac{113}{27} A_1 c^3 
+ \tfrac{1913}{486} A_1 A_3  
\right.
\nonumber \\ && \hspace*{-7mm} 
\left.
- \tfrac{1409}{108} A_1^2 b c\!
- \tfrac{433}{162} A_1^2 b^2 
- \tfrac{2093}{324} A_1^2 c^2 
- \tfrac{665}{324} A_1^2 A_2\!
+ \tfrac{1409}{324} A_1^3 b 
+ \tfrac{109}{27} A_1^3 c\! 
- \tfrac{583}{972} A_1^4   
+ \tfrac{1231}{324} b c A_2\!
- \tfrac{1249}{162} b c^3  
\right.
\nonumber \\ && \hspace*{-7mm} 
\left.
- \tfrac{71}{162} b  A_3 
- \tfrac{1019}{162} b^2 c^2 
+ \tfrac{293}{81} b^2 A_2 
+ \tfrac{53}{18} b^3 c 
+ \tfrac{11}{9} b^4 
- \tfrac{107}{162} c A_3 
+ \tfrac{95}{324} c^2 A_2 
- \tfrac{95}{162} c^4 
+ \tfrac{104}{81} A_2^2 
- \tfrac{208}{81} A_4 
\right)  
\nonumber \\ && \hspace*{-7mm}
- \left[ \pi \LS{4}{1}{\tfrac{2 \pi}{3}}  
   - \tfrac{3}{4} \LS{5}{2}{\tfrac{2 \pi}{3}}\right] 
\left( A_1 -b -c \right)
\left(
\tfrac{23}{12} A_1^3 
- \tfrac{20}{3} A_1^2 b 
- \tfrac{23}{4} A_1^2 c 
+ \tfrac{40}{3} A_1 b c 
- \tfrac{3}{2} A_1 b^2
\right.
\nonumber \\ && \hspace*{-7mm} 
\left.
- \tfrac{5}{12} A_1 A_2
+ \tfrac{37}{6} A_1 c^2 
- \tfrac{37}{3} b c^2 
+ \tfrac{3}{2} b^2 c 
+ 7 b^3 
+ \tfrac{5}{12} c A_2 
+ \tfrac{17}{3}  b A_2 
- \tfrac{3}{2} A_3 
- \tfrac{5}{6} c^3
\right)  
\biggr]
+ {\cal O}(\ep^5)
\biggr \} \; ,
\end{eqnarray}

\subsection{The $_4F_3$ function}

Finally, we present some results for the hypergeometric function 
\begin{equation}
 {}_{4}F_3 \left(\begin{array}{c|}
1 + a_1 \ep, 1 + a_2\ep, 1 + a_3\ep, 1+a_4\ep \\
\frac{3}{2} + b \ep, 2+c_1\ep, 2+c_2\ep \end{array} ~z \right) \; .
\label{4F3_app} 
\end{equation}

There is an interesting special case, $c_1=a_1$ and $c_2=a_2$,
when this function (\ref{4F3_app}) reduces to a combination 
of two $_3F_2$ functions of the type (\ref{3F2_app}) with $C=2$
(see, e.g., Eq.~(4) in p.~497 of \cite{PBM3}),
\begin{eqnarray}
&& \hspace*{-7mm}
\ep (a_2 -a_1) ~{}_{4}F_3 \left(\begin{array}{c|}
1+a_1 \ep, 1+ a_2 \ep, 1+ a_3 \ep, 1 + a_4 \ep  \\
\frac{3}{2}+ b \ep, 2+ a_1 \ep, 2 + a_2 \ep \end{array} ~z \right) 
\nonumber \\ && \hspace*{-7mm}
= (1\!+\!a_2 \ep )
~{}_{3}F_2 \left(\begin{array}{c|} 
\! 1\!+\! a_1 \ep, 1\!+\! a_3 \ep, 1\!+\! a_4 \ep \! \\
\frac{3}{2} + b \ep, 2 + a_1 \ep \end{array} ~z \right)
\!-\! (1\!+\!a_1 \ep )
~{}_{3}F_2 \left(\begin{array}{c|} 
\! 1\!+\! a_2 \ep, 1\!+\! a_3 \ep, 1 \!+\! a_4 \ep \! \\
\frac{3}{2} + b \ep, 2 + a_2 \ep \end{array} ~z \right) \; .
\hspace*{4mm}
\end{eqnarray}

As in the previous cases, using the procedure described
in Appendix~A.1, we find a few terms
of the $\ep$-expansion for the general values of $a_i$,
$b$ and $c_i$,
\begin{eqnarray}
&& \hspace*{-7mm}
 {}_{4}F_3 \left(\begin{array}{c|} 1 + a_1 \ep, 1 + a_2\ep, 
1 + a_3\ep, 1+a_4\ep \\ \frac{3}{2} + b \ep,
 2 + c_1 \ep, 2 + c_2 \ep \end{array} ~\frac{1}{4} \right)
 = 2 (1\!+\!2 b \ep) (1 \!+\! c_1 \ep) (1 \!+\! c_2 \ep)
\biggl\{
\tfrac{2}{3} \pi \Ls{2}{\tfrac{\pi}{3}} - \tfrac{4}{3} \zeta_3 
\nonumber \\ && \hspace*{-7mm} 
+ \ep \biggl[
\pi \Ls{3}{\tfrac{2\pi}{3}}
(b  - A_1 + C_1 )
+ \tfrac{2}{3} \left[ \Ls{2}{\tfrac{\pi}{3}} \right]^2
(  2 A_1 - 5 b - 2 C_1 )
+ \zeta_4 \biggl(  
\tfrac{101}{12} b - \tfrac{269}{36}  A_1  +  7 C_1 
\biggr)
\biggr]
\nonumber \\ && \hspace*{-7mm} 
+ \ep^2 \biggl[
-\chi_5 \tfrac{1}{8} 
(  2 A_1  - 5 b -  2 C_1 )
(   A_1 - b - C_1)
+ \pi \Ls{4}{\tfrac{2\pi}{3}} b (   A_1   - b - C_1)
\nonumber \\ && \hspace*{-7mm} 
+ \pi \zeta_2 \Ls{2}{\tfrac{\pi}{3}} \tfrac{1}{54}
\biggl(109 b^2 - 23 b A_1 + 23 b C_1 -  2 A_2 +  2 C_2 \biggr)
\nonumber \\ && \hspace*{-7mm} 
+ \zeta_2 \zeta_3 
\biggl(
\tfrac{1063}{216} b C_1
+ \tfrac{1}{54}  C_1 A_1 
- \tfrac{479}{216} b  A_1
+ \tfrac{1}{108} C_1^2
- \tfrac{13}{36} A_1^2
- \tfrac{1}{3} A_2
+ \tfrac{197}{216} b^2
\biggr)
\nonumber \\ && \hspace*{-7mm} 
+ \zeta_5
\biggl(
  \tfrac{1103}{72} b A_1
- \tfrac{127}{54}  C_1 A_1
+ \tfrac{53}{108} C_1^2
- \tfrac{71}{27} C_2
- \tfrac{29}{54} A_2
- \tfrac{47}{36} A_1^2 
+ \tfrac{2395}{216} b C_1
- \tfrac{8599}{216} b^2
\biggr)
\nonumber \\ && \hspace*{-7mm} 
+  \pi \Ls{4}{\tfrac{\pi}{3}}
\biggl(
\tfrac{10}{81}  C_1 A_1\!
- \! \tfrac{31}{162} b C_1\!
- \! \tfrac{241}{162} b A_1\!
+ \! \tfrac{1}{9} A_1^2\!
+ \! \tfrac{2}{27} A_2\!
+ \! \tfrac{427}{162} b^2\!
+ \! \tfrac{4}{27} C_2\!
- \! \tfrac{1}{81} C_1^2\!
\biggr)
\biggr]
+ {\cal O}(\ep^3)
\biggr \} \; ,
\hspace*{7mm}
\end{eqnarray}
where $A_j\equiv\sum_{i=1}^4 a_i^j$, $C_j\equiv\sum_{i=1}^2 c_i^j$,
while $\chi_5$ is defined in Eq.~(\ref{sigma_32}).


\end{document}